\documentclass{aa}
\usepackage{natbib,graphicx}
\usepackage{rotating}
\usepackage{amssymb, amsmath}
\usepackage[varg]{txfonts}
\usepackage{color}
\def\mathnew{\mathsurround=0pt}
\def\simov#1#2{\lower .5pt\vbox{\baselineskip0pt \lineskip-.5pt
\ialign{$\mathnew#1\hfil##\hfil$\crcr#2\crcr\sim\crcr}}}

\def\arcsec{\hbox{$^{\prime\prime}$}}

\def\MeV{Me\kern-0.11em V}
\def\keV{ke\kern-0.11em V}
\def\arcsec{\hbox{$^{\prime\prime}$}}

\def\ha{H$\alpha$}
\def\lya{Ly$\alpha$}
\def\Ca{C_{{\rm H}\alpha}}
\def\Cr{C_{R_C}}
\def\Cl{C_{\rm line}}
\def\Cca{C_{{\rm cont,H}\alpha}}
\def\Ccr{C_{{\rm cont,}R_C}}
\def\Ta{T_{{\rm H}\alpha}}
\def\Tr{T_{R_C}}

\definecolor{grey}{rgb}{0.5,0.6,0.7}

\begin{document}

\title{An optical view of the filament region of Abell~85
\thanks{Based on observations obtained with: 1)~MegaPrime/MegaCam, a
joint project of CFHT and CEA/DAPNIA, at the Canada-France-Hawaii
Telescope (CFHT) which is operated by the National Research Council
(NRC) of Canada, the Institut National des Sciences de l'Univers of
the Centre National de la Recherche Scientifique (CNRS) of France, and
the University of Hawaii, 2)~the European Southern Observatory, 3)~the
Anglo-Australian Observatory. This work is also based in part on data
products produced at TERAPIX and the Canadian Astronomy Data Centre as
part of the Canada-France-Hawaii Telescope Legacy Survey, a
collaborative project of NRC and CNRS. We have also made use of the
NASA/IPAC Extragalactic (NED) Database.}}

\author{G. Bou\'e \inst{1} \and
F. Durret \inst{1} \and
C. Adami \inst{2} \and
G.A.~Mamon \inst{1} \and
O.~Ilbert \inst{3,2} \and
V.~Cayatte \inst{4} 
}

\offprints{G. Bou\'e \email{boue@iap.fr}}

\institute{
Institut d'Astrophysique de Paris (UMR 7095: CNRS \& Universit\'e Pierre 
et Marie Curie), 98bis Bd Arago, F--75014 Paris, France
\and
LAM, Traverse du Siphon, F--13012 Marseille, France
\and
Institute for Astronomy, 2680 Woodlawn Dr., University of Hawaii,
Honolulu, Hawaii 96822, USA
\and
Observatoire de Paris, section Meudon, LUTH, CNRS-UMR 8102, Universit\'e Paris
7, 5 Pl. Janssen, F--92195 Meudon, France
}

\date{Accepted . Received ; Draft printed: \today}

\authorrunning{Bou\'e et al.}

\titlerunning{Filament of Abell 85}

\abstract
{}
{We present an optical investigation of the Abell~85 cluster filament
($z=0.055$) previously interpreted in X-rays as groups falling on to
the main cluster. We compare the distribution of galaxies with the X-ray
filament, and investigate the galaxy luminosity functions in several
bands and in several regions.  We search for galaxies where star
formation may have been triggered by interactions with intracluster gas
or tidal pressure due to the cluster potential when entering the
cluster. }
{Our analysis is based on images covering the South tip of Abell 85
and its infalling filament, obtained with 
 CFHT MegaPrime/MegaCam (1$\times$1 deg$^2$ field) in four bands
($u^*, g', r', i'$) and ESO 2.2m WFI
(38$\times$36~arcmin$^2$ field) in a narrow band filter corresponding
to the redshifted \ha\ line and in an $R_C$ broad band filter.  
The LFs are estimated by statistically
subtracting a reference field.  Background contamination is minimized
by cutting out galaxies redder than the observed red sequence in the
$g'-i'$ versus $i'$ colour-magnitude diagram.  }
{ The galaxy distribution shows a significantly flattened cluster, whose principal
  axis is slightly offset from the X-ray filament.  The analysis of the
  broad band galaxy luminosity functions shows that the filament region
  is well populated. The filament is also independently detected as a
  gravitationally bound structure by the Serna \& Gerbal (1996)
  hierarchical method.  101 galaxies are detected in the \ha\ filter,
  among which 23 have spectroscopic redshifts in the cluster, 2 have
  spectroscopic redshifts higher than the cluster and 58 have
  photometric redshifts that tend to indicate that they are background
  objects. One galaxy that is not detected in the \ha\ filter probably
  because of the filter low wavelength cut but shows \ha\ emission in
  its SDSS spectrum in the cluster redshift range has been added to our
  sample. The 24 galaxies with spectroscopic redshifts in the cluster
  are mostly concentrated in the South part of the cluster and along the
  filament.  }
{We find a number of galaxies showing evidence for star formation in the
filament, and all our results are consistent with the previous
hypothesis that the X-ray filament in Abell~85 is a gravitationally
bound structure made of groups falling on to the main cluster. }


\maketitle

\section{Introduction}\label{sec:intro}

Cosmological simulations of the large-scale structure of the Universe 
display the filamentary nature of the large-scale galaxy distribution
(e.g. \citealp{Springel+05}), and the observed large-scale galaxy
distribution is consistent with this picture, even though filaments are more
difficult to see in redshift 
  space (e.g. \citealp{Pimbblet05}). X ray observations of the nearby rich
  cluster \object{Abell~85} 
  highlight a filament of hot gas extending towards the South East to near
  one virial radius  \citep{Durret+98,Durret+03,DLNF05}.


One would obviously like to know if this X-ray filament can be traced in the
  galaxy distribution. If so, one would expect that this filament would be a
  preferential route for the infall of galaxies onto the cluster, even within
  the virialised region of the cluster.  The
influence of infall is not always well understood, except for a few
clusters, such as e.g. Coma \citep{Adami+07}.
One would also like to know if the
  filament region follows the same morphology-density relation as seen in
  clusters \citep{Dressler80,Dressler+97}, or whether the filament constitutes 
a special environment. Similarly, do
  the galaxies in the filament dipslay the same specific rates of star
  formation as seen in other cluster regions of the same density, or is the
  star formation enhanced or quenched? Indeed,
star formation
can be triggered when the groups of the filament enter the cluster or dense
areas, due to environmental effects such as ram pressure from the
intracluster gas or tidal pressure due to the cluster potential
(\citealp{Bekki99}).

The galaxy population in the X-ray filament is easily 
traced in maps of the projected distribution of galaxies up to a given
apparent magnitude limit and with a selection in redshifts to remove obvious
cluster outliers. Alternatively,
galaxy luminosity functions in several wavebands are a good tool to
sample the history of the faint galaxy population
(e.g. \citealp{Adami+07} and references therein) including star
formation history, evolutionary processes and environmental
effects. In particular, the faint-end slopes of galaxy luminosity
functions (LFs) in clusters of galaxies have been observed in some
cases to vary with clustercentric distance and are expected to be
influenced by physical processes (mergers, tides) affecting cluster
galaxies (as summarized e.g. by \citealp{Boue+08}, hereafter B08).

The \ha\ line is a good indicator of star formation and has been
detected in a number of galaxies in nearby clusters.  The first
pioneering work on this topic was due to \citet{MWI88}, \citet{MW93},
and \citet{MWP98}, who performed the first \ha\ surveys in a sample of
clusters with an objective prism. Based on this survey, \citet{MWP98}
and \citet{MW00} analyzed tidally induced star formation in several
clusters; they found spatial variations, both within a cluster and
from one cluster to another: starburst emission in spirals increases
from regions of lower to higher density, and from clusters with lower
to higher central galaxy space density.  \cite{MW05} were then able to
show that the frequency of emission line galaxies (ELGs) is similar
for field and cluster galaxies of all types, and that for galaxies of
a given morphological type the fraction of ELGs is independent of
environment.  A large \ha\ survey was performed on several nearby
clusters by \cite{BIVG02} and \cite{Gavazzi+02,Gavazzi+06}. They
analyzed several trends with radius and found in particular that
luminous galaxies show a decrease in their average \ha\ equivalent
width in the inner $\sim 1$ virial radius, while low-luminosity
galaxies do not show this trend.  Large \ha\ surveys have also allowed
to estimate \ha\ luminosity functions and star formation rates in some
of these clusters (\citealp{IglesiasParamo+02,Umeda+04}).
Observations of clusters in \ha\ have also revealed some interesting
features. For example, a few \ha\ tails and filaments as well as
intracluster HII regions have been detected in a few clusters such as
Abell~1795 (\citealp{CSF05}), Coma (\citealp{YKYF07}) or Abell~3627
(\citealp{SDV07}).  A starbursting compact group was also found to be
falling on to Abell~1367, where complex trails of ionized gas behind
the galaxies were detected (\citealp{Cortese+06}).

We present here a detailed optical analysis of the filament region of
Abell~85.  This cluster is at a redshift of 0.055 and shows a very
complex structure in X-rays, with a main cluster, a South blob and an
extended filament (discovered in X-rays) at least 4~Mpc in
length. Based on ROSAT PSPC and XMM-Newton data, evidence was found
for several merging episodes, one of these still ongoing, as suggested
by the interpretation of the X-ray filament as groups falling on to
the main cluster \citep{Durret+98,Durret+03,DLNF05}. However, the
optical properties of the galaxies composing the X-ray filament have
not been analyzed until now; they may give us clues on the physical
properties of this filament and on the likelihood of the merging
scenario described above.

We have obtained two sets of data: ESO 2.2m WFI
38$\times$36~arcmin$^2$ images in a narrow band filter corresponding
to the wavelength of H$\alpha$ at the cluster redshift and in a broad
band $R_C$ filter to subtract the continuum contribution, covering the
South half of Abell~85 and its filament, and deep 1$\times$1 deg$^2$
field images obtained at CFHT with MegaPrime/MegaCam in four bands
($u^*$$g'$$r'$$i'$) covering the South tip of Abell 85 and the
infalling filament.  Both sets of data sample the filament feeding the
cluster from the Southeast, and the impact region where the filament
is believed to be hitting the cluster itself (this impact region is
indeed hotter in X-rays). The virial radius, defined as the radius where
  the mean mass density is 100 times the critical density of the Universe, is
  $0\fdg65 $, derived by extrapolating the radius of overdensity 500 given by
  \cite{DLNF05}. Thus, our Megacam images (not centered on the cluster)
cover part of Abell~85 and more distant regions, well beyond the
virial radius. The Sloan Digital Sky Survey (SDSS) covers the region
of Abell~85.  We have retrieved all the redshifts available in the
SDSS to build a large redshift catalogue for the region of Abell~85,
as well as all the galaxy spectra in the region covered by our WFI
data.

The paper is organized as follows. We present our Megacam and WFI data
and data reduction in Section 2. In Section 3, we describe our results
on H$\alpha$ imaging and discuss the spatial distribution and
properties of H$\alpha$ emitting galaxies, together with properties
derived from the SDSS data.  In Section 4, we present our results
obtained for the LF in the four broad photometric bands.  In Section
5, we discuss our results concerning the LFs in terms of large scale
environmental effects on the cluster galaxy populations.  Final
conclusions are drawn in Section 6.

We assume a distance of 242.2 Mpc to Abell 85 ($H_0$ = 71 km s$^{-1}$
Mpc$^{-1}$, $\Omega _m$ = 0.27 and $\Omega_ {\Lambda}$ = 0.73). The
distance modulus is 36.92 and the scale is 1.055 kpc arcsec$^{-1}$. We
give magnitudes in the AB system.  
At this distance, the Megacam field of view corresponds to
3.8$\times$3.8 Mpc$^2$, while    
the virial radius  is 2.5~Mpc.

\section{Observations and data reduction}

\subsection{Megacam data reduction}

Abell 85 was observed at CFHT with the large field MegaPrime/MegaCam
camera in October 2004, program 04BF02, P.I. F.~Durret (see
Table~\ref{tab:observation}).  The deep 1$\times$1 deg$^2$ field
images obtained at CFHT with MegaPrime/MegaCam in four bands
($u^*$$g'$$r'$$i'$) cover the South tip of Abell 85 and the infalling
filament. These images were reduced by the Terapix pipeline using the
standard reduction tool configuration. We refer the reader to
http://terapix.iap.fr/ for reduction details.

\begin{table*}
\caption{Observation characteristics: coordinates,
exposure times in seconds (seeing in arcseconds), observation dates,
and program Id.}
\begin{tabular}{l*6{c}}
\hline
Image center coordinates &
$u^*$ & $g'$ & $r'$ & $i'$ & Dates & Program Id \\
00h41m56.5s $-09^\circ53'56''$ & 13440 (1.32\arcsec) & 
$7\,480$ (1.42\arcsec) &
$4\,550$ (1.45\arcsec) & $4\,200$ (0.87\arcsec)  & 10/2004 
& 04BF02 \\
\hline
Image center coordinates & $R_C$ & \ha & & & Dates & Program Id \\
00h41m39.8s $-09^\circ35'15''$ & 2900  (0.93\arcsec) & 5800
(0.85\arcsec) & & & 11/2004 &
074.A-0029B \\
\hline
\end{tabular}
\label{tab:observation}
\end{table*}

Object extraction was made using the SExtractor package \citep{BA96}
in double-image mode.  The CFHTLS pipeline at the Terapix data center
creates a $\chi^2$ image based upon the quadratic sum of the images in
the different wavebands.  Objects are then detected on this image.  In
contrast with the CFHTLS images, our set of $u^*$$g'$$r'$$i'$ images
for Abell~85 presents important differences in their PSFs (see
Table~\ref{tab:observation}).  For this reason, we chose a different
approach from that of the Terapix data center: instead of considering
the $\chi ^2$ image as the reference image, we use the band with the
best seeing in our data: $i'$. Detections were performed in this band
and object characteristics were measured in all bands. The detections
and measures were made using the CFHTLS parameters, among which an
absolute detection threshold of 0.4 ADUs, a minimal detection area of
3 pixels and a 7$\times$7 pixel gaussian convolution filter of 3
pixels of FWHM. In the output catalog, we only kept objects with
semi-minor axes larger than 1 pixel and mean surface brightness within
the half-light radius greater than $\mu=26.25$ in order to remove
artefacts.

\begin{figure}
\centering
\includegraphics[width=8cm,angle=270]{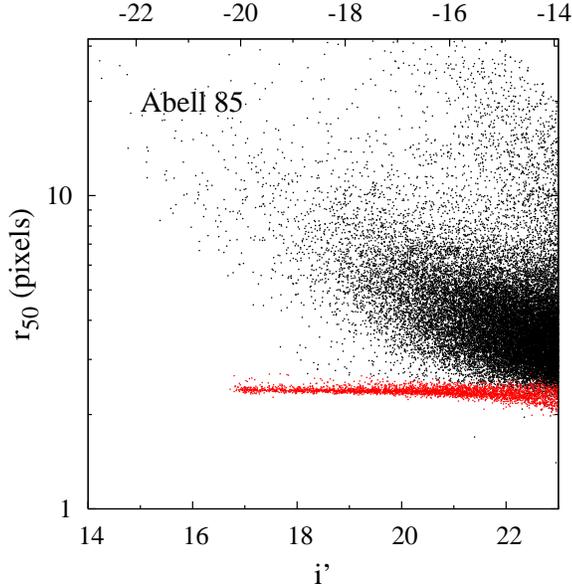}
\caption[]{Star-galaxy separation for the Megacam data: half-light
radius versus apparent magnitude \emph{(bottom axis)} and absolute
magnitude \emph{(top axis)} of the detections on the Abell~85
image. Objects assumed to be stars are plotted in red.  }
\label{fig:stargal}
\end{figure}

\begin{figure}
\centering
\includegraphics[width=7.3cm,angle=270]{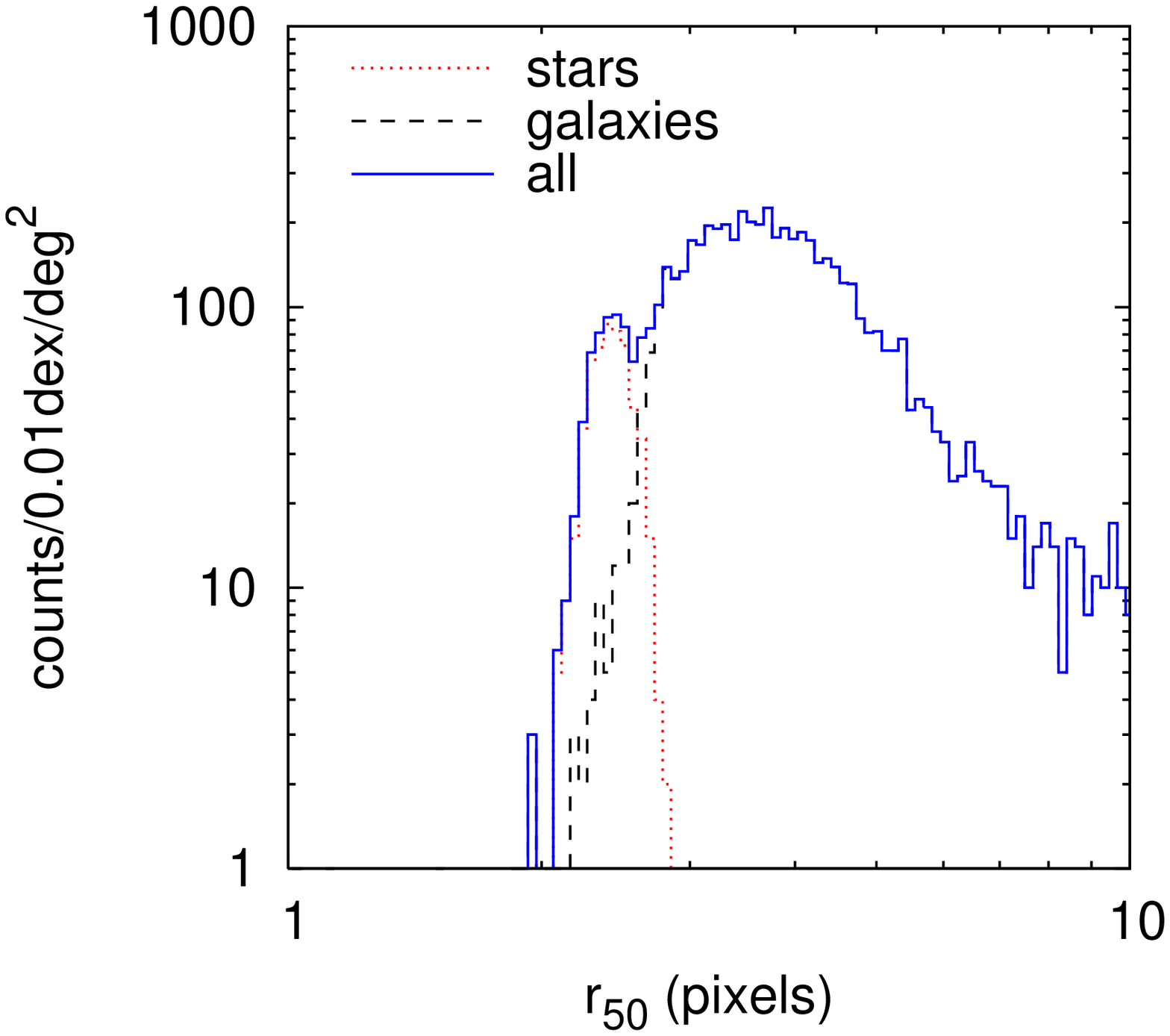}
\includegraphics[width=8cm,angle=270]{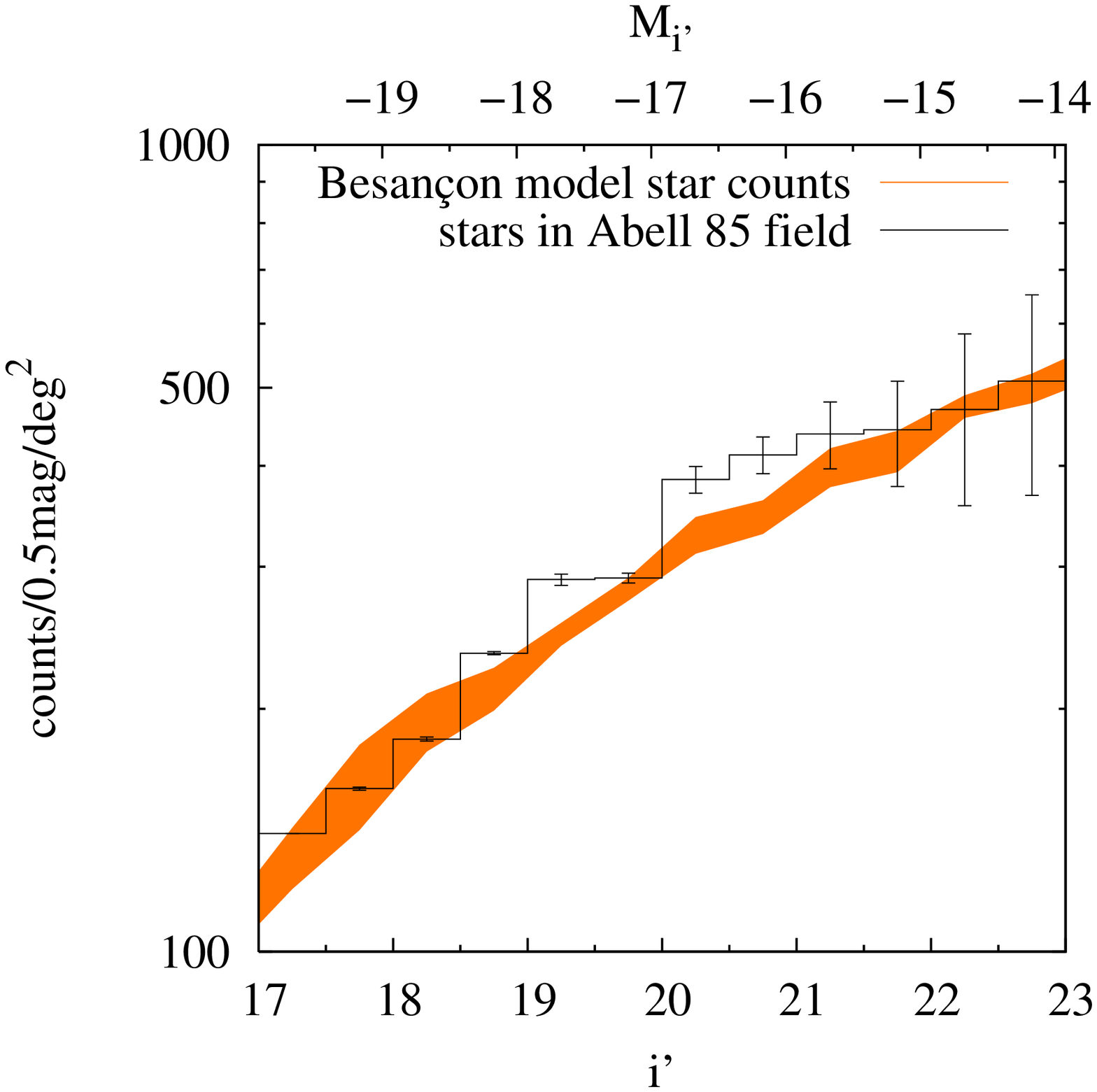}
\caption[]{Megacam data. \emph{Top:} histogram of all detections for
Abell~85 in the range 22$<i'<$22.5 with the distributions of stars and
galaxies obtained using our star-galaxy separation. \emph{Bottom:}
comparison of star counts obtained with the Besan\c{c}on model
\citep{RRDP03} (\emph{orange shaded region}) and from the Abell 85
image (\emph{black histogram}) using our star-galaxy
separation.
}
\label{fig:besancon}
\end{figure}

A detailed description of the way the completeness levels and
reliabilities of our detections were estimated using simulations, as
well as the star-galaxy separation (see Figs.~\ref{fig:stargal} and
\ref{fig:besancon}), magnitude corrections for Galactic extinction and
estimate of the useful area covered by the Megacam images (0.785
deg$^2$) can be found in B08.


For the computation of the luminosity functions, we used the CFHTLS Deep
(D1, D2, D3 and D4, i.e. 4 MegaCam fields) and Wide (W1, W2 and W3, 59
MegaCam fields) as comparison field data, as described in B08.  Note
that we re-extracted object catalogues from each of the $4\times 4$ Deep Field
(DF) images by making the detections in the $i'$ band, as for Abell~85.

Regions of the CFHTLS observed more than once (common areas of 19
Megacam fields) were considered to estimate the magnitude
uncertainties as a function of magnitude in an external way (see B08,
Fig.~4).

\subsection{WFI data reduction}

\begin{figure}
\centering
\includegraphics[width=8cm]{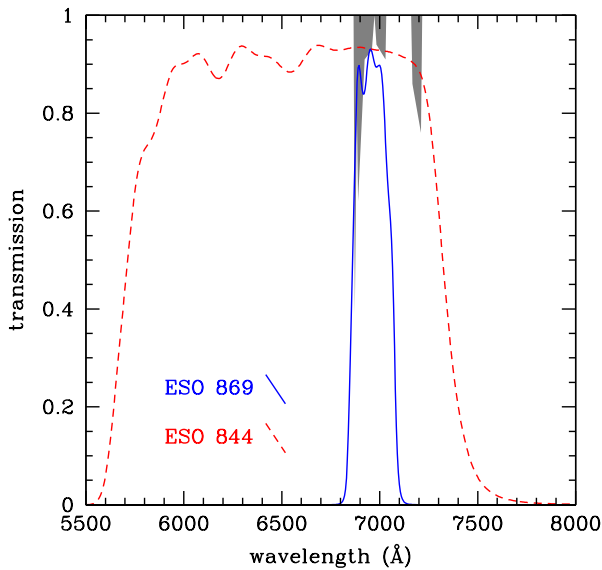}
\caption{Filter transmission for the narrow band (\emph{solid line}) and broad
band (\emph{dotted line}) filters used with the ESO 2.2m and WFI camera.
Also shown \emph{in grey} 
is the position of telluric absorption inferred from a twilight spectrum.}
\label{fig:WFI_filters}
\end{figure}

Imaging observations were performed in service mode with the ESO 2.2m
telescope and the WFI camera (program 074.A-0029B, P.I. F. Durret)
during the nights of 31/10/2004 to 02/11/2004.  The images cover an
area of 38$\times$36~arcmin$^2$ with a pixel scale of 0.238~arcsec/px,
covering the South half of Abell~85 and its filament.  They were taken
in a narrow band filter corresponding to the wavelength of H$\alpha$
at the cluster redshift (ESO~\#869) and in a broad band $R_C$ filter
(ESO~\#844) to subtract the continuum contribution.  The response
curves of these two filters are shown in Fig.~\ref{fig:WFI_filters}.

The images were corrected for bias and flat field in the usual way.
They were then combined and astrometrically corrected by E.~Bertin, in
order to obtain a final image in each filter.  These images were
calibrated photometrically based on observations in the $R_C$ filter
of the SA~113, LATPHE and RU~149 standard star fields from the 
\cite{Landolt92} list observed during the same nights.

Object detections were made on the $R_C$ image with the SExtractor
software. The star--galaxy separation was performed as for the Megacam
data. We tested our photometric calibration by cross-identifying 680
stars with magnitudes $17 <R_C <21$ and comparing their $R_C$
magnitudes with their $r'$ magnitudes measured in the Megacam
image. We find an average value $\left\langle r'- R_C
\right\rangle=0.210$. \cite{Fuku95} 
give $r'-R_C$= 0.25, 0.22 and 0.17 for elliptical, Scd and Im
galaxies respectively.  Therefore our WFI $R_C$ band and Megacam $r'$
band photometric calibrations are in good agreement.

We then ran SExtractor in double image mode on the \ha\ image, based on
the detections made in the $R_C$ band. The magnitudes of the same 680
stars were then measured, giving the average magnitude difference
$\left\langle{\rm H}\alpha - R_C \right\rangle=2.37$~mag between the
\ha\ and the $R_C$ filters, with a dispersion of $\pm 0.13$~mag.  This
relation allowed us to calibrate the \ha\ image.

Note that the [NII]$\lambda$6548,6584 lines are also included in the
\ha\ filter. However, they should not contribute more than $\sim$20\%
to the total emission line flux (e.g. \citealp{CGBI04}).

The \ha\ observations can encounter two potential problems.  First, a
strong telluric line is present near 6900~\AA, that is towards the
left wing of the ESO~\#869 filter (see Fig.~\ref{fig:WFI_filters}), and its
absorption is non 
negligible up to about 6911~\AA, which corresponds to the wavelength
of \ha\ redshifted by 0.053. The redshift interval for cluster
membership was estimated to be about [0.0451-0.0657] by
\citet{Durret+98}. Therefore the telluric line may lead to
underestimate the contribution of emission line galaxies in the
[0.045-0.053] redshift range. In the region covered by our WFI data,
there are 373 measured redshifts in our complete redshift catalogue of
1705 objects (see Section 2.4). Out of these, 220 have redshifts in the
cluster range, and 173 are in the [0.053-0.0657] interval. Therefore,
220-173=47 galaxies are potentially affected by the telluric line,
representing 21\% of the cluster galaxies. On the other hand,
the effect of telluric
absorption at 6900 and 7200~\AA\ on broad band imaging can be
considered as negligible, since these telluric lines are expected to
affect the $R_C$ filter by less than 10\%, given the breadth of the
  $R_C$ filter.

Second, the ESO~\#869 filter is centered on wavelength 6963~\AA\ and
has a width of 
207~\AA.\footnote{http://www.ls.eso.org/lasilla/sciops/2p2/E2p2M/WFI/filters}
Considering again that the redshift range for Abell~85 is
[0.0451-0.0657], the wavelength of the redshifted \ha\ line for a
galaxy at the lower redshift limit of 0.0451 is 6858.8~\AA; at this
wavelength, the filter transmission is about 40\%. For a redshift of
0.049, the wavelength of the redshifted \ha\ line becomes 6884.38~\AA\,
and at this wavelength the filter transmission is about
85\%. Therefore, we can consider that the galaxies that will be
affected by this filter cut are those with redshifts between 0.045 and
0.049. 
The correction for incompleteness in our \ha\ detections due to this
filter cut is quite uncertain because the exact position of the filter
shoulders shifts in wavelength 
with ambient
temperature. Besides, as discussed below, our redshift catalogue is
based in part on SDSS data for which we have no accurate completeness
estimate. We will therefore not attempt to correct for incompleteness.

\subsection{AAOmega spectroscopic data}

Spectroscopy was obtained with the Anglo-Australian Telescope in
November 2006 (P.I. G.A. Mamon). Objects were selected inside the
Megacam area between $g'$=18 and 21.5 (computed inside the AAOmega
2~arcsec diameter fiber area). The exposure times were 4700 sec for
the brightest targets and 6500 sec for the faintest targets. Details
on the spectroscopic run will be given in a forthcoming paper. These
spectra were only used here to increase our redshift catalogue.

\subsection{SDSS spectroscopic data and full redshift catalogue}

Since the region of Abell~85 was covered by the SDSS, 
many redshifts are available for this area in the NED database
in addition to the \cite{DFLS98} redshift catalogue. Additional
spectroscopic redshifts obtained at the AAT were added to the
redshift catalogue extracted from the NED database, in order to have a
redshift catalogue as complete as possible to cross correlate with
\ha\ detections. The complete catalogue contains 1705 objects, out of which
506 galaxies have redshifts in the [0.0451-0.0657] cluster interval;
out of the latter 220 are in the WFI field.


We extracted all the spectra available in the Sixth Data Release of the SDSS
(SDSS-DR6) 
within the area covered by our
WFI image, and selected those with an equivalent width in the \ha\ line
larger than 3\AA. We found 12 objects in this sample, among which 6
are in the cluster redshift range. Five of these galaxies are detected
in our \ha\ image, and the only one which is not is
ACO85J004127.86-092329.54 is at a redshift of 0.0494, which is
probably cut by the \ha\ filter, since its spectrum unambigously shows
\ha\ emission (see its spectrum in the Appendix).  Quantities derived
from the \ha\ flux for this galaxy would only correspond to the SDSS
spectroscopic aperture, while for the other objects these quantities
are integrated throughout the galaxy, so we will not compute them.
The spectra of the 6 SDSS galaxies with \ha\ emission in their spectra and
within the cluster redshift range are shown in the Appendix.

\section{Large-scale galaxy distribution}

\begin{figure}[ht]
\centering
\includegraphics[width=9cm]{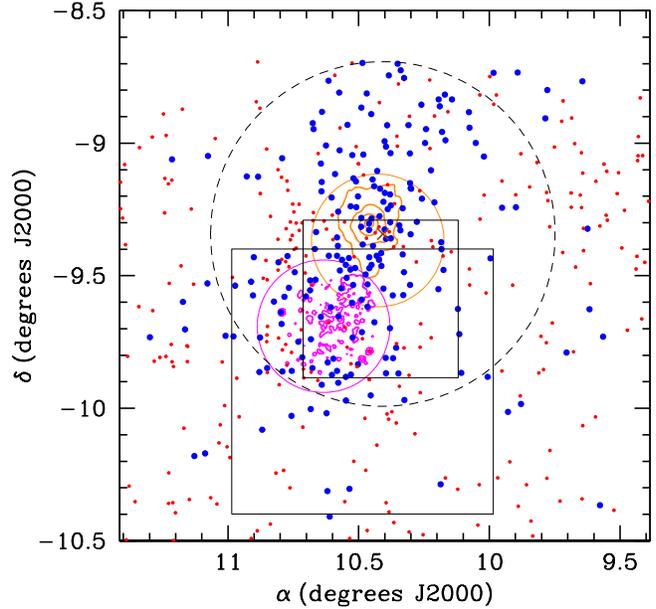}
\caption{Large-scale distribution of SDSS-DR6 galaxies around Abell~85.
\emph{Large blue} and \emph{small red filled circles} represent cluster
members (within 
$1500 \, \rm km \, s^{-1}$ from the cluster mean velocity) and outliers,
respectively. 
The \emph{huge dashed circle} represents the virial radius.
The \emph{small} and \emph{large squares} show the WFI and MegaCam fields,
respectively.
The region $-8\fdg7 < \delta < -8\fdg5$ was not covered by the SDSS.
XMM-Newton log-spaced X-ray contours and fields
are shown in \emph{orange} (main cluster) and
 \emph{purple} 
 (filament, with contour levels 1.1, 4 and 11 times lower than the
 lowest one of the main cluster)
}
\label{fig:lscale}
\end{figure}

The large-scale distribution of SDSS-DR6 galaxies around Abell~85 is shown in  
Fig.~\ref{fig:lscale}.
%
Around the cluster, the distribution of the SDSS galaxies having a
spectroscopic redshift $\pm 0.005$ around that of Abell~85 and inside
the virial radius has an axis ratio of $0.58 \pm 0.05$ along position
angle PA=$164^\circ \pm 4^\circ$, counted anticlockwise from north). These
measurements were done using the 2nd order moments of the particle
distribution (e.g., \citealp{Bertin06}), and the errors were estimated
using 1000 bootstraps. Therefore, the cluster is highly and
significantly flattened.  Even if the SDSS is not complete (in
particular because of fiber crowding), the regions with few SDSS
galaxies at the cluster redshift are filled with background SDSS
galaxies, so the flattening of the cluster galaxy distribution does not
appear to be caused by incompleteness.  Moreover, Fig.~\ref{fig:lscale} shows
that the flattening of the cluster is aligned with the X-ray filament
discovered by \cite{Durret+98}, in particular with the axis linking the
cluster X-ray centre to the extended region of the filament at
$\alpha=10\fdg55$ and $\delta=-9\fdg55$ (roughly at 0.4 virial radii
from the cluster X-ray centre).

\section{Galaxy luminosity functions}

We now present and discuss the properties of the galaxy
luminosity functions in the various Megacam bands and in different
regions of the cluster.  A full description of the method applied to
derive luminosity functions (LFs) and the importance of applying a
colour cut to eliminate background galaxies can be found in B08.

\begin{figure}
\centering
\includegraphics[width=7cm,angle=270]{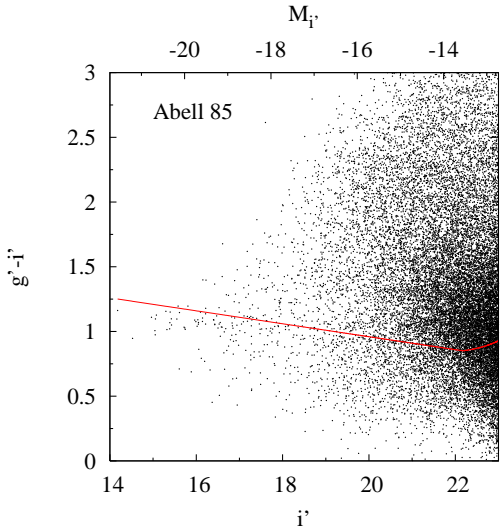}
\caption{($g'-i'$) versus $i'$ colour-magnitude plot for the objects
detected with Megacam.
The red line shows the limit colour-magnitude relation applied to select
possible cluster members when drawing the luminosity functions.  }
\label{fig:selfun}
\end{figure}

Fig.~\ref{fig:selfun} shows the galaxy distribution in a
colour-magnitude diagram. The reddest galaxies of the cluster lie in
the red sequence (its upper limit, as estimated in B08, is marked in
red).  All galaxies above this line are redder and are assumed to be
field objects.

We computed LFs both for the whole Megacam field of view and for 16
subfields. The subfields define a regular square grid of
15$\times$15~arcmin$^2$ each and allow a good compromise between
spatial resolution and uncertainties in individual magnitude bins. We
used 1~magnitude bins to limit the uncertainties. Several subregions
are then defined including a certain number of subfields with common
properties.

\begin{table*}
\centering
\caption{Galaxy luminosity function in four bands: power-law fits on
global image}
\begin{tabular}{l*5{c}}
\hline
Filter & $u^*$ & $g'$ & $r'$ & $i'$ \\ 
\hline
magnitude range &
$20.5\leq u^*     \leq23.0$ &
$19.0\leq g'\leq23.0$ & 
$18.5\leq r'\leq23.0$ &
$18.0\leq i'\leq22.5$ \\
$\alpha\pm 1\sigma$ &
    $-1.70\pm0.48$ &
    $-1.46\pm0.15$ &
    $-1.52\pm0.12$ &
    $-1.45\pm0.11$ \\ 
\hline
\end{tabular}
\label{tab:LF}
\end{table*}

\begin{figure}
\centering
\includegraphics[width=8cm,angle=270]{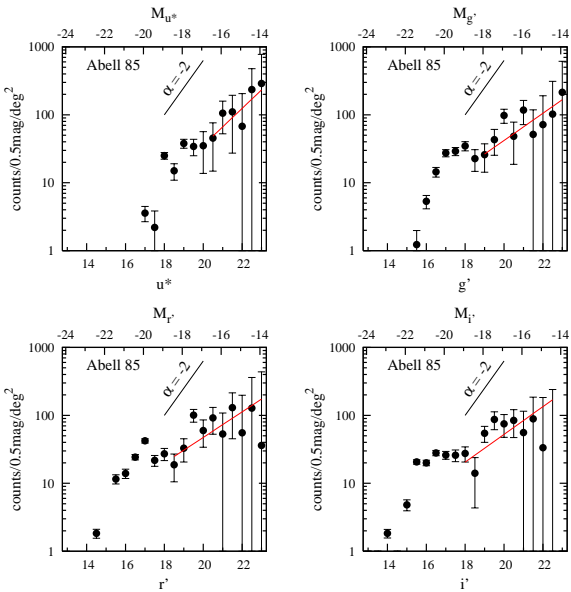}
\caption[]{Global luminosity functions for the total area covered by
Megacam (i.e. the filament and ``impact'' region of Abell 85) in the
four bands with the best power law fits shown in red. }
\label{fig:glf85}
\end{figure}

\begin{figure}
\centering
\includegraphics[width=8cm,angle=270]{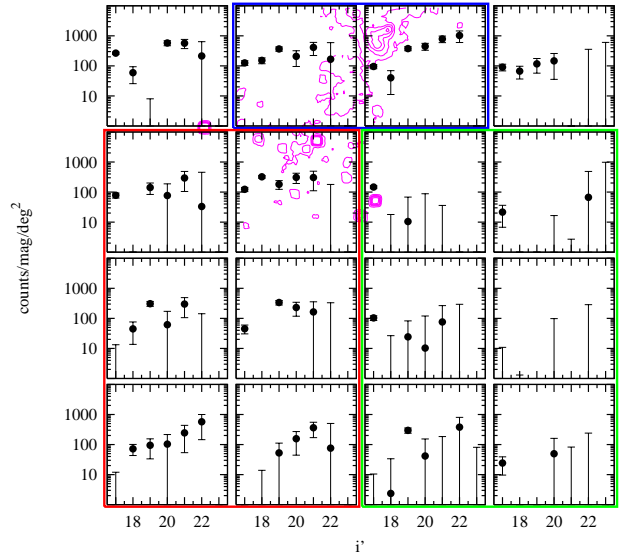}
\caption[]{Luminosity functions for the filament region of Abell~85 in
the $i'$ band. Each subfield is 15$\times$15~arcmin$^2$. X-ray
intensity contours with logarithmic steps from XMM-Newton data are
superimposed. Three main areas are defined: in \emph{blue} (the South
tip of the cluster and ``impact'' region), \emph{red} (containing the
filament) and \emph{green} (where only a few objects are detected).  }
\label{fig:llf85}
\end{figure}

\begin{figure}
\centering
\includegraphics[width=8cm,angle=270]{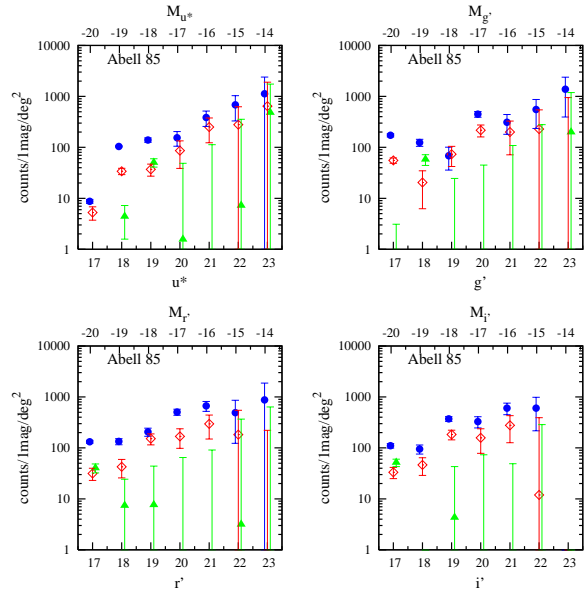}
\caption[]{Luminosity functions in the four bands for Abell 85 in the
three main areas described in Fig.~\ref{fig:llf85}. The colours of the
points are: blue for the North rectangle of Fig.~\ref{fig:llf85}
(South tip of the cluster and ``impact'' region), red for the
Southeast rectangle, and green for the Southwest rectangle.}
\label{fig:lglf85}
\end{figure}

The overall LF in the four photometric bands is displayed in
Fig.~\ref{fig:glf85} (we remind the reader that the cluster center is
located outside the surveyed field). At bright magnitudes, the LFs
have comparable shapes in the $g'$, $r'$ and $i'$ bands, while there
are fewer galaxies in the $u^*$ band.

The LF in 15$\times$15~arcmin$^2$ subfields is displayed in
Fig.~\ref{fig:llf85}, showing that the LFs of Abell~85 are not similar
over the whole field. Some subfields are very poorly populated, while
others exhibit rising LFs. As expected, the subfields towards the
Southeast of the image (coinciding with the X-ray filament and its
continuation) are more densely populated than the Southwest, implying
that the cluster (or the filament) extends far beyond the virial
radius.

We can therefore divide the cluster into three main regions: the North
zone where the cluster still dominates, the Southeast rectangle and
the Southwest rectangle (respectively in blue, red and green in
Fig.~\ref{fig:llf85}).  The LFs in these three subregions, in the four
photometric bands, are displayed in Fig.~\ref{fig:lglf85}.  As
expected, we can see significantly populated LFs in the zones that
also show cluster X-ray emission (the North and South east regions,
and the filament) while there are hardly any galaxies in the Southwest
region. Note that the southern edge of the North region corresponds to a
distance to the cluster center of about 1.9 Mpc, less than the virial
radius (2.5~Mpc).

Though there are significantly positive points in the two
best-populated areas (the North and filament zones), the LFs do not
display very well-defined power-law slopes.
We notice, however, that the shapes of the LFs in the North and
Southeast regions (respectively the blue and red regions in
Fig.~\ref{fig:lglf85}) are quite similar. The Southwest area is
obviously much less populated, since there are very few significant
points in its LFs.

\section{Dynamical properties of the filament}

\begin{figure}
\centering
\includegraphics[width=\hsize]{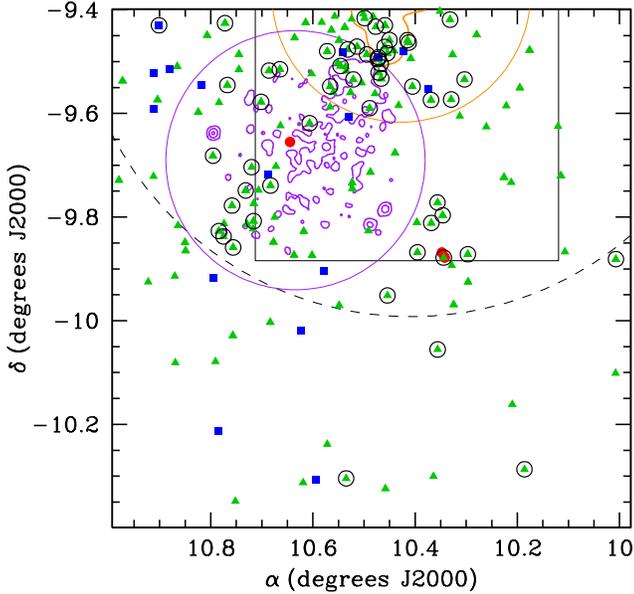}
\caption[]{Same as figure~\ref{fig:lscale} for galaxies in the cluster
  redshift range 
[0.0451-0.0657], highlighting the different galaxy types and dynamical
substructures they belong to.  
The symbols are the following: \emph{filled red circles} for
ellipticals, \emph{filled green triangles} for intermediate type spirals, and
\emph{blue squares} for late type spirals. \emph{Large black circles} show the
galaxies belonging to the dynamically bound substructure.
\label{fig:types}}
\end{figure}

We estimated galaxy types from their colours in the four Megacam bands,
based on the Le Phare photometric redshift technique developed by
S.~Arnouts and O.~Ilbert (\citealp{Zucca+06}).  Galaxies are thus
divided into four types: type 1 for ellipticals, type 2 for early type
spirals, type 3 for intermediate spirals and type 4 for late type
spirals.  Note that this classification was applied only to the
spectroscopic sample, by fitting the template with a fixed redshift, and
is therefore quite robust.  This classification corresponds to the
Coleman et al. (1980) templates, adjusted to the colours as described by
\citet{Zucca+06} (also see
http://www.ifa.hawaii.edu/~ilbert/these.pdf.gz, page 50 and page 142).



The distributions of galaxies in the cluster redshift range
[0.0451-0.0657] with different symbols for the various types are shown
in Fig.~\ref{fig:types}.  Very few type 2 (early type) galaxies were
found in our sample.  Late type spirals appear to follow more or less
the filament, or at least to be located in the East half of the image,
whereas intermediate type spirals are distributed throughout.

Such a distribution could be explained if late type spirals arrive on
to the cluster from the East, more or less along the filament and are
then transformed into earlier type spirals which then gravitate inside
the cluster (see e.g. \citealp{Adami+99} and references therein). This
scenario, although speculative, would agree with the fact that a VLA
survey of Abell~85 has shown that the majority of the galaxies
detected in HI were in the Eastern half of the cluster (Bravo-Alfaro
et al. in preparation). The galaxies located in the Western half of the
cluster would then have lost their HI through ram pressure
stripping. More detailed modelling is required however to fully
understand this phenomenon.


Out of our global redshift catalogue of 1705 objects with
spectroscopic redshifts, we created a catalogue of 181 galaxies with
redshifts in the [0.0451-0.0657] range and located within our Megacam
image, and applied to this catalogue the \cite{SG96} hierarchical
method (hereafter SG).  This method allows to extract galaxy subgroups
from a catalogue containing positions, magnitudes and redshifts, based
on the calculation of their (negative) binding energies.  The output
is a list of galaxies belonging to the selected group, as well as the
information on the binding energy of the group itself. We assumed an
$M/L_R = 200$; note however that, as shown e.g. by
\cite{ABDM05} for Coma galaxies, results derived from the SG method
are not sensitive to the exact choice of $M/L$ (taking $M/L_R=400$
instead of 200 did not change our results).

The SG method confirms the existence of a dynamically bound structure
roughly following the filament, the brightest galaxy in the filament
being the elliptical seen in Fig.~\ref{fig:types}.  The mass (i.e. the
sum of the galaxy masses) of this dynamically bound system is about
$6.6\times10^{12}$~M$_\odot$, in the range of groups of galaxies.
Therefore, the physical existence of the filament is consistent with the 
output of the SG analysis.

A second rather loose substructure is also found by the SG method
towards the Southwest. This substructure also includes an elliptical
galaxy, which is brighter by about 0.5~mag than the other 8 galaxies
in that zone, and has a mass of about $3.9\times10^{11}$~M$_\odot$.


\begin{table}
\caption{Completeness of redshift catalogue in the \ha\ image field.}
\centering
\begin{tabular}{rrrr}
\hline
$R_C$ & $N_{\hbox{with $z$}}$ & $N_{\rm total}$ & completeness (\%) \\
\hline
13-14 &  1~~ &   1 & 100~~~ \\
14-15 &  8~~ &   8 & 100~~~ \\
15-16 & 22~~ &  22 & 100~~~ \\
16-17 & 52~~ &  55 &  94.5 \\
17-18 & 68~~ &  74 &  91.9 \\
18-19 & 78~~ & 160 &  48.8 \\
19-20 & 34~~ & 366 &   9.3 \\
20-21 & 21~~ & 875 &   2.4 \\
21-22 & 12~~ & 1532&   0.8 \\
22-23 &  2~~~& 1553&   0.1 \\
\hline
\end{tabular}
\label{tab:compl}
\end{table}

One of the limitations of the SG method is the completeness of the
redshift catalogue. We give in Table~\ref{tab:compl} the number of
galaxies with redshifts in the field of the \ha\ image and the total
number of galaxies in that region in magnitude bins of 1~mag.  This
table shows that the redshift catalogue is more than 90\% complete up
to $R_C=18$, and still almost 50\% complete in the  $R_C=[18,19]$
magnitude bin; above $R_C=19$ the completeness decreases rapidly.  The
substructures found with the SG method are therefore very likely to be
true, since gravitational effects are dominated by massive
(i.e. bright) galaxies, but the numbers of objects included in each
substructure and the corresponding substructure masses should only be
considered as indicative.

\section{Emission line galaxies}

\subsection{Sample definition}

In order to select \ha\ emitting galaxies, we first created a ``pure''
emission line image (implicitly assuming that, in average, the spectra
of stars are flat in the wavelength region covered by the filters),
based on the relation found in Section 2.2: $\left\langle{\rm H}\alpha - R_C
\right\rangle=2.37$~mag.  The ``pure'' \ha\ image was therefore obtained by
subtracting $10^{-0.4\times 2.37} \times I(R_C) = 0.113\,I(R_C)$ to the
\ha\ image, where $I(R_C)$ is the intensity of the $R_C$ image.

However, when we tried to find \ha\ emitting objects by drawing a
colour-magnitude diagram of the ``net'' \ha\ flux as a function of the
$R_C$ magnitude, and selecting galaxies with net \ha\ fluxes higher
than 4 (or even more) times the dispersion, at least 20 or 30\% of the
selected objects were spurious (i.e. they were not visible on the
``pure'' emission line image). 

We were thus led to modify our extraction of \ha\ detections in two ways:

1) Since the sensitivity of the WFI camera is not perfectly uniform
throughout the field, we modeled the spatial variations of the parameter
$n$ (defined as the ratio of the \ha\ to $R_C$ fluxes) around its mean
value of 0.113 with Zernike polynomials of order 0 to 8 (which include
terms up to the 4th power in angular distance). This fit was based on
the all stars present in the image.

2) Since the seeings were not exactly the same in the \ha\ and $R_C$
images (see Table~1), we smoothed the $R_C$ subimages to match the
seeing of the \ha\ subimages.  Our smoothing filter was a gaussian of
FWHM equal to the root difference of the squares of the two FWHMs.  We
then subtracted the smoothed $R_C$ subimages multiplied by their
respective values of $n$ to the corresponding \ha\ subimages to create
``pure'' \ha\ subimages.

\begin{figure}
\centering
\includegraphics[width=5cm,clip=true]{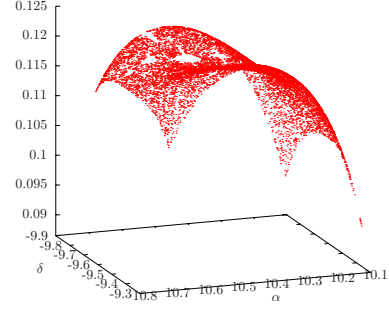}
\caption{Fit by Zernike polynomials 
of the \ha\ to $R_C$ flux ratio 
for stars throughout the
WFI field.
}
\label{fig:carte_n}
\end{figure}
The Zernike fit reduced the rms residuals in $n$ from 0.0169 to 0.0163 (an
$F$ test indicates that this small reduction in rms for the 7 extra fit
parameters is highly significant, given the 492 stars that the fit was
performed on), while 
the individual stars displayed variations of $n$ roughly 35 times
larger than the typical measurement errors ($n$ is 
significantly correlated to the $g-i$ colour
of the star). Nevertheless, the Zernike fit 
shown in Fig.~\ref{fig:carte_n}, indicates 10\% relative large-scale
variations of the \ha\ to $R_C$ ratio.


The detection of \ha\ emitters was then made by running SExtractor on
these ``pure'' \ha\ subimages, with a 2$\sigma$ limit detection (note
that this corresponds to a detection level of 3.3$\,\sigma$ with the 
usual definition, since SExtractor applies a smoothing)  and requesting
at least 12 pixels connected to each other to detect an object. 105
galaxies were detected in this way. The ``pure'' \ha\ subimages
corresponding to these 105 galaxies were visually inspected and 5
objects were discarded, three because they were too close to a bright
star, one was close to a CCD edge and one was a large galaxy with only
very weak diffuse \ha\ emission (if any).

For the 101 remaining objects, considered as detected in the \ha\
filter, fluxes were then measured by running again SExtractor on the
\ha\ and smoothed
$R_C$ subimages and measuring their respective fluxes in the
area where \ha\ emission was detected in the ``pure'' \ha\
subimages. The calculations of \ha\ fluxes and equivalent widths and
of the errors on these quantities are explained in Section 6.3.

\subsection{Cluster membership}

Cluster membership is obviously a crucial issue in this study. We
constructed a redshift catalogue of 1705 objects in the area of
Abell~85 (see Section 2.3), among which 373 are in the WFI field. Out of
the 101 galaxies detected in the \ha\ filter, 25 have redshifts: 23 are
in the [0.0451-0.0723] cluster redshift range (note that this range
was chosen to include the galaxy with a redshift of 0.0723, slightly
above the cluster range), and the remaining two galaxies have
redshifts 0.2331 and 0.4355.

\begin{figure}
\centering
\includegraphics[width=6cm,angle=0]{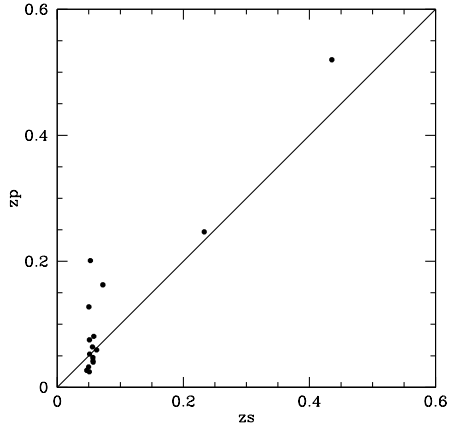}
\caption{Photometric redshifts estimated with the Le Phare software
versus spectroscopic redshifts for the 17 galaxies detected in \ha\
and having both spectroscopic and photometric redshifts.} 
\label{fig:zszp}
\end{figure}

In order to test whether the 76 galaxies detected in \ha\ but with no
spectroscopic redshift were likely to belong to the cluster, we
estimated photometric redshifts (hereafter $z_{\rm phot}$). This was only
possible for the objects covered by our Megacam images in four broad
bands (see Fig.~\ref{fig:zones}). A plot of photometric versus spectroscopic
redshifts for the 17 galaxies with both spectroscopic and photometric
redshifts is shown in Fig.~\ref{fig:zszp}. This figure shows that
galaxies belonging to the cluster from their spectroscopic redshifts 
are all at $z_{\rm phot}<0.22$.


Among the 76 \ha\ emitting galaxies with no spectroscopic redshift,
58 have photometric redshifts. The remaining 18 galaxies have no
photometric redshift either because they are outside the Megacam field
or because they are too close to a bright star, in a region masked in
the Megacam image analysis. Out of the 58 galaxies with photometric
redshifts, all have $z_{\rm phot}>0.2$. 

\begin{figure}
\centering
\includegraphics[width=6cm,angle=0]{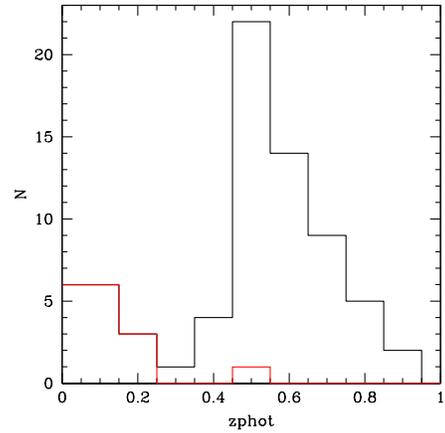}
\includegraphics[width=6cm,angle=0]{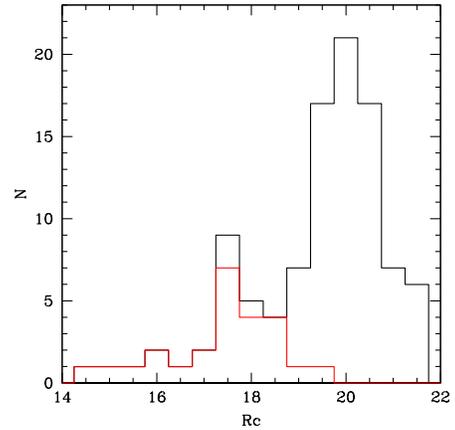}
\includegraphics[width=6cm,angle=0]{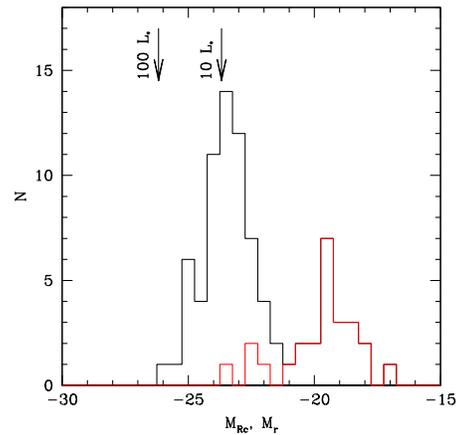}
\caption{Top: histograms of the photometric redshifts for the 50
galaxies detected in \ha\ without spectroscopic redshifts (black) and
for the 17 galaxies which also have spectroscopic redhifts (red).
Middle: histogram of the apparent magnitudes of these objects. Bottom:
histogram of the corresponding absolute magnitudes if the galaxies are
at the distances given by their photometric redshifts. The vertical
 arrows indicate the absolute magnitudes corresponding to 10L$_*$ and
100L$_*$.}
\label{fig:histos}
\end{figure}


\begin{figure}
\centering
\includegraphics[width=\hsize]{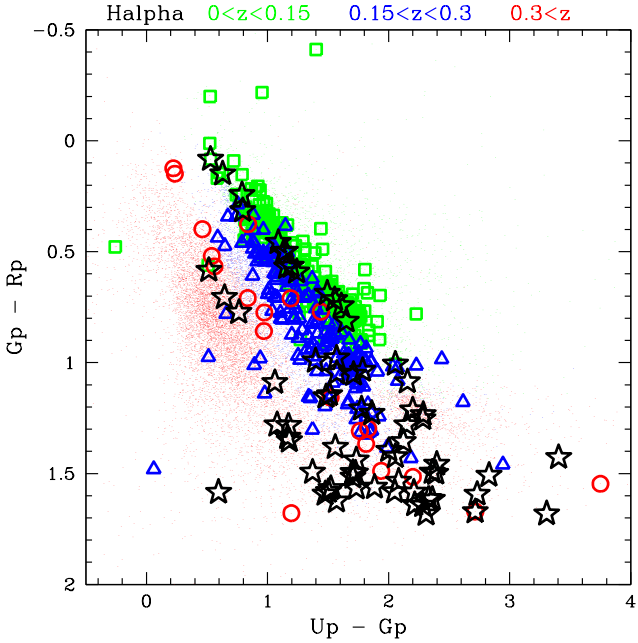}
\caption{Colour-colour diagramme for galaxies of known redshifts
(\emph{green squares} for $z < 0.15$, \emph{blue triangles} for $0.15 <
z < 0.3$ and \emph{red circles} for $z > 0.3$). The \ha\ detections are
shown as \emph{black stars}.  Galaxies with only photometric redshifts
are shown as \emph{green}, \emph{blue}, and \emph{red points}, for
$z_p<0.5$, $0.15 < z_p < 0.3$, and $z_p>0.3$, respectively.  }

\label{fig:colcol_oliv}
\end{figure}

\begin{figure}
\centering
\includegraphics[width=\hsize]{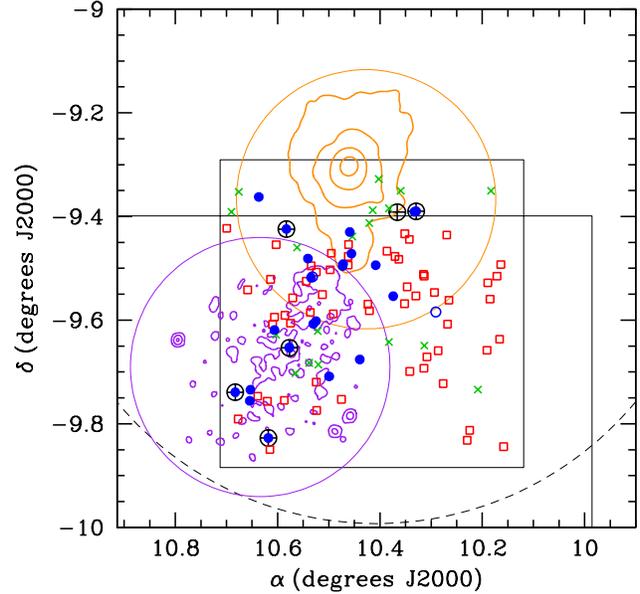}
\caption{Same as figure~\ref{fig:lscale} for 
the 101 galaxies detected in the \ha\
filter, plus the SDSS galaxy.
The \emph{blue circles} show our \ha\ detections for
galaxies with spectroscopic redshifts that place them in (\emph{filled
  circles}) or out (\emph{open circles}) of the
cluster.
Galaxies that we detect in \ha\ without spectroscopic
redshifts are shown as
\emph{red open squares} or \emph{green crosses}, depending on whether
they have photometric redshifts (always at $z>0.2$) or no photometric
redshifts, respectively.
The galaxies detected in \ha\ by the SDSS are shown as 
\emph{large circles with plus sign signs} (in particular
ACO85J004127.86-092329.54, which we did not detect in \ha).
}
\label{fig:zones}
\end{figure}

Among the 58 galaxies detected in \ha\ and with only photometric
redshifts, it is a little surprising to find no galaxy with
$z_{\rm phot}<0.2$, though photometric redshifts should be considered as
mainly giving a statistical information. Our recent experience when
applying photometric redshifts to the Abell~85 and Abell~496 four band
Megacam catalogues before obtaining spectroscopic redshifts (for
galaxies with typical magnitudes $r=20-21$) has shown that roughly 50\%
of the galaxies selected with $z_{\rm phot}<0.2$ were actually at
$z_{\rm spec}<0.2$. Note however that this fraction of 50\% would have
been reduced if we had used the galaxies observed in the SDSS to
``train'' the photometric redshift software.  On the other hand, less
than 15\% of the galaxies with $z_{\rm phot}>0.2$ were in fact found to be
cluster members in this study.


Can our \ha\ detections be contaminated by background objects?
The filter transmission of the ESO~\#869 filter is 5\% at $\lambda_1
\sim$ 6830~\AA\ and $\lambda _2 \sim 7100$\AA.  
Table~\ref{tab:lines} shows the range of redshifts for which different
astronomical lines can be detected in our \ha\ filter.
\begin{table}[ht]
\centering
\caption{Range of redshifts where lines enter our H$\alpha$ filter}
\begin{tabular}{lcc}
\hline
Line & Wavelength ($\rm\AA$) & redshifts\\
\hline
Ly$\alpha$ & 1216 & 4.62--4.84\\
OII & 3737 & 0.83--0.90\\
OIII & 5007 & 0.36--0.42\\
H$\beta$ & 4861 & 0.40--0.46\\
H$\alpha$ & 6563 & 0.04--0.08\\
NII & 6584 & 0.04--0.08\\ 
\hline
\end{tabular}
\label{tab:lines}
\end{table}
Besides the \ha-[NII] lines at $z\sim 0.06\pm0.02$, the emission detected in
our narrow band filter can also be due to H$\beta$ ($z=0.43\pm0.03$)
or
[OIII]$\lambda$4959,5007 at $0.39\pm0.03$, [OII]$\lambda$3727 at
$0.83<z<0.90$ or Ly$\alpha$ at $4.62<z<4.84$.  


\cite{Murayama+07} have
estimated the mean density of \lya\ emitters at redshift $z\sim
5.7$. For the mean redshift at which we would detect \lya\ galaxies
($z=4.73$), with our cosmological parameters, the distance modulus is
48.3. The faintest galaxy in our sample for which we detect line
emission in the ESO~\#869 filter has a magnitude
$R_C$=21. Therefore at $z=4.73$ this would correspond to a galaxy of
absolute magnitude $M_{R_C}=-27.3$, i.e. $280\,L*$ according to the
Schechter fit to the SDSS luminosity function by \cite{Blanton+03}, 
an unlikely value.  We
therefore believe that the contamination of our sample by \lya\
galaxies is negligible.

\cite{Takahashi+07} have detected 3176 [OII]$\lambda$3727 emitting
galaxies at z$\sim$1.2 in a volume of $5.5\times 10^5$~Mpc$^3$. The
corresponding density is $5.77\times 10^{-3}$ Mpc$^{-3}$.  In the ESO~\#869
filter, [OII] emitting galaxies will be detected if their redshifts
are between 0.8324 and 0.9049.  With our cosmology, these redshifts
correspond to luminosity distances of 5279 and 5857~Mpc
respectively. The mean angular distance at the mean redshift of 0.8687
is 1594~Mpc. Therefore the volume filled by [OII] galaxies in the
field of our WFI image is $2.0\times 10^5$~Mpc$^3$, leading to a total
number of galaxies of 1149. However, as mentioned above, we detect \ha\
emission only for galaxies brighter than $R_C=21$, corresponding
roughly to $i\sim 20.6$. We retrieved the \cite{Takahashi+07}
catalogue and found that no galaxy is brighter than $i=20.6$ (the
brightest one is at $i=21.15$; this is a magnitude measured in a
3~arcsec aperture, but since these are distant objects their total
magnitudes are probably not very different). Therefore the number of
[OII] contaminants in our \ha\ filter is expected to be negligible too.

As for H$\beta$ or [OIII] emitters at respective redshifts of about
0.38 and 0.42, the histogram of the photometric redshifts for the 58
galaxies detected in \ha\ displayed in Fig.~\ref{fig:histos} (top)
shows a bimodal distribution, with many galaxies having photometric
redshifts around $z_{\rm phot}\sim 0.5$. This peak mainly corresponds to
galaxies fainter than $R_C\sim 19$ (see Fig.~\ref{fig:histos},
middle), for which there are much fewer spectroscopic redshifts (the
completeness of our full redshift catalogue drops to less than 50\%
for $R_C >18$, see Table~4) and therefore for which photometric
redshifts are less well ``trained''. A number of galaxies in this peak
probably correspond to a population of rather faint galaxies at $z\sim
0.4$ emitting in H$\beta$ or [OIII], while a few may in reality be at
smaller distances and belong to Abell~85.  Note that if these galaxies
are all really at the distances corresponding to their photometric
redshifts, the histogram of their absolute magnitudes (see
Fig.~\ref{fig:histos}, bottom) shows that most of them are
intrinsically bright to very bright objects with absolute magnitudes
between M$_r=-22$ and $-25$ (absolute magnitudes were computed
including k-correction, following Appendix A.1 of \citealp{Ilbert+05}). 
This would imply a high
number of very intrinsically bright ($\approx 7\,L*$) galaxies in the
field (they are 
spread throughout the WFI field so they do not seem to correspond to a
background cluster).  We are therefore inclined to believe that some
of the \ha\ emitters that we detect and which have photometric
redshifts larger than 0.2 may in reality be more nearby objects and
could even possibly be members of Abell~85; the other objects are
probably H$\beta$ or [OIII] emitters at $z\sim 0.4$.

Fig.~\ref{fig:colcol_oliv} shows a colour-colour diagram for all the
galaxies with measured redshifts. The 23 \ha\ objects with redshifts in
the cluster are those coinciding with the green squares.

In view of all these arguments, we will consider as detected in \ha\
and members of the Abell~85 cluster the 23 galaxies with spectroscopic
redshifts in the [0.0451-0.0723] cluster range (plus galaxy
ACO85J004127.86-092329.54 from the SDSS).  We will also present our
data on the 76 other galaxies detected in the \ha\ filter, keeping in
mind that a good fraction of them is unlikely to belong to the cluster.

The full catalogues of these 101 galaxies are given in Tables~1--4 of
the Appendix, together with postage stamp images of each galaxy in the
$r'$, $u^*$, $R_C$ and ``pure'' \ha\ bands.  Note that we added to
Table~1 of the Appendix the SDSS galaxy ACO85J004127.86-092329.54.

\subsection{\ha\ characteristics of the emission line galaxies}

From the fluxes measured by SExtractor in the \ha\ and $R_C$ subimages
for each detected galaxy, as described in Section 3.1, we now compute
net \ha\ fluxes.  The overlap of the \ha\ and $R_C$ bands makes \ha\
net (i.e. continuum subtracted) flux calculations not
straightforward. Let $\Ca$ and $\Cr$ be the counts per unit time
measured in the \ha\ and $R_C$ bands respectively.  We can split these
counts into two contributions: (1) the continuum $\Cca$ and $\Ccr$
respectively and (2) the emission line $\Cl$ corrected for the
transmission factor of each filter:
\begin{eqnarray}
\Ca &=& \Cca + \Ta(\lambda_i)\Cl\ , \\
\Cr &=& \Ccr + \Tr(\lambda_i)\Cl\ .
\end{eqnarray}
where $\lambda_i$ is the wavelength of the \ha\ emission line redshifted to
the redshift of the $i^{\rm th}$ galaxy, and $\Ta(\lambda)$ and
$\Tr(\lambda)$ are the transmission factors of the two filters. $\Cca$
and $\Ccr$ are related to each other by
\begin{eqnarray}
\Cca &=& H\int\Ta(\lambda)\,d\lambda, \\
\Ccr &=& H\int\Tr(\lambda)\,d\lambda
\end{eqnarray}
where $H$ is the height of the continuum (assumed to be constant). We
can define the ratio between the integrated $R_C$ and \ha\ filter
transmissions with
\begin{equation}
n = \frac{\Cca}{\Ccr} = \frac{\int \Ta(\lambda)\,d\lambda}
{\int \Tr(\lambda)\,d\lambda}.
\end{equation}
From Equations (1)--(4) we obtain the integrated \ha\ flux F (in erg
cm$^{-2}$ s$^{-1}$) and the \ha\ equivalent width EW (in \AA):
\begin{eqnarray}
F &=&\displaystyle 
{\rm ZP}\times\Cl 
= {\rm ZP} \frac{\Ca-n\Cr}{\Ta(\lambda_i)-n\Tr(\lambda_i)}, \\
{\rm EW} &=&\displaystyle
\frac{\Cl}{H} = 
\frac{\Ca-n\Cr}{\Ta(\lambda_i)\Cr-\Tr(\lambda_i)\Ca}
\int\Tr(\lambda)\,d\lambda,
\end{eqnarray}
where ${\rm ZP}$ is the zero point in erg cm$^{-2}$. The results are
comparable but not identical to the equations of \cite{BIVG02} and 
\cite{Gavazzi+02}. \ha\ fluxes and equivalent widths are given for all the
objects of our sample in Table~1--2 of the Appendix.


Since our catalogue contains galaxies without redshifts, we decided
to correct the fluxes by constant factors $\Ta(\lambda_i)=0.9$ and
$\Tr(\lambda_i)=0.9$. Thus, \ha\ luminosities and consequent star formation
rates were computed from \ha\ fluxes assuming that all the galaxies
are at the cluster redshift.

\subsection{Spatial distribution of the emission line galaxies}

The spatial distribution of the 101 galaxies detected in \ha\ and of the
SDSS galaxy ACO85J004127.86-092329.54 is displayed in
Fig.~\ref{fig:zones}.  As seen in this figure, most of the galaxies with
redshifts in the cluster appear to be concentrated either in the
periphery of the cluster, including the south blob (the group seen in
X-rays just south of the main cluster at coordinates $\alpha \sim
10.45^o$, 
$\delta \sim -9.45^o$) and along the filament. If we
consider that the filament corresponds to the south east zone at
$\alpha>10.45^o$ and $\delta < -9.5^o$, we can note that only 3
objects fall west of the filament in this southern area. We have seen
when describing the galaxy luminosity functions in Section~4 that the
Southeast part of the cluster has many more galaxies than the
Southwest. Our hope when making these narrow band filter observations
was to detect a high number of \ha\ emitters in the filament region,
where star formation could have been enhanced by the movement of groups
towards the cluster. Although more spectroscopically confirmed cluster
members would be needed to claim such a result, it seems that \ha\
emitters in the cluster redshift range are more concentrated along the
filament.

\subsection{Star formation in the filament region of Abell~85}

From the \ha\ fluxes, we estimated the \ha\ luminosities $L$(\ha) and,
following Eq.~(2) in \cite{Kennicutt98}, the star formation rates (SFR)
in the galaxies detected in the \ha\ filter: SFR(M$_\odot$ yr$^{-1}$) =
7.9$\times 10^{-42}\ L$(\ha) (with $L$(\ha) in erg
s$^{-1}$). Luminosities are estimated assuming that the galaxies belong
to the cluster; they are obviously underestimated for background objects
(for which it is not \ha\ emission that we are measuring but H$\beta$ or
[OIII]).  Note that, due to the filter cut for galaxies with
redshifts smaller than about 0.053, star formation rates for objects
with redshifts towards the lower limit of the cluster range are only
lower limits.

\ha\ luminosities and star formation rates are given in Table~1 of the
Appendix. We can see that star formation rates are very small, smaller
than 0.04~M$_\odot$/yr except for one object having
0.24~M$_\odot$/yr. These values are comparable to or lower than those
found e.g. by \cite{James+04} in a large sample of nearby galaxies and
by \cite{Cortese+06} in the starbursting group falling into
Abell~1367.

We have looked for variations of the \ha\ equivalent width along the
direction of the filament but did not find any obvious trend.

An important {\sl caveat} is that corresponding to completeness.  As
discussed in Section~2, the narrow band filter and a strong telluric
absorption line make detections difficult for galaxies at redshifts
lower than about 0.0515. Therefore our \ha\ imaging is probably
missing galaxies towards the lower range of the cluster redshift.
Note also that the SDSS spectroscopic survey is probably not complete
either.

\section{Discussion}

\subsection{Very large scale structure}

\begin{figure}
\centering
\includegraphics[width=7cm,angle=0]{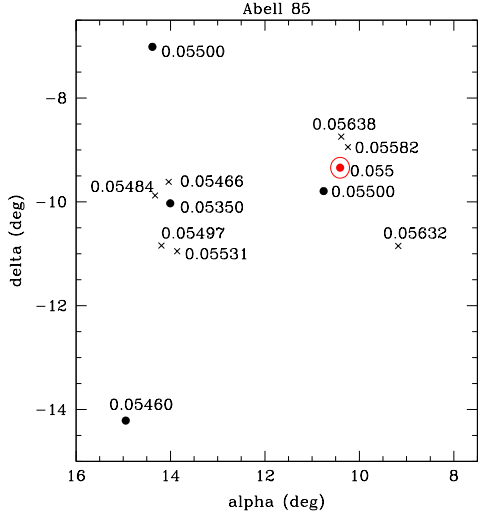}
\caption[]{Large scale structure surrounding Abell~85 taken from NED.
The red circle shows the position of the cluster center, black circles
are Abell clusters and crosses are other clusters found in the SDSS.
Clusters and groups were selected to have a redshift within $\pm
0.005$ of that of the cluster, in a region of 8.5$\times$8.5 deg$^2$
(32$\times$32 Mpc$^2$). Abell~85 is not exactly at the center of the figure
because there are no groups or clusters towards the West and North of the
cluster.}
\label{fig:lss}
\end{figure}

The infall directions are well known for Abell 85
\citep{Durret+98,DLNF05}. A previous merger is likely to have occured
3--4~Gyr ago from the Northwest, as derived from the presence of an
``arc'' of hotter X-ray gas regions, infall from the Southeast is
presently been observed along the filament, and there is a third
possible infall direction from the Northeast. The filament coming from
the Southeast is obviously quite massive since it is detected in
X-rays; from its X-ray temperature, it was suggested that it is
probably made of groups falling on to the cluster \citep{Durret+03}.
This agrees with our map of LFs showing very populated subfields
towards the Southeast direction (see Figs.~\ref{fig:llf85} and
\ref{fig:lglf85}) and with our analysis based on the \ha\ image and on
the SG analysis. 

The distribution of groups and clusters with redshifts within $\pm$0.005
(i.e. $\pm$1500~km~s$^{-1}$) of that of Abell~85 (extracted from the NED
database) at very large scale is displayed in Fig.\ref{fig:lss}.  Close
to the cluster, a concentration of structures along a roughly
North-Northwest to South-Southeast direction is observed. Further away,
a concentration of structures towards the East-Southeast is seen, as
well as two clusters even further away towards the Northeast and
Southeast. Such a large concentration of clusters at redshifts very
close to that of Abell~85
agrees with the general idea that this cluster has undergone
and is still undergoing several merging events. It also indicates that
there may be even more interactions going on than previously believed,
and that although one of the preferential regions for merging activity
is indicated by the filament, it may not be the only one.

\subsection{Are galaxies in the filament of Abell~85 more prone to
bursts of star formation?}

We first note that our images extend to 2 virial radii, hence include the
virialized part of the cluster, including the X-ray filament, plus a
mainly infalling region beyond.
The sampled regions are
therefore populated with galaxies that are just beginning to undergo
interactions with dense regions (the cluster itself or the dense parts
of the filament as seen in X-rays). The faint-end of the overall LF of
these fields could be somewhat steeper in the $u^*$ band than in the
other bands, but the error bar on the slope is large.



Our \ha\ detections may be preferentially concentrated in the filament
and in the impact region, but the statistics are too low to draw a firm
conclusion.  Note besides that we do not detect any \ha\ filamentary
structures (such as those found by \citealp{Cortese+06}) which could be
in the wake of galaxies having crossed the cluster.


The filament of Abell~85, detected in X-rays and consistent with the
present optical data is composed of galaxies and gas. Since the gas
temperature was found to be $2\pm 1$~keV, it cannot be identified with
the cooler filaments predicted by numerical simulations of large scale
structures (e.g. \citealp{Dave+01}). This agrees with our previous
interpretation that the Abell~85 filament is made of groups.

\section{Conclusions}

We have analyzed an \ha\ image and broad band images covering the
South of Abell~85 and its filament, which was previously discovered in
X-rays (\citealp{Durret+98}).  The Galaxy Luminosity Functions in the
South area of the cluster Abell~85 (including the impact region, where
the groups constituting the filament hit the main cluster) and in the
filament show the existence of a rich population of galaxies.
%
%
The overall galaxy distribution in A85 is flattened with a principal
axis that is parallel to the axis separating the filament seen in X-rays
with the cluster center.
All our results are consistent with the previous
interpretation of the filament being made of groups falling onto the
main cluster.

\begin{acknowledgements}
The authors are very grateful to the CFHT and Terapix teams for their
efficiency in the Megacam data reduction, and to E.~Bertin for his
help in reducing the WFI data. We thank M.~Montessuit for her
contribution to the photometric calibration of the WFI data, I.~Chilingarian
for his twilight spectrum, and
A.~Boselli and R.~Demarco for discussions. 
We also thank Matthew Colless for permission to use 6dFGS-DR3, in advance
  of publication.
We acknowledge financial
support from CNRS through the PNG and INSU, and from OPTICON.

\end{acknowledgements}

\bibliography{AA_2007_8972_arXiv}




\Online

\begin{appendix} 

A full catalogue of the 101 galaxies detected in \ha\ is given
Tables~1--4 of the present Appendix, together with the galaxy
ACO85J004127.86-092329.54 not detected in our \ha\ image due to the
filter cut but with \ha\ emission in its SDSS spectrum (in Table~1).

Tables~1--3 respectively correspond to galaxies with spectroscopic
redshifts, galaxies with photometric redshifts, and galaxies with no
redshift information. They include the following columns: (1)~number,
(2)~full IAU name, (3)~spectroscopic redshift taken from NED, (4)~\ha\ 
flux in erg~cm$^{-2}$~s$^{-1}$ and error measured in the aperture in
which the detection is made in the $R_C$ band, (5)~\ha\ equivalent
width and error in \AA , (6)~\ha\ luminosity in in erg~s$^{-1}$,
(7)~star formation rate in M$_\odot$/yr, (8)~factor $n$ by which the
$R_C$ image was multiplied before being subtracted from the \ha\
image.

Tables~4 and 5 give: (1)~number, (2)~full IAU name, (3)-(7) $R_C$, $r'$,
$u^*$, $g'$ and $i'$ band magnitudes (SExtractor MAG\_AUTO). 

Images are displayed for each galaxy in the $i'$, $u^*$, $R_C$ and
``pure'' \ha\ bands in Figs.~1--101.  Since the Megacam image does not
fully cover the WFI image, some galaxies have no Megacam ($u^*$ and
$i'$) data. 

The spectra of the 6 SDSS galaxies with EW(\ha)$>$3~\AA\ are displayed
in Figs. 102--107.

\clearpage

\begin{table*}
\centering
\caption{Catalogue of the 25 galaxies detected in \ha\ and with
spectroscopic redshifts (the first 23 belong to the cluster) , also
including the SDSS galaxy ACO85J004127.86-092329.54.}
\begin{tabular}{rrrrrrrr}
\hline
Num & Name~~~~~~~~ &$z_{\rm spec}$ & F(\ha)~~~~~~  & EW(\ha) & $L$(\ha) & SFR & $n$~~~ \\
   &              &           & erg cm$^{-2}$ s$^{-1}$~~~ & \AA~~~~~ & erg s$^{-1}$ & M$_\odot$ yr$^{-1}$ & \\
\hline
24 & ACO85J004119.01-092323.5 & 0.0498 & 1.19e-16$\pm$7.2e-18 & 13.3$\pm$0.8 & 8.32e+38 & 0.00658 & 0.118765 \\
26 & ACO85J004119.83-092327.0 & 0.0490 & 1.16e-16$\pm$7.8e-18 & 4.2$\pm$0.3 & 8.15e+38 & 0.00644 & 0.118765 \\
35 & ACO85J004129.77-093313.2 & 0.0624 & 4.13e-16$\pm$9.3e-18 & 129.3$\pm$3.3 & 2.90e+39 & 0.02288 & 0.115607 \\
40 & ACO85J004138.01-092938.0 & 0.0497 & 6.91e-17$\pm$3.8e-18 & 13.9$\pm$0.8 & 4.85e+38 & 0.00383 & 0.116822 \\
45 & ACO85J004145.45-094033.1 & 0.0569 & 2.33e-16$\pm$1.0e-17 & 17.2$\pm$0.7 & 1.64e+39 & 0.01294 & 0.11534 \\
47 & ACO85J004149.38-092818.3 & 0.0723 & 1.07e-16$\pm$5.7e-18 & 20.4$\pm$1.1 & 7.49e+38 & 0.00592 & 0.117371 \\
48 & ACO85J004150.17-092547.6 & 0.0579 & 7.60e-17$\pm$1.3e-17 & 1.7$\pm$0.3 & 5.34e+38 & 0.00422 & 0.118765 \\
52 & ACO85J004153.27-092930.4 & 0.0525 & 4.35e-15$\pm$3.8e-17 & 47.9$\pm$0.5 & 3.05e+40 & 0.24093 & 0.116822 \\
53 & ACO85J004153.51-092943.8 & 0.0510 & 8.74e-17$\pm$3.1e-17 & 3.1$\pm$1.1 & 6.13e+38 & 0.00485 & 0.116686 \\
58 & ACO85J004159.84-094230.9 & 0.0517 & 9.68e-17$\pm$9.4e-18 & 15.8$\pm$1.6 & 6.79e+38 & 0.00537 & 0.116009 \\
65 & ACO85J004206.02-093606.4 & 0.0529 & 6.71e-17$\pm$5.8e-18 & 5.7$\pm$0.5 & 4.71e+38 & 0.00372 & 0.114943 \\
66 & ACO85J004207.26-093626.0 & 0.0468 & 5.36e-16$\pm$1.2e-17 & 49.1$\pm$1.1 & 3.76e+39 & 0.02972 & 0.114943 \\
69 & ACO85J004208.36-093104.6 & 0.0566 & 9.33e-17$\pm$7.7e-18 & 15.0$\pm$1.3 & 6.55e+38 & 0.00517 & 0.116009 \\
72 & ACO85J004209.81-092852.2 & 0.0558 & 6.26e-16$\pm$8.7e-18 & 54.6$\pm$0.8 & 4.39e+39 & 0.03471 & 0.116959 \\
78 & ACO85J004218.47-093912.1 & 0.0515 & 7.02e-16$\pm$1.5e-17 & 39.3$\pm$0.9 & 4.93e+39 & 0.03893 & 0.115207 \\
79 & ACO85J004218.55-093910.2 & 0.0515 & 5.35e-16$\pm$1.4e-17 & 59.4$\pm$1.6 & 3.75e+39 & 0.02964 & 0.115207 \\
80 & ACO85J004219.89-092527.5 & 0.0512 & 1.69e-16$\pm$5.1e-18 & 17.0$\pm$0.5 & 1.19e+39 & 0.00939 & 0.118064 \\
86 & ACO85J004225.54-093708.9 & 0.0565 & 3.56e-17$\pm$3.7e-18 & 6.6$\pm$0.7 & 2.50e+38 & 0.00198 & 0.115075 \\
90 & ACO85J004228.38-094938.3 & 0.0501 & 7.95e-17$\pm$6.5e-18 & 9.8$\pm$0.8 & 5.58e+38 & 0.00441 & 0.116959 \\
92 & ACO85J004232.84-092144.2 & 0.0569 & 7.66e-17$\pm$7.2e-18 & 11.9$\pm$1.1 & 5.38e+38 & 0.00425 & 0.117509 \\
94 & ACO85J004236.76-094403.8 & 0.0643 & 9.99e-17$\pm$6.0e-18 & 16.0$\pm$1.0 & 7.01e+38 & 0.00554 & 0.115741 \\
95 & ACO85J004237.07-094520.5 & 0.0569 & 5.74e-17$\pm$6.2e-18 & 7.6$\pm$0.8 & 4.03e+38 & 0.00318 & 0.115875 \\
99 & ACO85J004243.90-094420.8 & 0.0508 & 1.20e-16$\pm$9.4e-18 & 10.8$\pm$0.9 & 8.44e+38 & 0.00667 & 0.114943 \\
\hline
   & ACO85J004127.86-092329.54 & 0.0494 &                       &  4.5 $\pm$  0.3 &&&\\ 
\hline
17 & ACO85J004109.80-093503.0 & 0.2331 & 5.23e-17$\pm$5.8e-18 & 7.8$\pm$0.9 & 3.67e+38 & 0.00290 & 0.114811 \\
67 & ACO85J004207.71-093059.3 & 0.4355 & 2.86e-16$\pm$1.4e-17 & 37.7$\pm$1.9 & 2.00e+39 & 0.01583 & 0.116009 \\
\hline
\end{tabular}
\label{tab:cat}
\end{table*}

\clearpage

\begin{table*}
\centering
\caption{Catalogue of the 58 galaxies detected in \ha\ and with
photometric redshifts.}
\begin{tabular}{rrrrrrrrrr}
\hline
Num & Name~~~~~~~~ &$z_{\rm phot}$ & $z_{\rm inf}$ & $z_{\rm sup}$ &
   F(\ha)~~~~~~ & EW(\ha) & $L$(\ha) & SFR & $n$~~~ \\ 
   &  &  &  &    & erg cm$^{-2}$ s$^{-1}$~~~  & ~~\AA~~~~~~ &  erg s$^{-1}$ & M$_\odot$ yr$^{-1}$ & \\
\hline
1 & ACO85J004037.99-095038.8 & 0.5571 & 0.5300 & 0.5800 & 5.63e-17$\pm$6.4e-18 & 75.1$\pm$9.0 & 3.95e+38 & 0.00312 & 0.0986193 \\
2 & ACO85J004039.27-092933.8 & 0.6215 & 0.6000 & 0.6500 & 7.98e-17$\pm$5.8e-18 & 60.1$\pm$4.6 & 5.60e+38 & 0.00442 & 0.108814 \\
3 & ACO85J004039.90-093812.5 & 0.4628 & 0.4400 & 0.4800 & 7.19e-17$\pm$6.8e-18 & 23.5$\pm$2.3 & 5.05e+38 & 0.00399 & 0.108225 \\
4 & ACO85J004041.11-093054.7 & 0.6063 & 0.5800 & 0.6300 & 6.90e-17$\pm$7.1e-18 & 40.0$\pm$4.3 & 4.84e+38 & 0.00383 & 0.109649 \\
6 & ACO85J004044.33-093334.4 & 0.6150 & 0.5900 & 0.6500 & 1.04e-16$\pm$7.5e-18 & 55.7$\pm$4.3 & 7.27e+38 & 0.00574 & 0.110619 \\
7 & ACO85J004045.38-093140.9 & 0.5045 & 0.4700 & 0.5300 & 5.86e-17$\pm$6.2e-18 & 27.0$\pm$2.9 & 4.11e+38 & 0.00325 & 0.111111 \\
8 & ACO85J004045.80-093929.1 & 1.0724 & 0.8700 & 1.1000 & 5.64e-17$\pm$6.6e-18 & 93.4$\pm$11.9 & 3.96e+38 & 0.00313 & 0.110011 \\
10 & ACO85J004053.86-094845.6 & 0.7453 & 0.7000 & 0.7900 & 7.45e-17$\pm$7.5e-18 & 121.3$\pm$13.5 & 5.23e+38 & 0.00413 & 0.109051 \\
11 & ACO85J004055.06-094954.9 & 0.4935 & 0.4700 & 0.5200 & 7.50e-17$\pm$7.5e-18 & 41.5$\pm$4.3 & 5.26e+38 & 0.00416 & 0.108578 \\
12 & ACO85J004103.65-093342.1 & 0.8230 & 0.7900 & 0.8600 & 4.03e-17$\pm$4.3e-18 & 63.9$\pm$7.2 & 2.83e+38 & 0.00223 & 0.114548 \\
13 & ACO85J004104.27-093627.8 & 0.7455 & 0.7200 & 0.7700 & 5.20e-17$\pm$4.6e-18 & 37.3$\pm$3.4 & 3.65e+38 & 0.00288 & 0.114155 \\
14 & ACO85J004104.67-092608.6 & 0.5393 & 0.4400 & 0.5700 & 5.72e-17$\pm$6.4e-18 & 21.6$\pm$2.5 & 4.01e+38 & 0.00317 & 0.116279 \\
15 & ACO85J004106.35-094321.1 & 0.5131 & 0.4900 & 0.5400 & 7.39e-17$\pm$8.5e-18 & 32.4$\pm$3.8 & 5.19e+38 & 0.00410 & 0.114025 \\
16 & ACO85J004108.66-093932.2 & 0.4685 & 0.4500 & 0.4900 & 4.27e-17$\pm$4.7e-18 & 28.8$\pm$3.2 & 3.00e+38 & 0.00237 & 0.114416 \\
18 & ACO85J004110.45-093247.7 & 0.5440 & 0.5000 & 0.6000 & 8.15e-17$\pm$5.7e-18 & 72.4$\pm$5.4 & 5.72e+38 & 0.00452 & 0.11534 \\
19 & ACO85J004113.96-094015.3 & 0.4717 & 0.4500 & 0.4900 & 6.64e-17$\pm$6.6e-18 & 28.8$\pm$2.9 & 4.66e+38 & 0.00368 & 0.114811 \\
21 & ACO85J004115.27-093053.9 & 0.6565 & 0.6100 & 0.7200 & 4.56e-17$\pm$6.0e-18 & 85.0$\pm$12.1 & 3.20e+38 & 0.00253 & 0.116144 \\
22 & ACO85J004115.38-094134.6 & 0.2652 & 0.2300 & 0.3000 & 7.21e-17$\pm$5.4e-18 & 9.1$\pm$0.7 & 5.06e+38 & 0.00400 & 0.114943 \\
23 & ACO85J004115.55-093041.3 & 0.5972 & 0.5600 & 0.6300 & 1.03e-16$\pm$8.7e-18 & 55.5$\pm$4.9 & 7.25e+38 & 0.00573 & 0.116144 \\
25 & ACO85J004119.11-093312.0 & 0.4669 & 0.4500 & 0.4900 & 7.95e-17$\pm$6.8e-18 & 20.1$\pm$1.8 & 5.58e+38 & 0.00441 & 0.115607 \\
27 & ACO85J004122.02-094156.0 & 0.8833 & 0.8400 & 0.9300 & 6.10e-17$\pm$7.1e-18 & 108.8$\pm$14.0 & 4.28e+38 & 0.00338 & 0.11534 \\
28 & ACO85J004122.17-092639.5 & 1.5751 & 1.4500 & 1.6500 & 2.26e-16$\pm$4.8e-18 & 369.5$\pm$10.8 & 1.59e+39 & 0.01256 & 0.117925 \\
29 & ACO85J004123.20-093208.6 & 0.6549 & 0.6200 & 0.6900 & 6.33e-17$\pm$6.7e-18 & 65.1$\pm$7.3 & 4.44e+38 & 0.00351 & 0.116009 \\
30 & ACO85J004124.32-092600.2 & 0.4699 & 0.4400 & 0.5000 & 8.01e-17$\pm$7.4e-18 & 32.5$\pm$3.1 & 5.62e+38 & 0.00444 & 0.118343 \\
31 & ACO85J004124.48-093405.3 & 0.3880 & 0.3600 & 0.4900 & 3.95e-17$\pm$3.8e-18 & 31.9$\pm$3.1 & 2.77e+38 & 0.00219 & 0.115473 \\
33 & ACO85J004127.13-092857.6 & 0.5987 & 0.5700 & 0.6300 & 5.48e-17$\pm$6.2e-18 & 50.9$\pm$6.1 & 3.85e+38 & 0.00304 & 0.117233 \\
34 & ACO85J004128.95-092837.1 & 0.9271 & 0.9100 & 0.9400 & 4.76e-17$\pm$6.0e-18 & 38.3$\pm$5.0 & 3.34e+38 & 0.00264 & 0.117371 \\
38 & ACO85J004132.70-092800.7 & 0.6133 & 0.5800 & 0.6400 & 6.78e-17$\pm$6.3e-18 & 43.0$\pm$4.2 & 4.76e+38 & 0.00376 & 0.117647 \\
42 & ACO85J004140.92-093454.9 & 0.6234 & 0.6000 & 0.6400 & 2.94e-16$\pm$7.2e-18 & 69.8$\pm$1.8 & 2.06e+39 & 0.01628 & 0.115207 \\
44 & ACO85J004141.67-093409.5 & 0.4903 & 0.4700 & 0.5100 & 1.05e-16$\pm$8.4e-18 & 29.6$\pm$2.4 & 7.40e+38 & 0.00585 & 0.11534 \\
49 & ACO85J004150.75-092714.8 & 0.7028 & 0.6900 & 0.7200 & 2.14e-16$\pm$5.3e-18 & 48.6$\pm$1.3 & 1.51e+39 & 0.01189 & 0.117925 \\
50 & ACO85J004150.88-092836.9 & 0.7051 & 0.6700 & 0.7500 & 6.62e-17$\pm$6.7e-18 & 49.3$\pm$5.3 & 4.65e+38 & 0.00367 & 0.117233 \\
51 & ACO85J004150.94-092938.1 & 0.4422 & 0.4200 & 0.4600 & 5.92e-17$\pm$5.9e-18 & 19.4$\pm$2.0 & 4.16e+38 & 0.00328 & 0.116686 \\
54 & ACO85J004154.04-094510.2 & 0.4613 & 0.4400 & 0.5000 & 4.30e-17$\pm$5.8e-18 & 39.1$\pm$5.5 & 3.02e+38 & 0.00238 & 0.116822 \\
55 & ACO85J004157.86-093516.5 & 0.7979 & 0.7800 & 0.8100 & 6.95e-17$\pm$4.7e-18 & 23.0$\pm$1.6 & 4.88e+38 & 0.00385 & 0.114943 \\
56 & ACO85J004158.81-092815.3 & 0.5625 & 0.5400 & 0.5900 & 6.36e-17$\pm$6.3e-18 & 56.1$\pm$5.9 & 4.46e+38 & 0.00353 & 0.117371 \\
57 & ACO85J004159.36-093010.9 & 0.4956 & 0.4700 & 0.5200 & 5.60e-17$\pm$6.3e-18 & 27.1$\pm$3.1 & 3.93e+38 & 0.00311 & 0.116414 \\
59 & ACO85J004202.99-093302.1 & 0.5143 & 0.4900 & 0.5400 & 3.21e-17$\pm$4.0e-18 & 28.2$\pm$3.6 & 2.25e+38 & 0.00178 & 0.11534 \\
62 & ACO85J004205.67-094627.1 & 0.5615 & 0.5300 & 0.5900 & 7.27e-17$\pm$7.5e-18 & 19.6$\pm$2.1 & 5.11e+38 & 0.00403 & 0.117371 \\
63 & ACO85J004205.79-093026.0 & 0.5200 & 0.5000 & 0.5400 & 8.15e-17$\pm$6.9e-18 & 29.2$\pm$2.6 & 5.72e+38 & 0.00452 & 0.116279 \\
64 & ACO85J004205.86-094310.1 & 0.4659 & 0.4600 & 0.4800 & 5.47e-16$\pm$5.2e-18 & 22.5$\pm$0.2 & 3.84e+39 & 0.03030 & 0.116279 \\
68 & ACO85J004208.27-092942.7 & 0.4708 & 0.4500 & 0.4900 & 8.43e-17$\pm$7.5e-18 & 27.8$\pm$2.5 & 5.92e+38 & 0.00467 & 0.11655 \\
70 & ACO85J004208.67-093506.5 & 0.5853 & 0.5600 & 0.6200 & 1.45e-16$\pm$9.4e-18 & 63.3$\pm$4.4 & 1.02e+39 & 0.00804 & 0.115075 \\
73 & ACO85J004210.63-093129.7 & 0.7336 & 0.6900 & 0.7800 & 6.80e-17$\pm$7.6e-18 & 48.8$\pm$5.7 & 4.77e+38 & 0.00377 & 0.115875 \\
76 & ACO85J004216.93-093325.6 & 0.4148 & 0.3900 & 0.4400 & 9.71e-17$\pm$7.7e-18 & 37.4$\pm$3.1 & 6.82e+38 & 0.00539 & 0.11534 \\
77 & ACO85J004217.94-093620.9 & 0.7158 & 0.6900 & 0.7400 & 8.65e-17$\pm$5.2e-18 & 75.9$\pm$4.9 & 6.07e+38 & 0.00480 & 0.115075 \\
81 & ACO85J004220.58-093526.4 & 0.4747 & 0.4300 & 0.5800 & 6.08e-17$\pm$6.2e-18 & 26.1$\pm$2.7 & 4.27e+38 & 0.00337 & 0.115075 \\
82 & ACO85J004220.87-094517.5 & 0.7646 & 0.7300 & 0.8100 & 7.51e-17$\pm$7.1e-18 & 55.9$\pm$5.6 & 5.27e+38 & 0.00416 & 0.116822 \\
83 & ACO85J004224.68-092716.2 & 0.4294 & 0.4000 & 0.4800 & 9.11e-17$\pm$5.8e-18 & 39.2$\pm$2.6 & 6.39e+38 & 0.00505 & 0.117233 \\
85 & ACO85J004225.48-093538.6 & 0.5634 & 0.5400 & 0.5900 & 8.63e-17$\pm$7.2e-18 & 51.5$\pm$4.5 & 6.06e+38 & 0.00478 & 0.115075 \\
87 & ACO85J004226.35-093629.5 & 0.4816 & 0.4600 & 0.5100 & 6.69e-17$\pm$6.1e-18 & 26.2$\pm$2.4 & 4.70e+38 & 0.00371 & 0.115075 \\
88 & ACO85J004227.26-093116.5 & 0.7001 & 0.6600 & 0.7800 & 6.72e-17$\pm$3.4e-18 & 27.4$\pm$1.4 & 4.72e+38 & 0.00373 & 0.115741 \\
89 & ACO85J004227.58-095059.3 & 0.8126 & 0.7700 & 0.8400 & 1.04e-16$\pm$7.1e-18 & 51.0$\pm$3.6 & 7.33e+38 & 0.00579 & 0.116822 \\
91 & ACO85J004228.83-094523.9 & 0.5059 & 0.4900 & 0.5300 & 3.71e-17$\pm$5.1e-18 & 21.7$\pm$3.1 & 2.60e+38 & 0.00206 & 0.11655 \\
93 & ACO85J004233.31-094448.0 & 0.5896 & 0.5600 & 0.6200 & 7.12e-17$\pm$7.3e-18 & 23.2$\pm$2.4 & 5.00e+38 & 0.00395 & 0.116144 \\
96 & ACO85J004238.03-093229.9 & 0.6229 & 0.5900 & 0.6500 & 8.08e-17$\pm$8.5e-18 & 53.9$\pm$5.9 & 5.67e+38 & 0.00448 & 0.115075 \\
98 & ACO85J004242.54-094726.4 & 0.4562 & 0.4500 & 0.4700 & 1.61e-15$\pm$8.0e-18 & 60.5$\pm$0.3 & 1.13e+40 & 0.08943 & 0.115075 \\
101 & ACO85J004247.85-092522.6 & 0.8305 & 0.8000 & 0.8600 & 6.50e-17$\pm$9.2e-18 & 96.4$\pm$14.8 & 4.56e+38 & 0.00361 & 0.114679 \\
\hline
\end{tabular}
\label{tab:catzphot}
\end{table*}

\clearpage

\begin{table*}
\centering
\caption{Catalogue of the 18 galaxies detected in \ha\ and with no
photometric redshifts.}
\begin{tabular}{rrrrrrr}
\hline
Num & Name~~~~~~~~ & F(\ha)~~~~~~ & EW(\ha) & $L$(\ha) & SFR & $n$~~~ \\
   &              & erg cm$^{-2}$ s$^{-1}$ & ~~\AA & erg s$^{-1}$ & M$_\odot$ yr$^{-1}$ & \\
\hline
 5 & ACO85J004043.91-092101.7 & 1.26e-16$\pm$4.7e-18 & 34.3$\pm$1.3 & 8.82e+38 & 0.00697 & 0.107296 \\
 9 & ACO85J004050.05-094402.5 & 6.24e-17$\pm$6.5e-18 & 45.7$\pm$4.9 & 4.38e+38 & 0.00346 & 0.110132 \\
20 & ACO85J004115.20-093856.8 & 3.97e-17$\pm$4.3e-18 & 20.6$\pm$2.3 & 2.79e+38 & 0.00220 & 0.114811 \\
32 & ACO85J004126.27-092101.5 & 1.02e-16$\pm$8.5e-18 & 42.2$\pm$3.7 & 7.17e+38 & 0.00566 & 0.11976 \\
36 & ACO85J004131.77-093832.1 & 3.09e-16$\pm$9.6e-18 & 55.1$\pm$1.8 & 2.17e+39 & 0.01714 & 0.115075 \\
37 & ACO85J004131.80-092303.7 & 6.66e-17$\pm$7.0e-18 & 56.2$\pm$6.3 & 4.67e+38 & 0.00369 & 0.11976 \\
39 & ACO85J004136.55-091939.5 & 5.66e-17$\pm$8.9e-18 & 14.5$\pm$2.3 & 3.98e+38 & 0.00314 & 0.120627 \\
41 & ACO85J004139.37-092316.3 & 6.85e-17$\pm$7.7e-18 & 63.0$\pm$7.5 & 4.81e+38 & 0.00380 & 0.120048 \\
43 & ACO85J004141.04-092444.9 & 7.36e-17$\pm$7.7e-18 & 57.0$\pm$6.3 & 5.17e+38 & 0.00408 & 0.119332 \\
46 & ACO85J004148.91-092618.7 & 7.45e-17$\pm$7.6e-18 & 64.1$\pm$7.0 & 5.23e+38 & 0.00413 & 0.118483 \\
60 & ACO85J004204.90-094108.6 & 9.07e-17$\pm$7.1e-18 & 24.4$\pm$2.0 & 6.36e+38 & 0.00503 & 0.115607 \\
61 & ACO85J004205.15-093715.5 & 2.39e-16$\pm$6.0e-18 & 155.4$\pm$4.5 & 1.68e+39 & 0.01326 & 0.114943 \\
71 & ACO85J004209.19-094056.6 & 9.47e-17$\pm$8.0e-18 & 57.1$\pm$5.1 & 6.64e+38 & 0.00525 & 0.115607 \\
74 & ACO85J004214.92-092735.4 & 5.99e-17$\pm$7.7e-18 & 26.5$\pm$3.5 & 4.20e+38 & 0.00332 & 0.117371 \\
75 & ACO85J004215.64-094209.2 & 4.77e-17$\pm$6.7e-18 & 34.7$\pm$5.0 & 3.35e+38 & 0.00265 & 0.115875 \\
84 & ACO85J004224.74-093741.3 & 3.37e-17$\pm$3.1e-18 & 37.9$\pm$3.6 & 2.36e+38 & 0.00187 & 0.115075 \\
97 & ACO85J004242.24-092108.7 & 6.00e-17$\pm$6.7e-18 & 105.8$\pm$13.0 & 4.21e+38 & 0.00333 & 0.11534 \\
100 &ACO85J004245.68-092327.8 & 4.90e-17$\pm$7.7e-18 & 72.7$\pm$12.2 & 3.44e+38 & 0.00272 & 0.115075 \\
\hline
\end{tabular}
\label{tab:catnozphot}
\end{table*}

\clearpage

\begin{table*}
\centering
\caption{Broad band magnitudes of the 101 objects detected in \ha}
\begin{tabular}{rrrrrrr}
\hline
Num & Name~~~~~~~~  &  $R_C$ & $r'$ & $u^*$ & $g'$ & $i'$ \\
\hline
1 & ACO85J004037.99-095038.8 & 21.02 & 21.48 & 25.24 & 23.04 & 20.48 \\
2 & ACO85J004039.27-092933.8 & 20.65 & 21.08 & 24.80 & 22.36 & 19.48 \\
3 & ACO85J004039.90-093812.5 & 19.34 & 19.68 & 23.72 & 21.28 & 18.96 \\
4 & ACO85J004041.11-093054.7 & 20.04 & 20.58 & 24.41 & 22.14 & 19.48 \\
5 & ACO85J004043.91-092101.7 & 18.95 & --- & --- & --- & --- \\
6 & ACO85J004044.33-093334.4 & 20.08 & 20.60 & 24.41 & 22.23 & 19.57 \\
7 & ACO85J004045.38-093140.9 & 19.81 & 20.23 & 23.54 & 21.74 & 19.41 \\
8 & ACO85J004045.80-093929.1 & 21.17 & 21.82 & 24.51 & 23.27 & 20.37 \\
9 & ACO85J004050.05-094402.5 & 20.19 & 20.70 & 24.38 & 22.21 & 19.70 \\
10 & ACO85J004053.86-094845.6 & 21.39 & 21.95 & 24.83 & 23.41 & 20.58 \\
11 & ACO85J004055.06-094954.9 & 20.13 & 20.53 & 24.26 & 22.08 & 19.72 \\
12 & ACO85J004103.65-093342.1 & 21.30 & 22.04 & 25.34 & 23.28 & 20.22 \\
13 & ACO85J004104.27-093627.8 & 20.46 & 21.08 & 24.45 & 22.32 & 19.40 \\
14 & ACO85J004104.67-092608.6 & 19.66 & 19.92 & 22.88 & 21.23 & 19.22 \\
15 & ACO85J004106.35-094321.1 & 19.76 & 20.14 & 23.47 & 21.69 & 19.40 \\
16 & ACO85J004108.66-093932.2 & 20.37 & 20.88 & 24.18 & 22.32 & 20.03 \\
17 & ACO85J004109.80-093503.0 & 17.73 & 17.95 & 20.93 & 19.16 & 17.45 \\
18 & ACO85J004110.45-093247.7 & 20.68 & 20.75 & 23.38 & 22.11 & 19.81 \\
19 & ACO85J004113.96-094015.3 & 19.56 & 19.95 & 23.64 & 21.48 & 19.23 \\
20 & ACO85J004115.20-093856.8 & 19.74 & --- & --- & --- & --- \\
21 & ACO85J004115.27-093053.9 & 21.57 & 21.88 & 25.68 & 23.54 & 20.79 \\
22 & ACO85J004115.38-094134.6 & 18.02 & 18.37 & 21.09 & 19.43 & 17.86 \\
23 & ACO85J004115.55-093041.3 & 19.99 & 20.34 & 23.58 & 21.88 & 19.42 \\
24 & ACO85J004119.01-092323.5 & 17.33 & --- & --- & --- & --- \\
25 & ACO85J004119.11-093312.0 & 19.08 & 19.17 & 23.03 & 20.77 & 18.48 \\
26 & ACO85J004119.83-092327.0 & 16.55 & --- & --- & --- & --- \\
27 & ACO85J004122.02-094156.0 & 21.26 & 21.52 & 24.10 & 22.87 & 20.29 \\
28 & ACO85J004122.17-092639.5 & 21.26 & 21.70 & 22.28 & 21.98 & 21.54 \\
29 & ACO85J004123.20-093208.6 & 20.68 & 20.98 & 26.28 & 22.60 & 19.88 \\
30 & ACO85J004124.32-092600.2 & 19.76 & 19.94 & 23.42 & 21.48 & 19.17 \\
31 & ACO85J004124.48-093405.3 & 20.21 & 20.30 & 21.76 & 21.06 & 19.99 \\
32 & ACO85J004126.27-092101.5 & 19.67 & --- & --- & --- & --- \\
33 & ACO85J004127.13-092857.6 & 20.39 & 20.70 & 24.30 & 22.19 & 19.58 \\
34 & ACO85J004128.95-092837.1 & 20.58 & 20.90 & 25.05 & 22.45 & 19.74 \\
35 & ACO85J004129.77-093313.2 & 19.21 & 19.33 & 19.94 & 19.38 & 19.10 \\
36 & ACO85J004131.77-093832.1 & 18.88 & --- & --- & --- & --- \\
37 & ACO85J004131.80-092303.7 & 20.26 & --- & --- & --- & --- \\
38 & ACO85J004132.70-092800.7 & 19.86 & 20.11 & 23.14 & 21.56 & 19.06 \\
39 & ACO85J004136.55-091939.5 & 18.95 & --- & --- & --- & --- \\
40 & ACO85J004138.01-092938.0 & 18.50 & 18.58 & 20.10 & 19.03 & 18.28 \\
41 & ACO85J004139.37-092316.3 & 20.87 & --- & --- & --- & --- \\
42 & ACO85J004140.92-093454.9 & 19.28 & 20.00 & 23.35 & 21.18 & 18.29 \\
43 & ACO85J004141.04-092444.9 & 20.56 & 20.94 & 24.27 & 22.45 & 19.96 \\
44 & ACO85J004141.67-093409.5 & 19.21 & 19.59 & 23.37 & 21.24 & 18.80 \\
45 & ACO85J004145.45-094033.1 & 16.79 & 16.98 & 18.63 & 17.49 & 16.69 \\
46 & ACO85J004148.91-092618.7 & 20.85 & 21.37 & 25.03 & 22.95 & 20.29 \\
47 & ACO85J004149.38-092818.3 & 17.90 & 18.26 & 19.99 & 18.81 & 17.95 \\
48 & ACO85J004150.17-092547.6 & 14.51 & 15.26 & 17.87 & 16.09 & 14.83 \\
49 & ACO85J004150.75-092714.8 & 19.13 & 19.63 & 22.44 & 20.69 & 18.34 \\
50 & ACO85J004150.88-092836.9 & 20.49 & 21.00 & 25.14 & 22.57 & 19.82 \\
51 & ACO85J004150.94-092938.1 & 19.51 & 19.78 & 23.50 & 21.39 & 19.09 \\
52 & ACO85J004153.27-092930.4 & 15.74 & 16.04 & 17.12 & 16.34 & 15.83 \\
53 & ACO85J004153.51-092943.8 & 14.90 & 15.27 & 17.53 & 15.97 & 14.88 \\
54 & ACO85J004154.04-094510.2 & 20.77 & 21.11 & 24.11 & 22.58 & 20.25 \\
55 & ACO85J004157.86-093516.5 & 19.56 & 20.05 & 23.09 & 21.13 & 18.94 \\
56 & ACO85J004158.81-092815.3 & 20.17 & 20.56 & 23.82 & 21.98 & 19.60 \\
57 & ACO85J004159.36-093010.9 & 20.13 & 20.56 & 24.35 & 22.14 & 19.77 \\
58 & ACO85J004159.84-094230.9 & 17.36 & 17.58 & 18.83 & 17.92 & 17.41 \\
59 & ACO85J004202.99-093302.1 & 20.19 & 20.68 & 24.73 & 22.33 & 19.89 \\
60 & ACO85J004204.90-094108.6 & 19.38 & 19.66 & 23.40 & 21.26 & 18.98 \\
\hline
\end{tabular}
\end{table*}
\addtocounter{table}{-1}
\begin{table*}
\centering
\caption{Broad band magnitudes of the 101 objects detected in \ha (continued)}
\begin{tabular}{rrrrrrr}
\hline
Num & Name~~~~~~~~  &  $R_C$ & $r'$ & $u^*$ & $g'$ & $i'$ \\
\hline
61 & ACO85J004205.15-093715.5 & 20.20 & 20.59 & 22.91 & 21.68 & 20.03 \\
62 & ACO85J004205.67-094627.1 & 19.31 & 19.63 & 22.13 & 20.96 & 18.95 \\
63 & ACO85J004205.79-093026.0 & 19.74 & 20.19 & 24.13 & 21.84 & 19.37 \\
64 & ACO85J004205.86-094310.1 & 17.49 & 17.92 & 21.40 & 19.16 & 16.69 \\
65 & ACO85J004206.02-093606.4 & 16.20 & 16.38 & 18.62 & 17.10 & 15.99 \\
66 & ACO85J004207.26-093626.0 & 18.15 & 18.36 & 19.13 & 18.51 & 18.20 \\
67 & ACO85J004207.71-093059.3 & 18.31 & 18.84 & 22.92 & 20.52 & 18.09 \\
68 & ACO85J004208.27-092942.7 & 19.57 & 20.04 & 23.84 & 21.65 & 19.26 \\
69 & ACO85J004208.36-093104.6 & 17.87 & 18.20 & 19.81 & 18.68 & 17.89 \\
70 & ACO85J004208.67-093506.5 & 19.77 & 20.26 & 23.24 & 21.74 & 19.36 \\
71 & ACO85J004209.19-094056.6 & 20.32 & 20.77 & 23.04 & 21.93 & 19.91 \\
72 & ACO85J004209.81-092852.2 & 17.65 & 17.94 & 18.97 & 18.17 & 17.73 \\
73 & ACO85J004210.63-093129.7 & 20.35 & 20.80 & 24.54 & 22.33 & 19.66 \\
74 & ACO85J004214.92-092735.4 & 19.74 & 20.12 & 23.84 & 21.67 & 19.37 \\
75 & ACO85J004215.64-094209.2 & 20.32 & 20.82 & 24.26 & 22.27 & 19.77 \\
76 & ACO85J004216.93-093325.6 & 19.91 & 20.05 & 23.30 & 21.13 & 19.06 \\
77 & ACO85J004217.94-093620.9 & 20.71 & 21.23 & 24.82 & 22.51 & 19.48 \\
78 & ACO85J004218.47-093912.1 & 17.50 & --- & --- & --- & --- \\
79 & ACO85J004218.55-093910.2 & 19.55 & --- & --- & --- & --- \\
80 & ACO85J004219.89-092527.5 & 17.20 & 17.37 & 19.03 & 17.87 & 17.08 \\
81 & ACO85J004220.58-093526.4 & 19.88 & 20.11 & 21.44 & 20.77 & 19.73 \\
82 & ACO85J004220.87-094517.5 & 20.43 & 20.65 & 24.51 & 22.20 & 19.59 \\
83 & ACO85J004224.68-092716.2 & 20.00 & 20.29 & 21.34 & 20.84 & 20.09 \\
84 & ACO85J004224.74-093741.3 & 20.48 & 20.68 & 24.35 & 22.29 & 19.91 \\
85 & ACO85J004225.48-093538.6 & 20.00 & 20.32 & 24.14 & 21.81 & 19.35 \\
86 & ACO85J004225.54-093708.9 & 17.47 & 17.54 & 19.91 & 18.29 & 17.15 \\
87 & ACO85J004226.35-093629.5 & 19.67 & 19.82 & 23.16 & 21.35 & 19.08 \\
88 & ACO85J004227.26-093116.5 & 19.43 & 20.12 & 23.63 & 21.37 & 18.91 \\
89 & ACO85J004227.58-095059.3 & 20.19 & 20.38 & 23.01 & 21.52 & 18.92 \\
90 & ACO85J004228.38-094938.3 & 17.66 & 17.79 & 19.65 & 18.38 & 17.47 \\
91 & ACO85J004228.83-094523.9 & 20.08 & 20.25 & 24.10 & 21.83 & 19.51 \\
92 & ACO85J004232.84-092144.2 & 18.37 & --- & --- & --- & --- \\
93 & ACO85J004233.31-094448.0 & 19.58 & 19.68 & 22.77 & 21.18 & 18.99 \\
94 & ACO85J004236.76-094403.8 & 18.69 & 18.84 & 20.09 & 19.18 & 18.55 \\
95 & ACO85J004237.07-094520.5 & 17.83 & 18.02 & 19.71 & 18.61 & 17.65 \\
96 & ACO85J004238.03-093229.9 & 20.36 & 20.75 & 24.49 & 22.39 & 19.69 \\
97 & ACO85J004242.24-092108.7 & 21.39 & --- & --- & --- & --- \\
98 & ACO85J004242.54-094726.4 & 17.42 & 17.80 & 21.30 & 19.02 & 16.26 \\
99 & ACO85J004243.90-094420.8 & 16.12 & 16.34 & 18.54 & 17.05 & 15.96 \\
100 & ACO85J004245.68-092327.8 & 21.07 & --- & --- & --- & ---R \\
101 & ACO85J004247.85-092522.6 & 21.13 & 22.01 & 25.46 & 23.21 & 19.93 \\
\hline
\end{tabular}
\label{tab:mags}
\end{table*}

\clearpage

\begin{figure}
\centering
\includegraphics[width=7cm,angle=0]{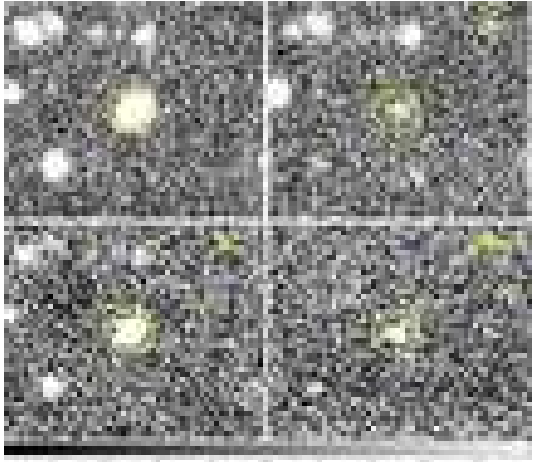}
\caption[]{Images of galaxy ACO85J004037.99-095038.8, in the $i'$ (top
left), $u^*$ (top right), $R_C$ (bottom left) and ``pure'' \ha\
(bottom right) bands.  The circles (with radius of 2\arcsec) indicate
the position of the galaxy detected in the $R_C$ band (they do not
indicate the region in which the \ha\ fluxes were measured).  Each
image is 23.3\arcsec$\times$19.5\arcsec.
}
\end{figure}



\begin{figure}
\centering
\includegraphics[width=7cm,angle=0]{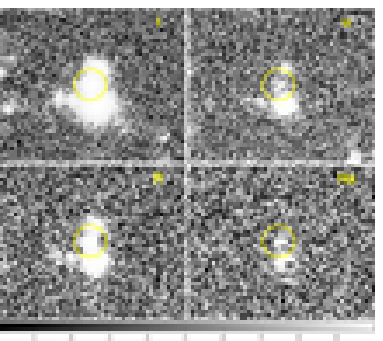}
\caption[]{Same as Fig..1 for galaxy ACO85J004039.27-092933.8}
\end{figure}

\begin{figure}
\centering
\includegraphics[width=7cm,angle=0]{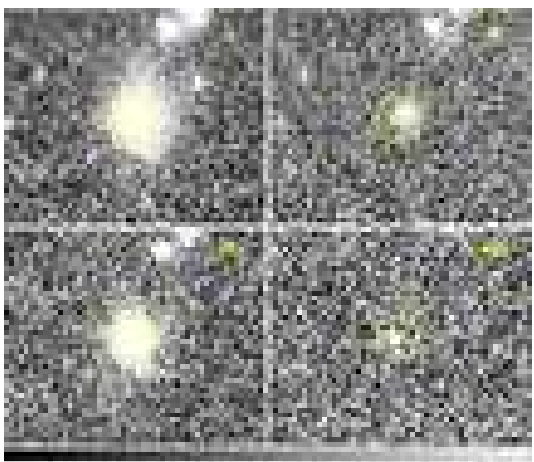}
\caption[]{Same as Fig..1 for galaxy ACO85J004039.90-093812.5}
\end{figure}

\begin{figure}
\centering
\includegraphics[width=7cm,angle=0]{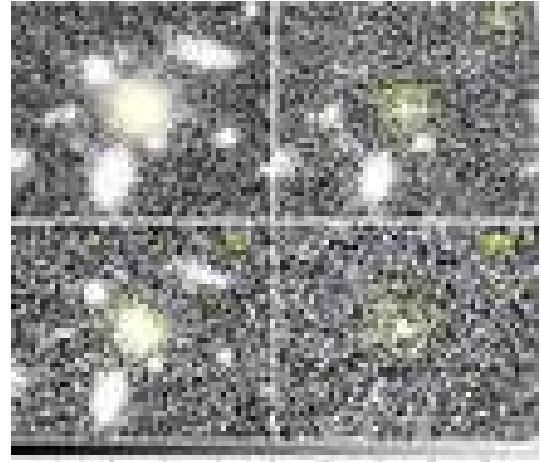}
\caption[]{Same as Fig..1 for galaxy ACO85J004041.11-093054.7}
\end{figure}

\begin{figure}
\centering
\includegraphics[width=7cm,angle=0]{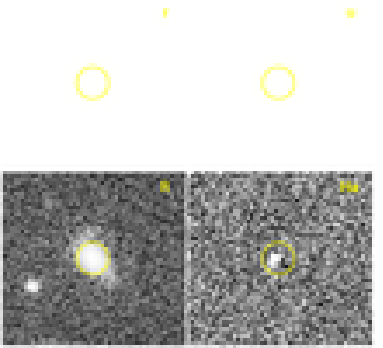}
\caption[]{Same as Fig..1 for galaxy ACO85J004043.91-092101.7}
\end{figure}

\begin{figure}
\centering
\includegraphics[width=7cm,angle=0]{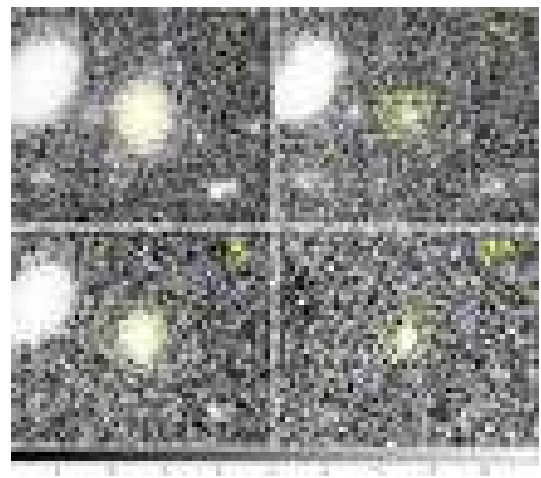}
\caption[]{Same as Fig..1 for galaxy ACO85J004044.33-093334.4}
\end{figure}
     
\clearpage

\begin{figure}
\centering
\includegraphics[width=7cm,angle=0]{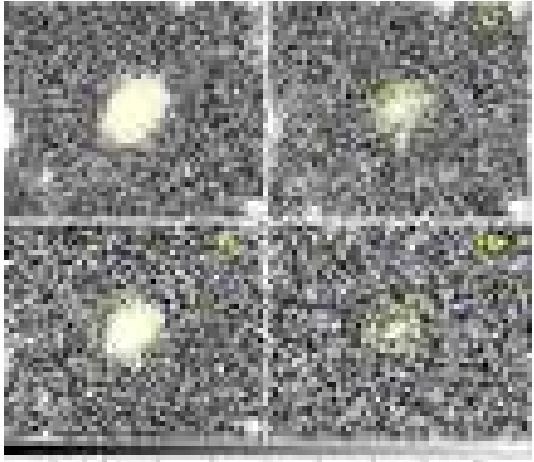}
\caption[]{Same as Fig..1 for galaxy ACO85J004045.38-093140.9}
\end{figure}

\begin{figure}
\centering
\includegraphics[width=7cm,angle=0]{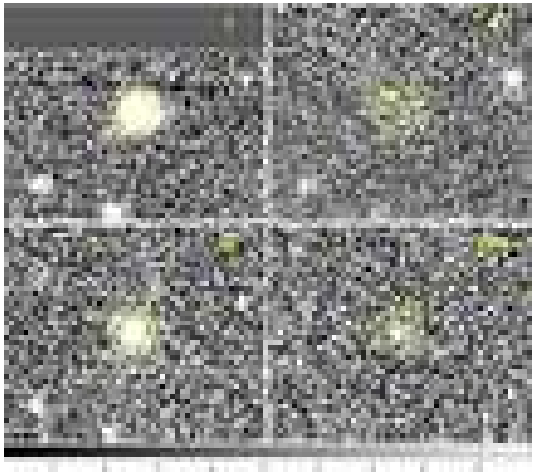}
\caption[]{Same as Fig..1 for galaxy ACO85J004045.80-093929.1}
\end{figure}

\begin{figure}
\centering
\includegraphics[width=7cm,angle=0]{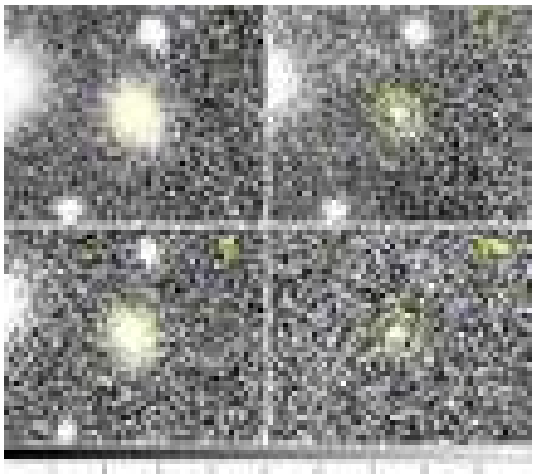}
\caption[]{Same as Fig..1 for galaxy ACO85J004050.05-094402.5}
\end{figure}

\begin{figure}
\centering
\includegraphics[width=7cm,angle=0]{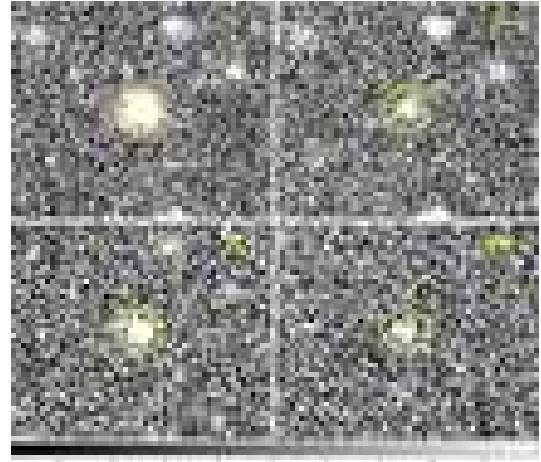}
\caption[]{Same as Fig..1 for galaxy ACO85J004053.86-094845.6}
\end{figure}

\begin{figure}
\centering
\includegraphics[width=7cm,angle=0]{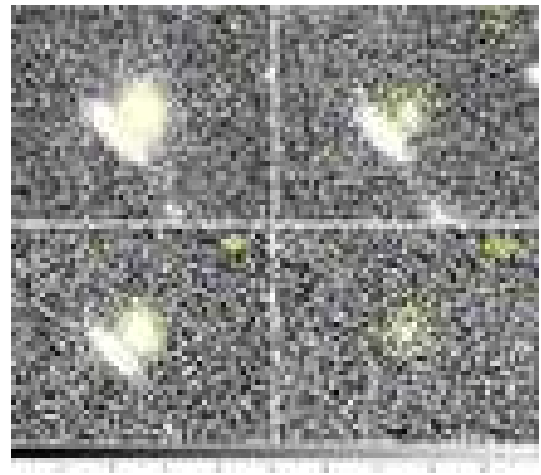}
\caption[]{Same as Fig..1 for galaxy ACO85J004055.06-094954.9}
\end{figure}

\begin{figure}
\centering
\includegraphics[width=7cm,angle=0]{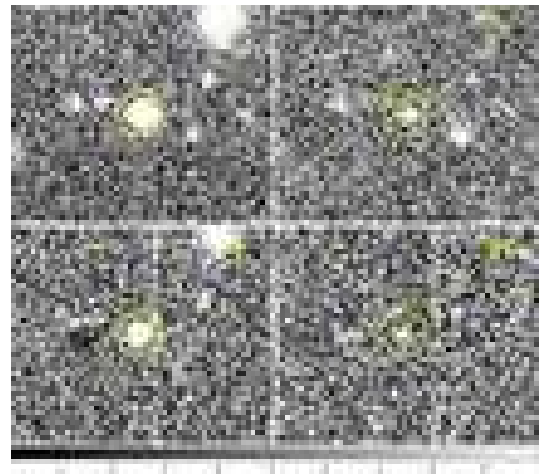}
\caption[]{Same as Fig..1 for galaxy ACO85J004103.65-093342.1}
\end{figure}
     
\clearpage

\begin{figure}
\centering
\includegraphics[width=7cm,angle=0]{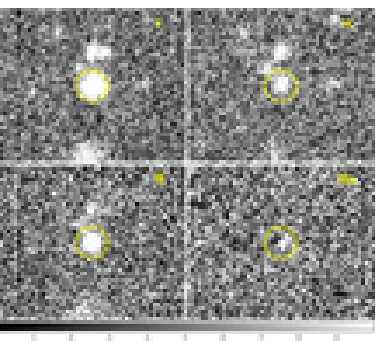}
\caption[]{Same as Fig..1 for galaxy ACO85J004104.27-093627.8}
\end{figure}

\begin{figure}
\centering
\includegraphics[width=7cm,angle=0]{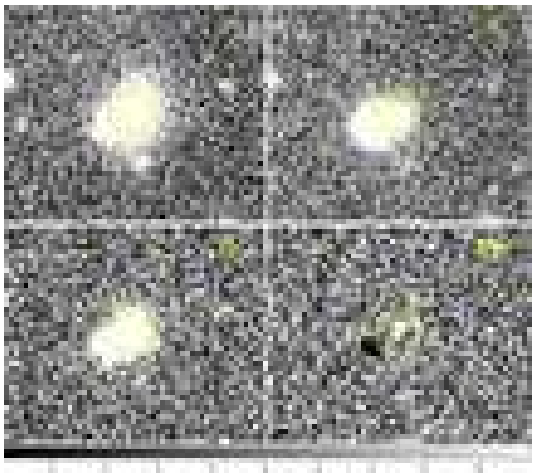}
\caption[]{Same as Fig..1 for galaxy ACO85J004104.67-092608.6}
\end{figure}

\begin{figure}
\centering
\includegraphics[width=7cm,angle=0]{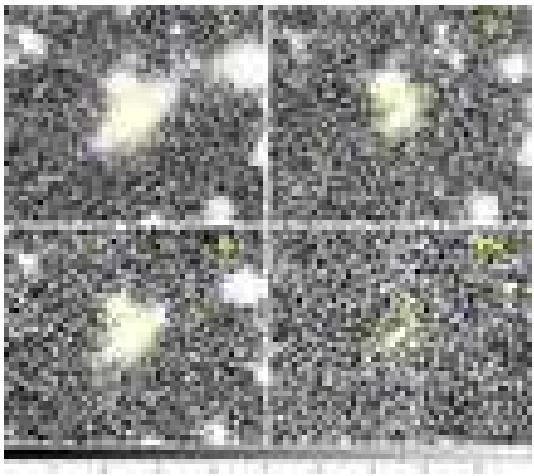}
\caption[]{Same as Fig..1 for galaxy ACO85J004106.35-094321.1}
\end{figure}

\begin{figure}
\centering
\includegraphics[width=7cm,angle=0]{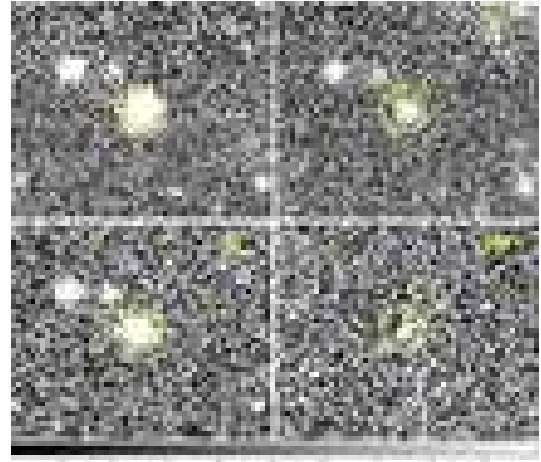}
\caption[]{Same as Fig..1 for galaxy ACO85J004108.66-093932.2}
\end{figure}

\begin{figure}
\centering
\includegraphics[width=7cm,angle=0]{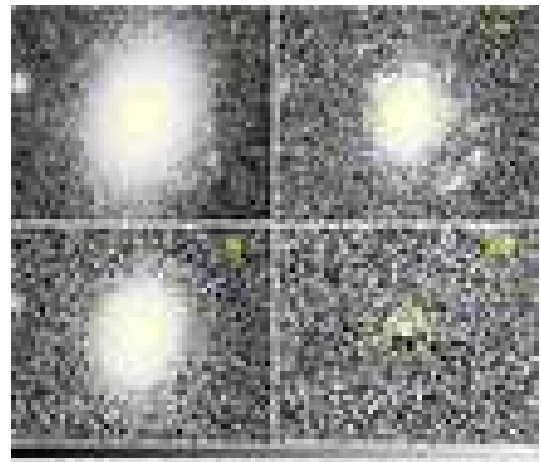}
\caption[]{Same as Fig..1 for galaxy ACO85J004109.80-093503.0}
\end{figure}

\begin{figure}
\centering
\includegraphics[width=7cm,angle=0]{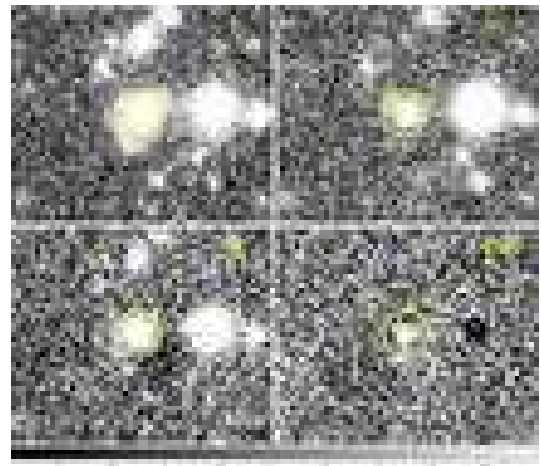}
\caption[]{Same as Fig..1 for galaxy ACO85J004110.45-093247.7}
\end{figure}
     
\clearpage

\begin{figure}
\centering
\includegraphics[width=7cm,angle=0]{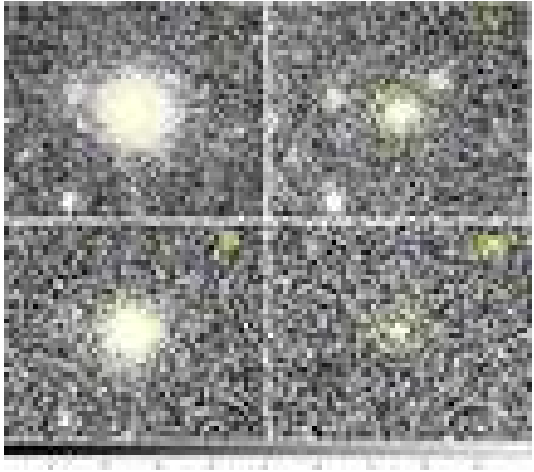}
\caption[]{Same as Fig..1 for galaxy ACO85J004113.96-094015.3}
\end{figure}

\begin{figure}
\centering
\includegraphics[width=7cm,angle=0]{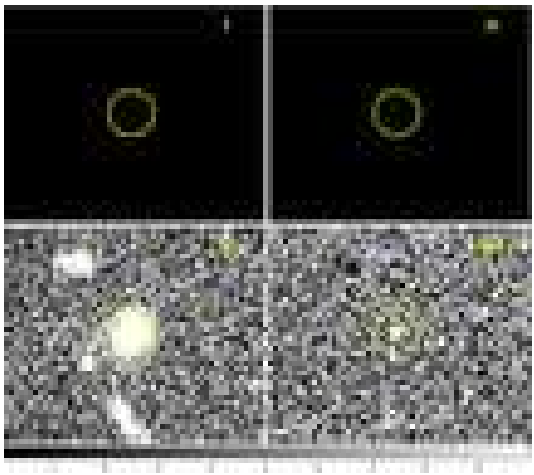}
\caption[]{Same as Fig..1 for galaxy ACO85J004115.20-093856.8}
\end{figure}

\begin{figure}
\centering
\includegraphics[width=7cm,angle=0]{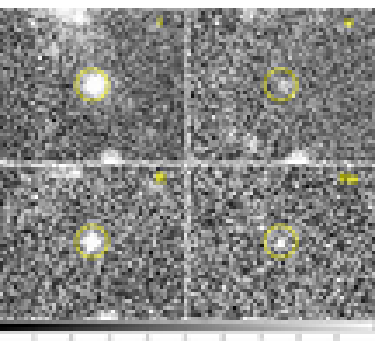}
\caption[]{Same as Fig..1 for galaxy ACO85J004115.27-093053.9}
\end{figure}

\begin{figure}
\centering
\includegraphics[width=7cm,angle=0]{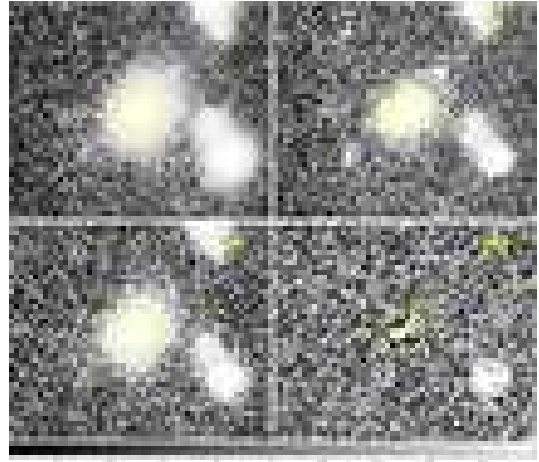}
\caption[]{Same as Fig..1 for galaxy ACO85J004115.38-094134.6}
\end{figure}

\begin{figure}
\centering
\includegraphics[width=7cm,angle=0]{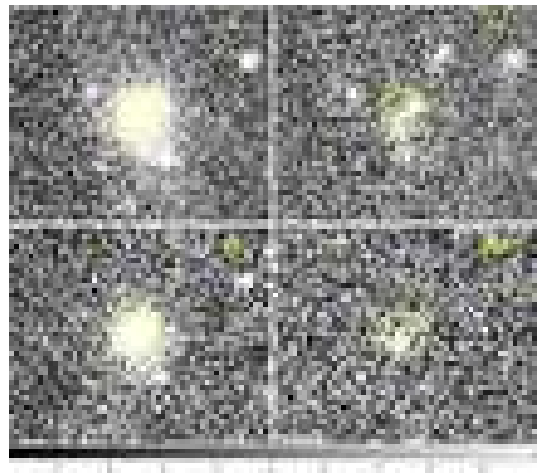}
\caption[]{Same as Fig..1 for galaxy ACO85J004115.55-093041.3}
\end{figure}

\begin{figure}
\centering
\includegraphics[width=7cm,angle=0]{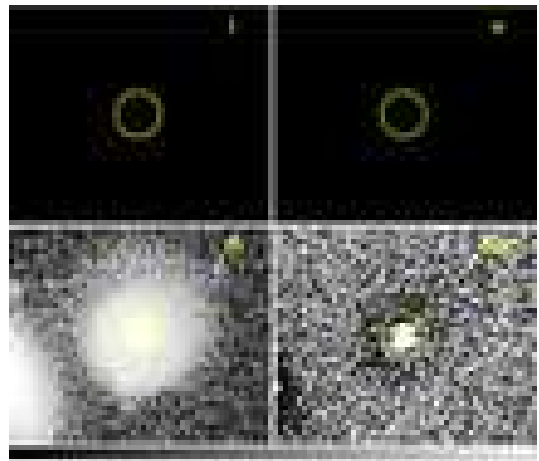}
\caption[]{Same as Fig..1 for galaxy ACO85J004119.01-092323.5}
\end{figure}
     
\clearpage

\begin{figure}
\centering
\includegraphics[width=7cm,angle=0]{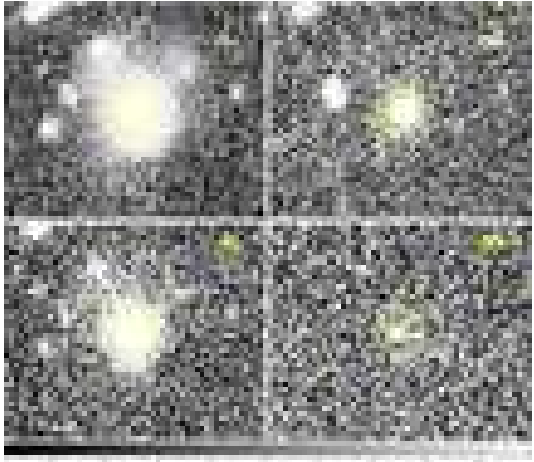}
\caption[]{Same as Fig..1 for galaxy ACO85J004119.11-093312.0}
\end{figure}

\begin{figure}
\centering
\includegraphics[width=7cm,angle=0]{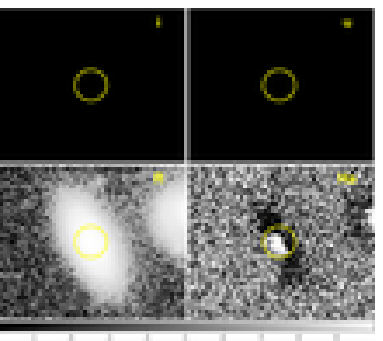}
\caption[]{Same as Fig..1 for galaxy ACO85J004119.83-092327.0}
\end{figure}

\begin{figure}
\centering
\includegraphics[width=7cm,angle=0]{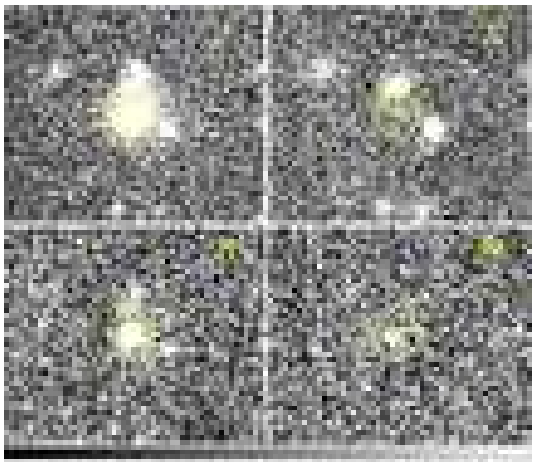}
\caption[]{Same as Fig..1 for galaxy ACO85J004122.02-094156.0}
\end{figure}

\begin{figure}
\centering
\includegraphics[width=7cm,angle=0]{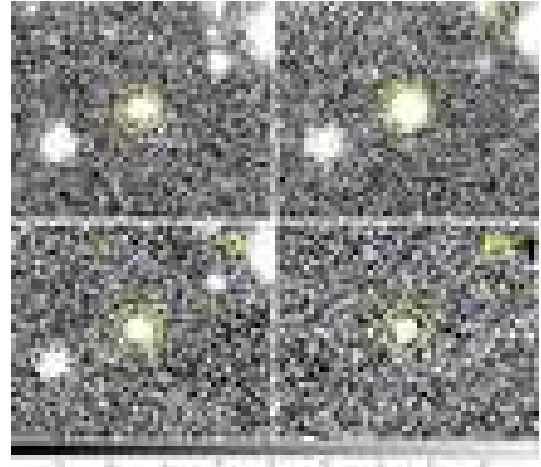}
\caption[]{Same as Fig..1 for galaxy ACO85J004122.17-092639.5}
\end{figure}

\begin{figure}
\centering
\includegraphics[width=7cm,angle=0]{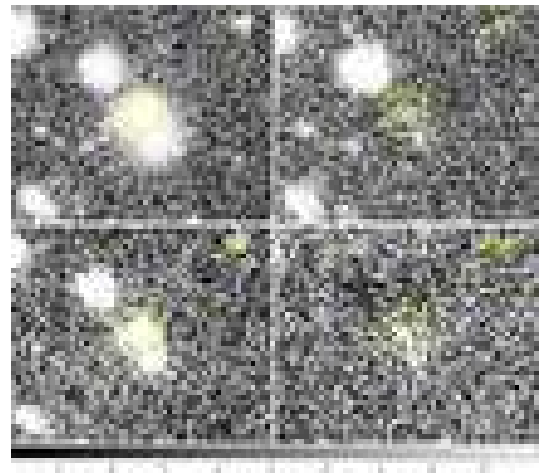}
\caption[]{Same as Fig..1 for galaxy ACO85J004123.20-093208.6}
\end{figure}

\begin{figure}
\centering
\includegraphics[width=7cm,angle=0]{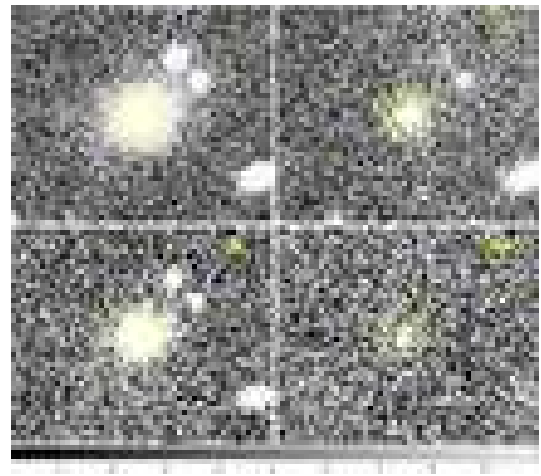}
\caption[]{Same as Fig..1 for galaxy ACO85J004124.32-092600.2}
\end{figure}
     
\clearpage

\begin{figure}
\centering
\includegraphics[width=7cm,angle=0]{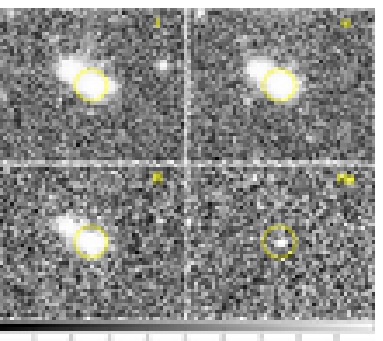}
\caption[]{Same as Fig..1 for galaxy ACO85J004124.48-093405.3}
\end{figure}

\begin{figure}
\centering
\includegraphics[width=7cm,angle=0]{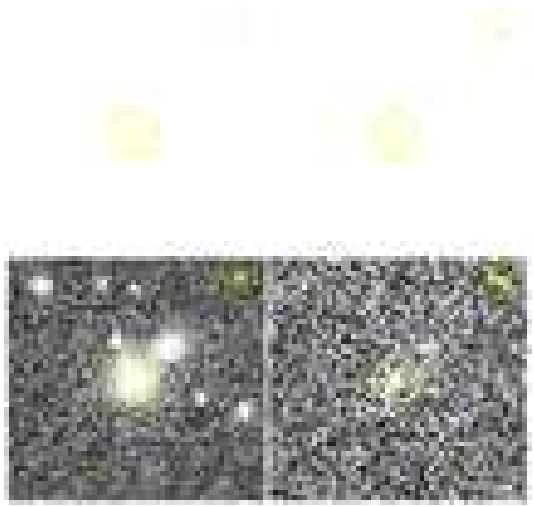}
\caption[]{Same as Fig..1 for galaxy ACO85J004126.27-092101.5}
\end{figure}

\begin{figure}
\centering
\includegraphics[width=7cm,angle=0]{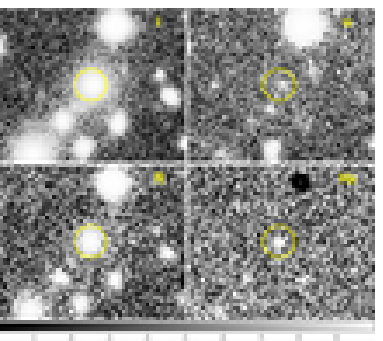}
\caption[]{Same as Fig..1 for galaxy ACO85J004127.13-092857.6}
\end{figure}

\begin{figure}
\centering
\includegraphics[width=7cm,angle=0]{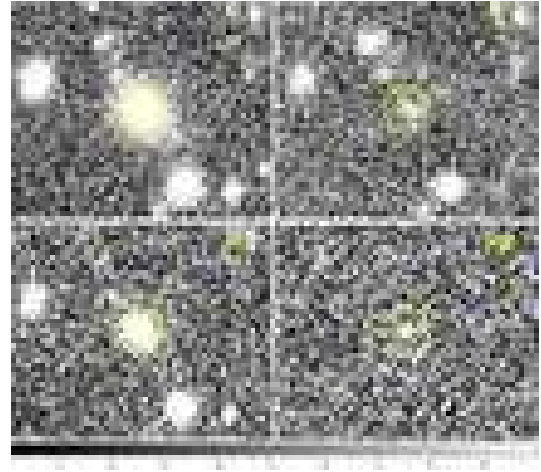}
\caption[]{Same as Fig..1 for galaxy ACO85J004128.95-092837.1}
\end{figure}

\begin{figure}
\centering
\includegraphics[width=7cm,angle=0]{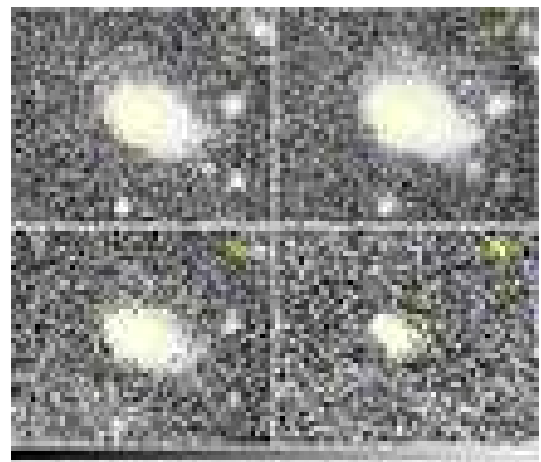}
\caption[]{Same as Fig..1 for galaxy ACO85J004129.77-093313.2}
\end{figure}

\begin{figure}
\centering
\includegraphics[width=7cm,angle=0]{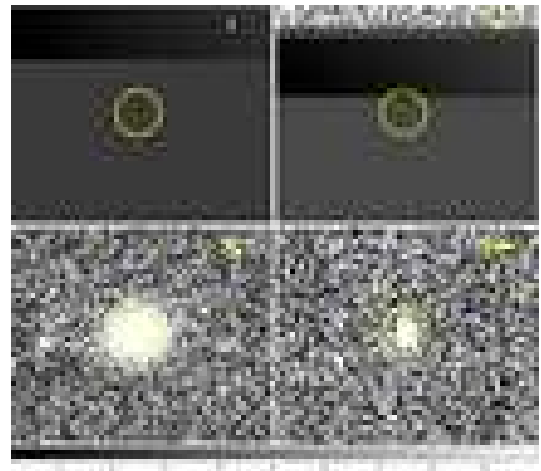}
\caption[]{Same as Fig..1 for galaxy ACO85J004131.77-093832.1}
\end{figure}
     
\clearpage

\begin{figure}
\centering
\includegraphics[width=7cm,angle=0]{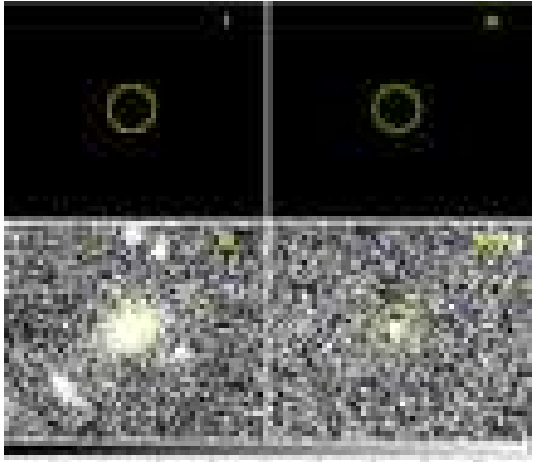}
\caption[]{Same as Fig..1 for galaxy ACO85J004131.80-092303.7}
\end{figure}

\begin{figure}
\centering
\includegraphics[width=7cm,angle=0]{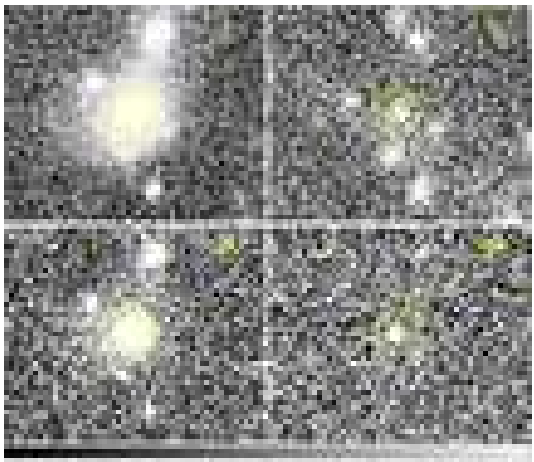}
\caption[]{Same as Fig..1 for galaxy ACO85J004132.70-092800.7}
\end{figure}

\begin{figure}
\centering
\includegraphics[width=7cm,angle=0]{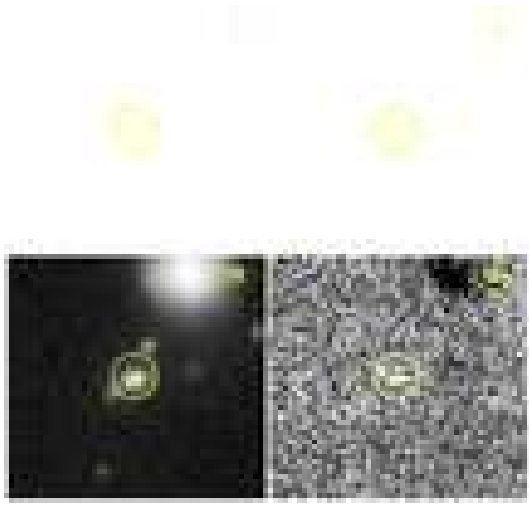}
\caption[]{Same as Fig..1 for galaxy ACO85J004136.55-091939.5}
\end{figure}

\begin{figure}
\centering
\includegraphics[width=7cm,angle=0]{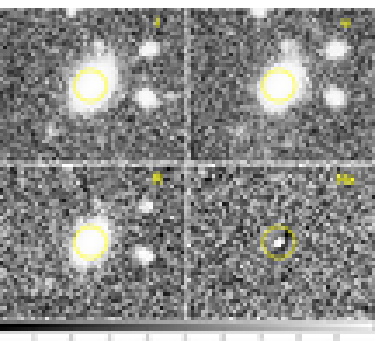}
\caption[]{Same as Fig..1 for galaxy ACO85J004138.01-092938.0}
\end{figure}

\begin{figure}
\centering
\includegraphics[width=7cm,angle=0]{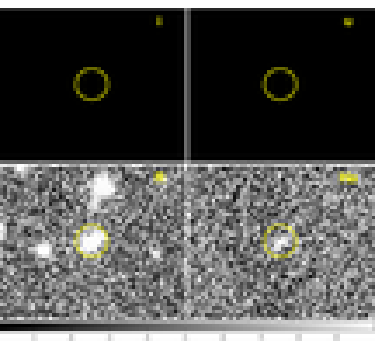}
\caption[]{Same as Fig..1 for galaxy ACO85J004139.37-092316.3}
\end{figure}

\begin{figure}
\centering
\includegraphics[width=7cm,angle=0]{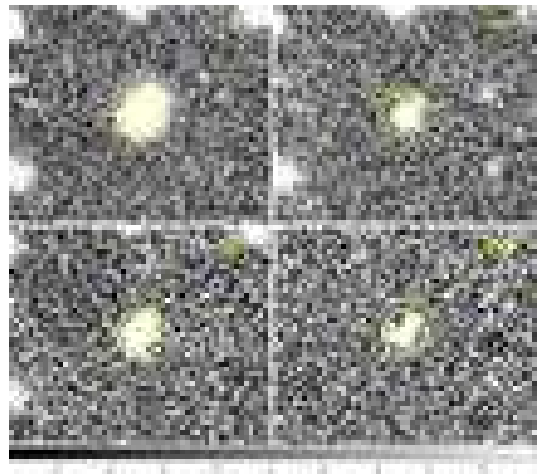}
\caption[]{Same as Fig..1 for galaxy ACO85J004140.92-093454.9}
\end{figure}
     
\clearpage

\begin{figure}
\centering
\includegraphics[width=7cm,angle=0]{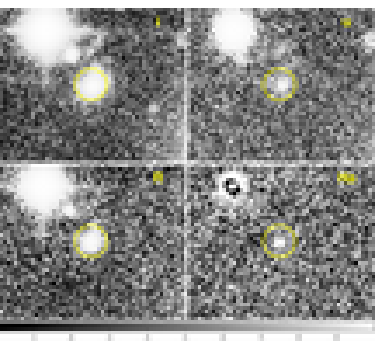}
\caption[]{Same as Fig..1 for galaxy ACO85J004141.04-092444.9}
\end{figure}

\begin{figure}
\centering
\includegraphics[width=7cm,angle=0]{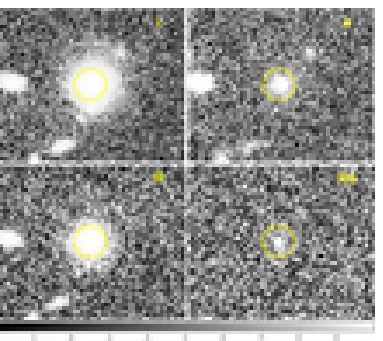}
\caption[]{Same as Fig..1 for galaxy ACO85J004141.67-093409.5}
\end{figure}

\begin{figure}
\centering
\includegraphics[width=7cm,angle=0]{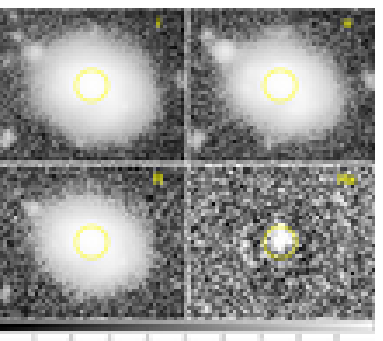}
\caption[]{Same as Fig..1 for galaxy ACO85J004145.45-094033.1}
\end{figure}

\begin{figure}
\centering
\includegraphics[width=7cm,angle=0]{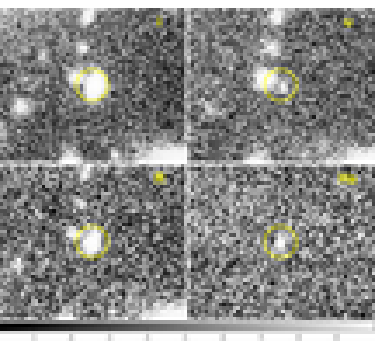}
\caption[]{Same as Fig..1 for galaxy ACO85J004148.91-092618.7}
\end{figure}

\begin{figure}
\centering
\includegraphics[width=7cm,angle=0]{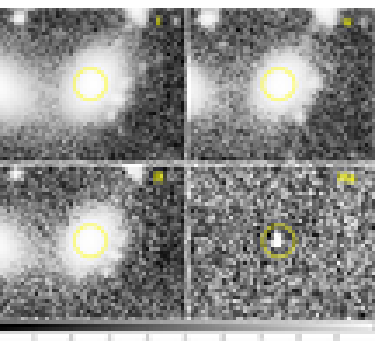}
\caption[]{Same as Fig..1 for galaxy ACO85J004149.38-092818.3}
\end{figure}

\begin{figure}
\centering
\includegraphics[width=7cm,angle=0]{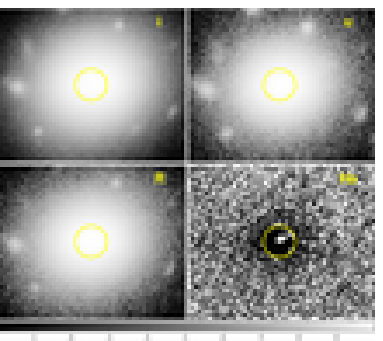}
\caption[]{Same as Fig..1 for galaxy ACO85J004150.17-092547.6}
\end{figure}
     
\clearpage

\begin{figure}
\centering
\includegraphics[width=7cm,angle=0]{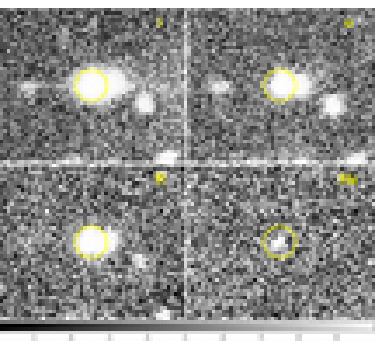}
\caption[]{Same as Fig..1 for galaxy ACO85J004150.75-092714.8}
\end{figure}

\begin{figure}
\centering
\includegraphics[width=7cm,angle=0]{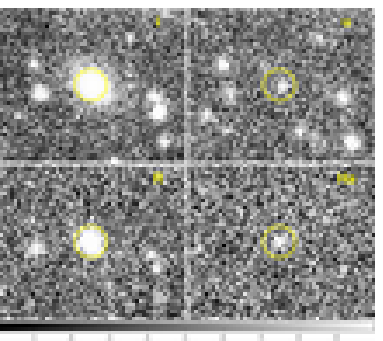}
\caption[]{Same as Fig..1 for galaxy ACO85J004150.88-092836.9}
\end{figure}

\begin{figure}
\centering
\includegraphics[width=7cm,angle=0]{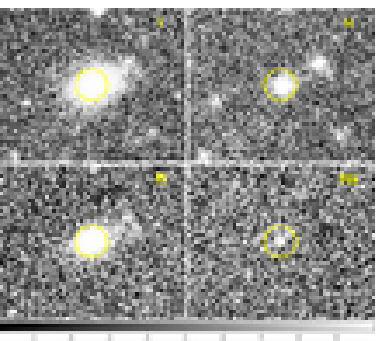}
\caption[]{Same as Fig..1 for galaxy ACO85J004150.94-092938.1}
\end{figure}

\begin{figure}
\centering
\includegraphics[width=7cm,angle=0]{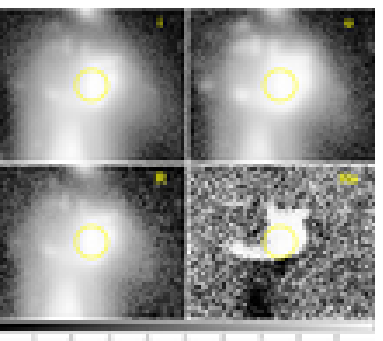}
\caption[]{Same as Fig..1 for galaxy ACO85J004153.27-092930.4}
\end{figure}

\begin{figure}
\centering
\includegraphics[width=7cm,angle=0]{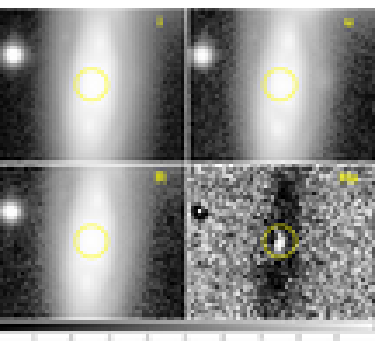}
\caption[]{Same as Fig..1 for galaxy ACO85J004153.51-092943.8}
\end{figure}

\begin{figure}
\centering
\includegraphics[width=7cm,angle=0]{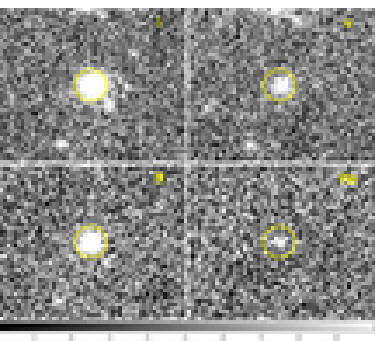}
\caption[]{Same as Fig..1 for galaxy ACO85J004154.04-094510.2}
\end{figure}
     
\clearpage

\begin{figure}
\centering
\includegraphics[width=7cm,angle=0]{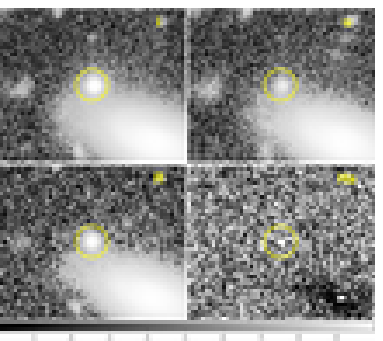}
\caption[]{Same as Fig..1 for galaxy ACO85J004157.86-093516.5}
\end{figure}

\begin{figure}
\centering
\includegraphics[width=7cm,angle=0]{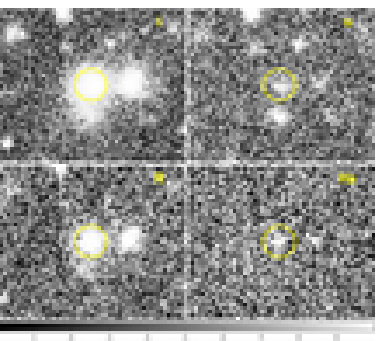}
\caption[]{Same as Fig..1 for galaxy ACO85J004158.81-092815.3}
\end{figure}

\begin{figure}
\centering
\includegraphics[width=7cm,angle=0]{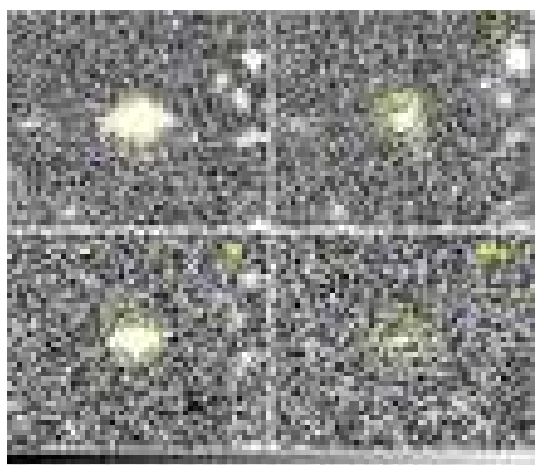}
\caption[]{Same as Fig..1 for galaxy ACO85J004159.36-093010.9}
\end{figure}

\begin{figure}
\centering
\includegraphics[width=7cm,angle=0]{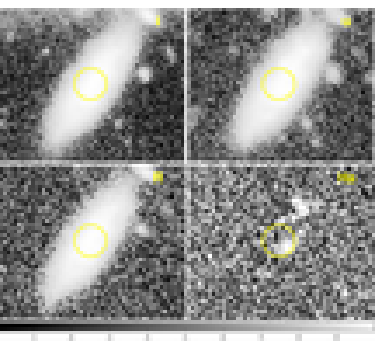}
\caption[]{Same as Fig..1 for galaxy ACO85J004159.84-094230.9}
\end{figure}

\begin{figure}
\centering
\includegraphics[width=7cm,angle=0]{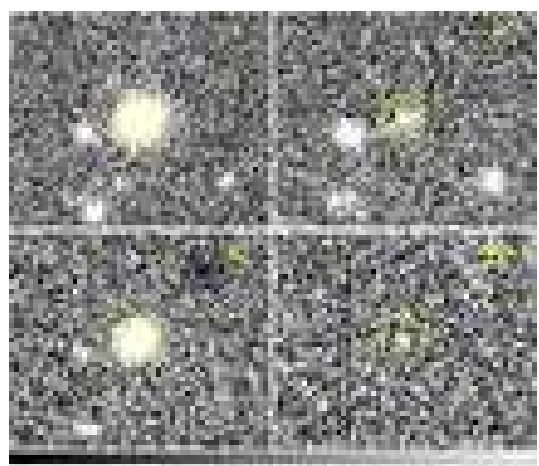}
\caption[]{Same as Fig..1 for galaxy ACO85J004202.99-093302.1}
\end{figure}

\begin{figure}
\centering
\includegraphics[width=7cm,angle=0]{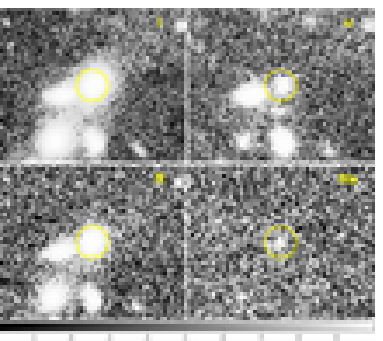}
\caption[]{Same as Fig..1 for galaxy ACO85J004204.90-094108.6}
\end{figure}
     
\clearpage

\begin{figure}
\centering
\includegraphics[width=7cm,angle=0]{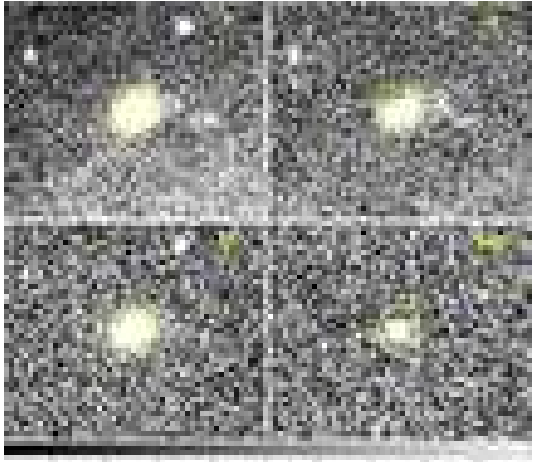}
\caption[]{Same as Fig..1 for galaxy ACO85J004205.15-093715.5}
\end{figure}

\begin{figure}
\centering
\includegraphics[width=7cm,angle=0]{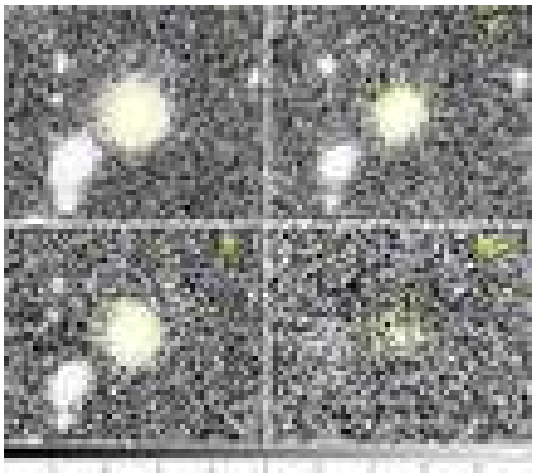}
\caption[]{Same as Fig..1 for galaxy ACO85J004205.67-094627.1}
\end{figure}

\begin{figure}
\centering
\includegraphics[width=7cm,angle=0]{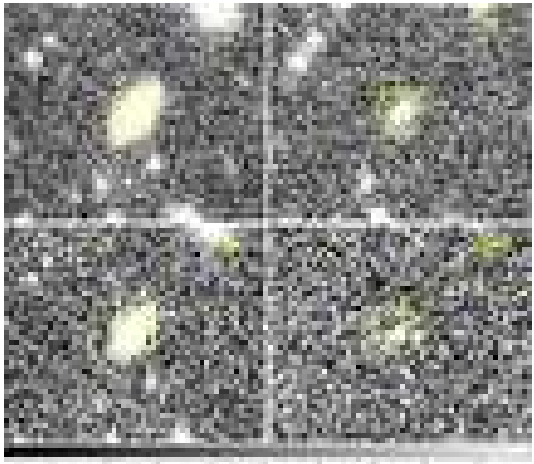}
\caption[]{Same as Fig..1 for galaxy ACO85J004205.79-093026.0}
\end{figure}

\begin{figure}
\centering
\includegraphics[width=7cm,angle=0]{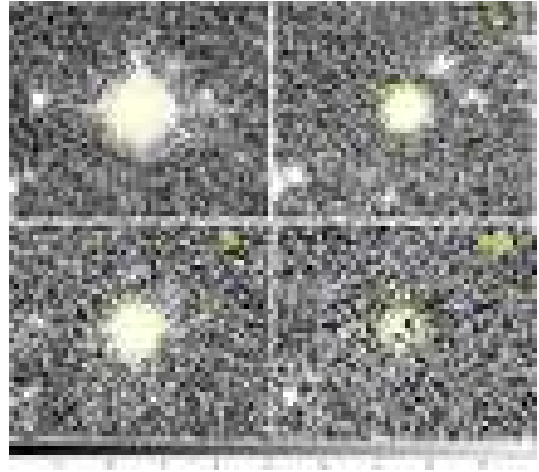}
\caption[]{Same as Fig..1 for galaxy ACO85J004205.86-094310.1}
\end{figure}

\begin{figure}
\centering
\includegraphics[width=7cm,angle=0]{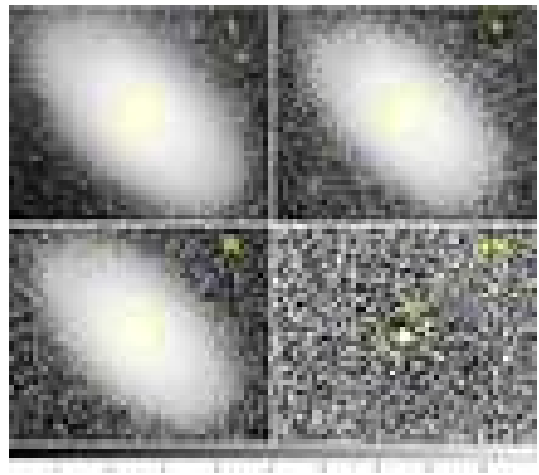}
\caption[]{Same as Fig..1 for galaxy ACO85J004206.02-093606.4}
\end{figure}

\begin{figure}
\centering
\includegraphics[width=7cm,angle=0]{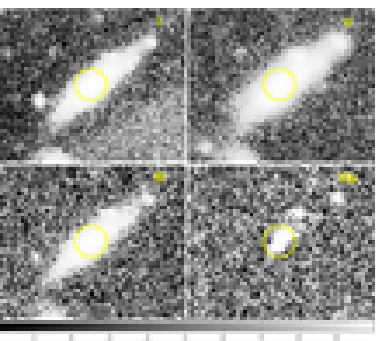}
\caption[]{Same as Fig..1 for galaxy ACO85J004207.26-093626.0}
\end{figure}
     
\clearpage

\begin{figure}
\centering
\includegraphics[width=7cm,angle=0]{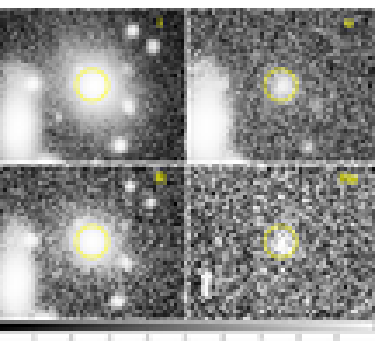}
\caption[]{Same as Fig..1 for galaxy ACO85J004207.71-093059.3}
\end{figure}

\begin{figure}
\centering
\includegraphics[width=7cm,angle=0]{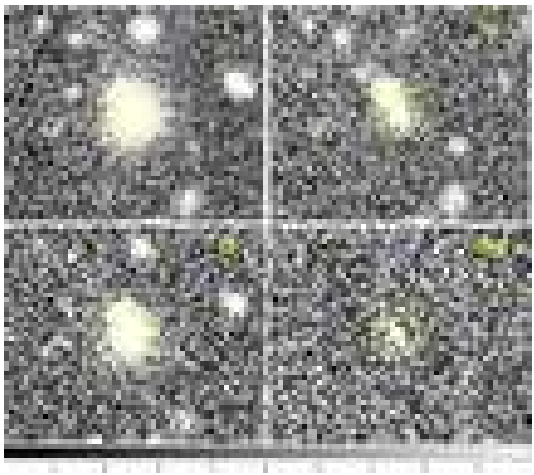}
\caption[]{Same as Fig..1 for galaxy ACO85J004208.27-092942.7}
\end{figure}

\begin{figure}
\centering
\includegraphics[width=7cm,angle=0]{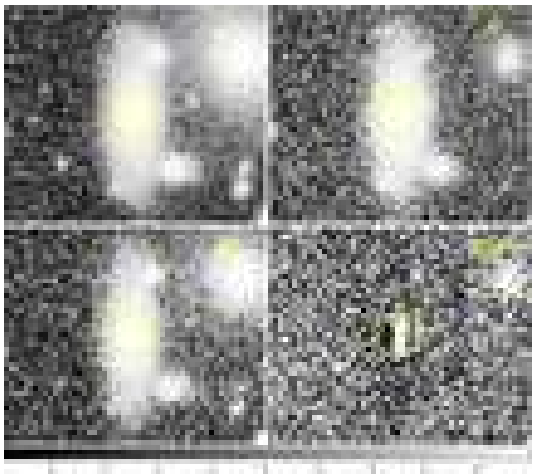}
\caption[]{Same as Fig..1 for galaxy ACO85J004208.36-093104.6}
\end{figure}

\begin{figure}
\centering
\includegraphics[width=7cm,angle=0]{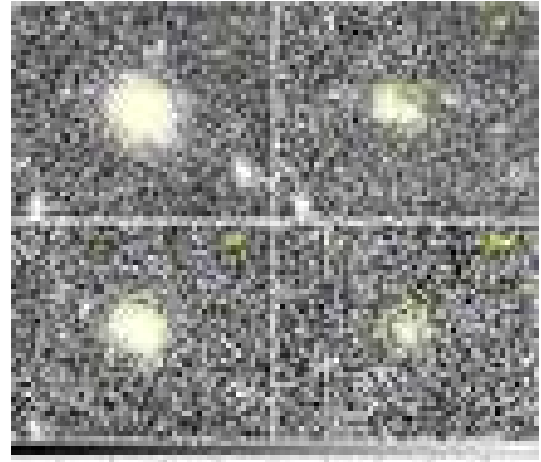}
\caption[]{Same as Fig..1 for galaxy ACO85J004208.67-093506.5}
\end{figure}

\begin{figure}
\centering
\includegraphics[width=7cm,angle=0]{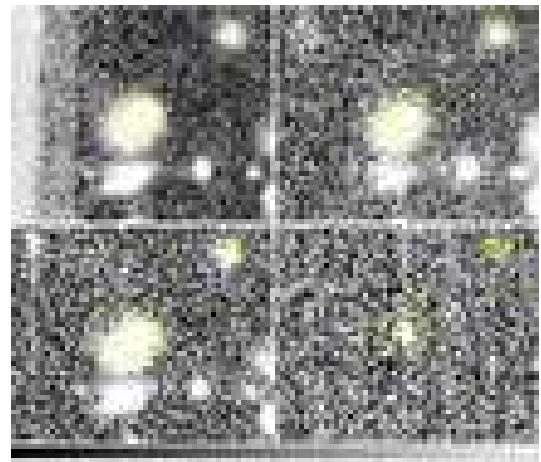}
\caption[]{Same as Fig..1 for galaxy ACO85J004209.19-094056.6}
\end{figure}

\begin{figure}
\centering
\includegraphics[width=7cm,angle=0]{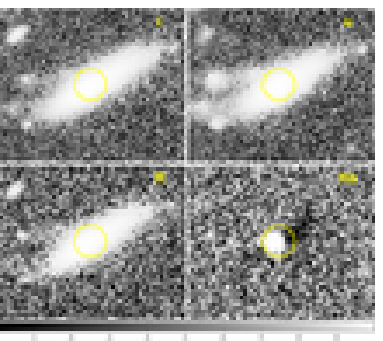}
\caption[]{Same as Fig..1 for galaxy ACO85J004209.81-092852.2}
\end{figure}
     
\clearpage

\begin{figure}
\centering
\includegraphics[width=7cm,angle=0]{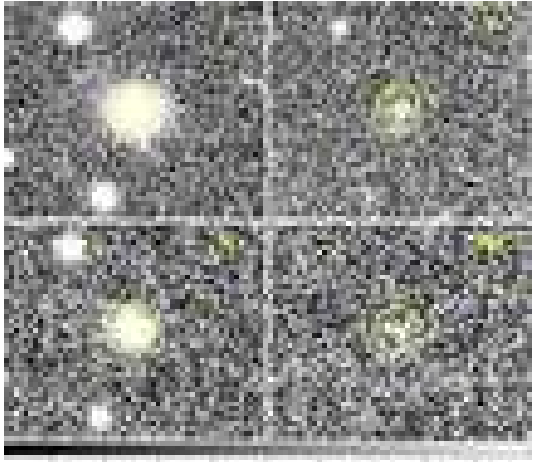}
\caption[]{Same as Fig..1 for galaxy ACO85J004210.63-093129.7}
\end{figure}

\begin{figure}
\centering
\includegraphics[width=7cm,angle=0]{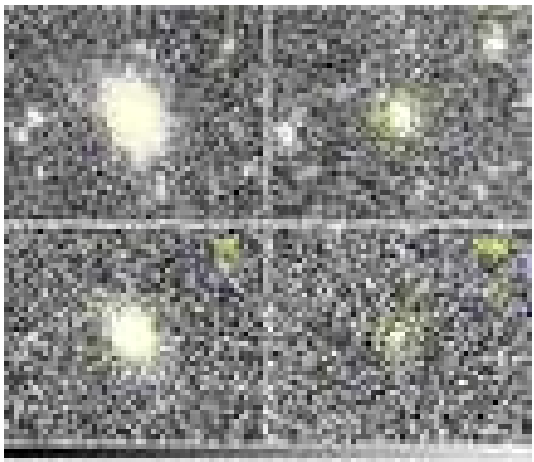}
\caption[]{Same as Fig..1 for galaxy ACO85J004214.92-092735.4}
\end{figure}

\begin{figure}
\centering
\includegraphics[width=7cm,angle=0]{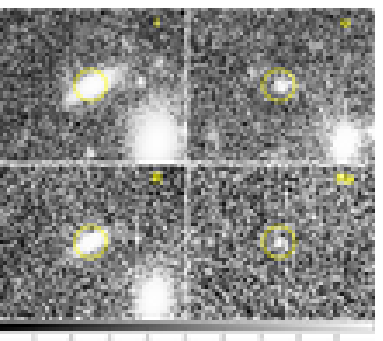}
\caption[]{Same as Fig..1 for galaxy ACO85J004215.64-094209.2}
\end{figure}

\begin{figure}
\centering
\includegraphics[width=7cm,angle=0]{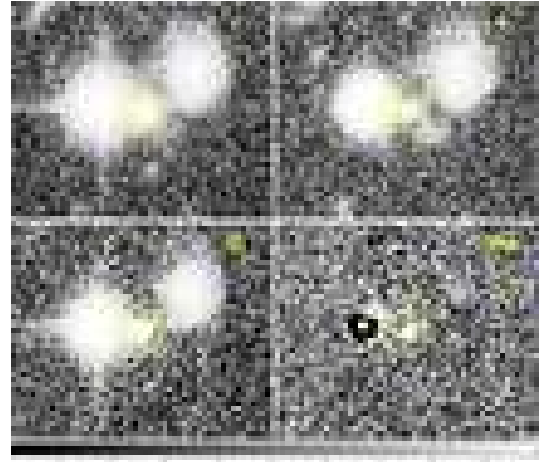}
\caption[]{Same as Fig..1 for galaxy ACO85J004216.93-093325.6}
\end{figure}

\begin{figure}
\centering
\includegraphics[width=7cm,angle=0]{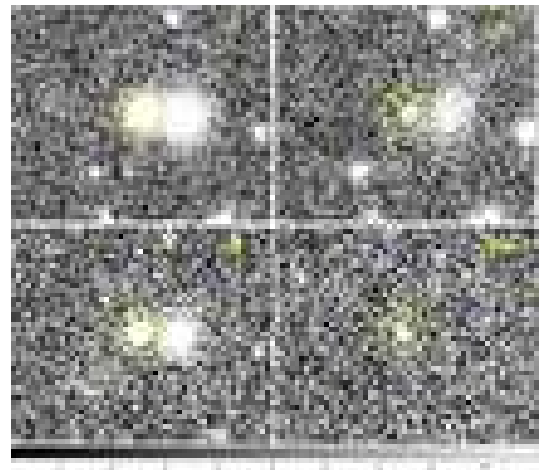}
\caption[]{Same as Fig..1 for galaxy ACO85J004217.94-093620.9}
\end{figure}

\begin{figure}
\centering
\includegraphics[width=7cm,angle=0]{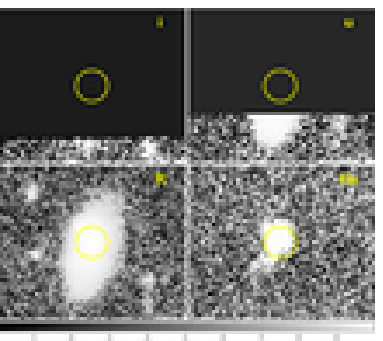}
\caption[]{Same as Fig..1 for galaxy ACO85J004218.47-093912.1}
\end{figure}
     
\clearpage

\begin{figure}
\centering
\includegraphics[width=7cm,angle=0]{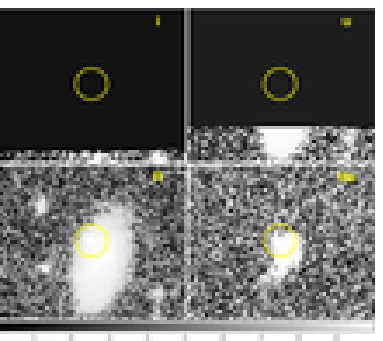}
\caption[]{Same as Fig..1 for galaxy ACO85J004218.55-093910.2}
\end{figure}

\begin{figure}
\centering
\includegraphics[width=7cm,angle=0]{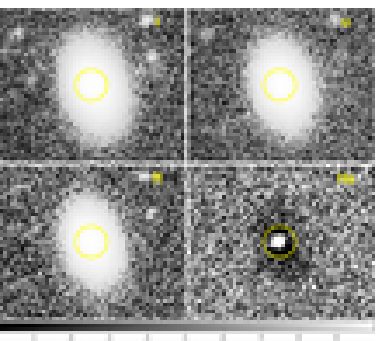}
\caption[]{Same as Fig..1 for galaxy ACO85J004219.89-092527.5}
\end{figure}

\begin{figure}
\centering
\includegraphics[width=7cm,angle=0]{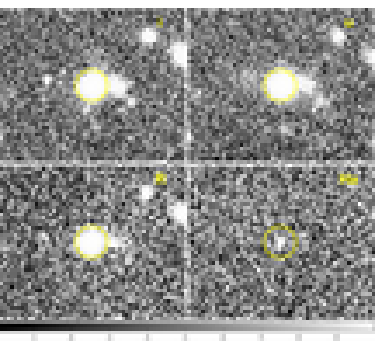}
\caption[]{Same as Fig..1 for galaxy ACO85J004220.58-093526.4}
\end{figure}

\begin{figure}
\centering
\includegraphics[width=7cm,angle=0]{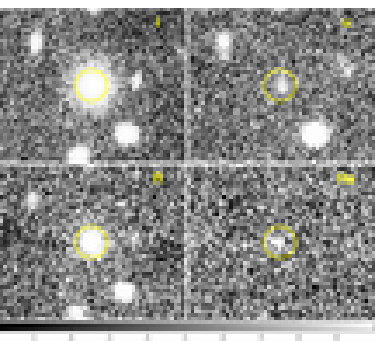}
\caption[]{Same as Fig..1 for galaxy ACO85J004220.87-094517.5}
\end{figure}

\begin{figure}
\centering
\includegraphics[width=7cm,angle=0]{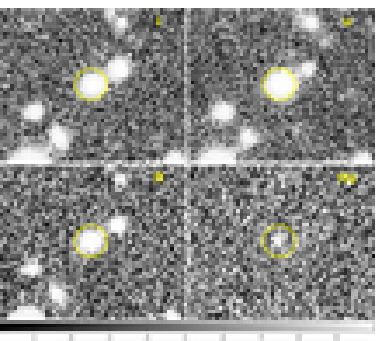}
\caption[]{Same as Fig..1 for galaxy ACO85J004224.68-092716.2}
\end{figure}

\begin{figure}
\centering
\includegraphics[width=7cm,angle=0]{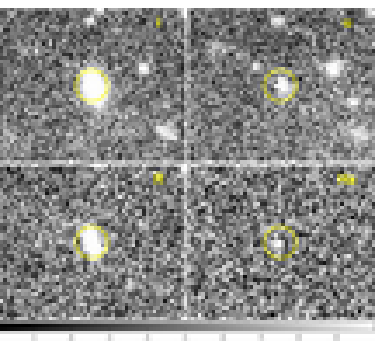}
\caption[]{Same as Fig..1 for galaxy ACO85J004224.74-093741.3}
\end{figure}
     
\clearpage

\begin{figure}
\centering
\includegraphics[width=7cm,angle=0]{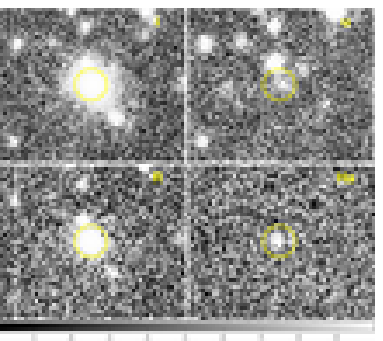}
\caption[]{Same as Fig..1 for galaxy ACO85J004225.48-093538.6}
\end{figure}

\begin{figure}
\centering
\includegraphics[width=7cm,angle=0]{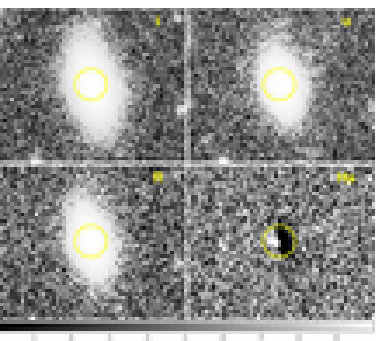}
\caption[]{Same as Fig..1 for galaxy ACO85J004225.54-093708.9}
\end{figure}

\begin{figure}
\centering
\includegraphics[width=7cm,angle=0]{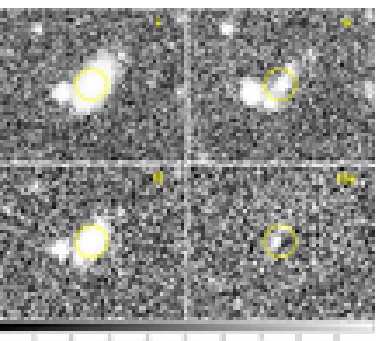}
\caption[]{Same as Fig..1 for galaxy ACO85J004226.35-093629.5}
\end{figure}

\begin{figure}
\centering
\includegraphics[width=7cm,angle=0]{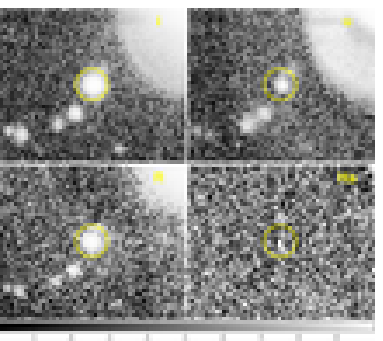}
\caption[]{Same as Fig..1 for galaxy ACO85J004227.26-093116.5}
\end{figure}

\begin{figure}
\centering
\includegraphics[width=7cm,angle=0]{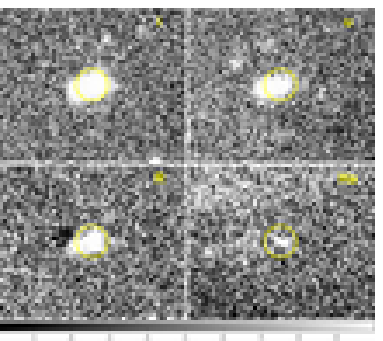}
\caption[]{Same as Fig..1 for galaxy ACO85J004227.58-095059.3}
\end{figure}

\begin{figure}
\centering
\includegraphics[width=7cm,angle=0]{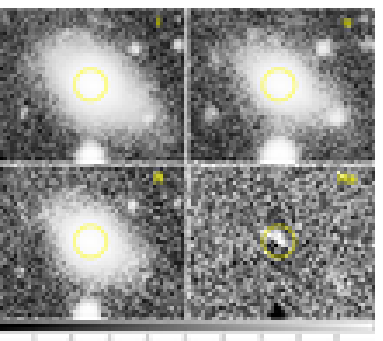}
\caption[]{Same as Fig..1 for galaxy ACO85J004228.38-094938.3}
\end{figure}
     
\clearpage

\begin{figure}
\centering
\includegraphics[width=7cm,angle=0]{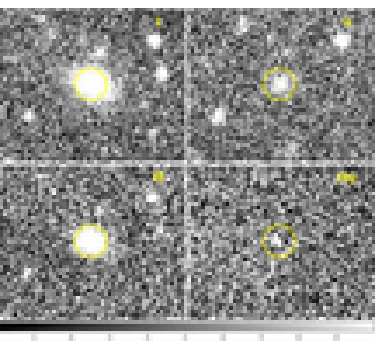}
\caption[]{Same as Fig..1 for galaxy ACO85J004228.83-094523.9}
\end{figure}

\begin{figure}
\centering
\includegraphics[width=7cm,angle=0]{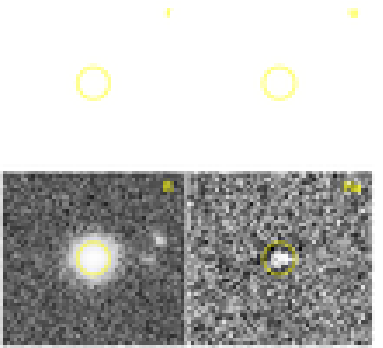}
\caption[]{Same as Fig..1 for galaxy ACO85J004232.84-092144.2}
\end{figure}

\begin{figure}
\centering
\includegraphics[width=7cm,angle=0]{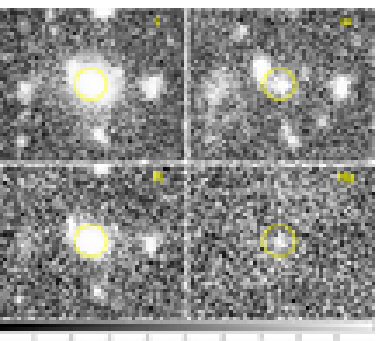}
\caption[]{Same as Fig..1 for galaxy ACO85J004233.31-094448.0}
\end{figure}

\begin{figure}
\centering
\includegraphics[width=7cm,angle=0]{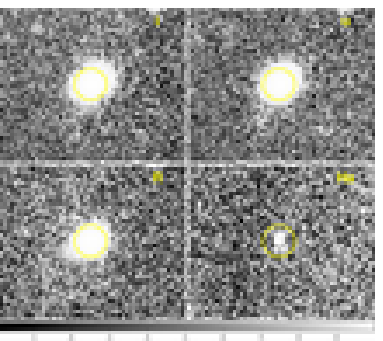}
\caption[]{Same as Fig..1 for galaxy ACO85J004236.76-094403.8}
\end{figure}

\begin{figure}
\centering
\includegraphics[width=7cm,angle=0]{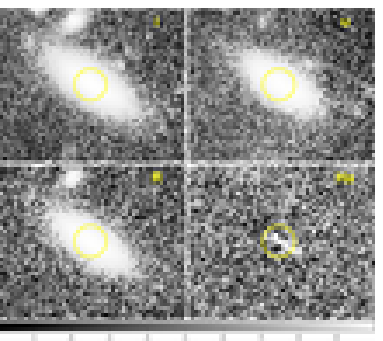}
\caption[]{Same as Fig..1 for galaxy ACO85J004237.07-094520.5}
\end{figure}

\begin{figure}
\centering
\includegraphics[width=7cm,angle=0]{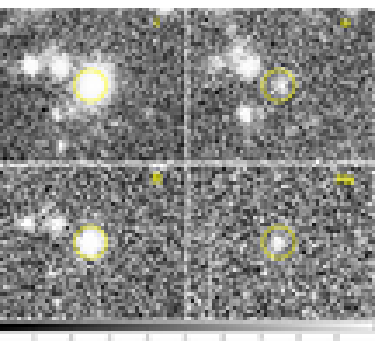}
\caption[]{Same as Fig..1 for galaxy ACO85J004238.03-093229.9}
\end{figure}
     
\clearpage

\begin{figure}
\centering
\includegraphics[width=7cm,angle=0]{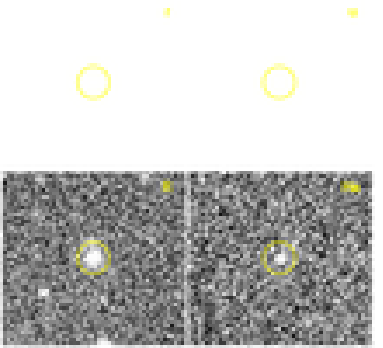}
\caption[]{Same as Fig..1 for galaxy ACO85J004242.24-092108.7}
\end{figure}

\begin{figure}
\centering
\includegraphics[width=7cm,angle=0]{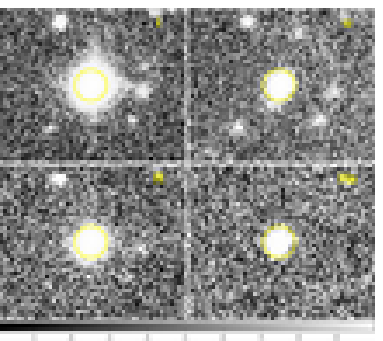}
\caption[]{Same as Fig..1 for galaxy ACO85J004242.54-094726.4}
\end{figure}

\begin{figure}
\centering
\includegraphics[width=7cm,angle=0]{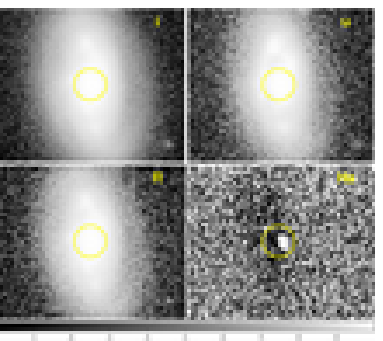}
\caption[]{Same as Fig..1 for galaxy ACO85J004243.90-094420.8}
\end{figure}

\begin{figure}
\centering
\includegraphics[width=7cm,angle=0]{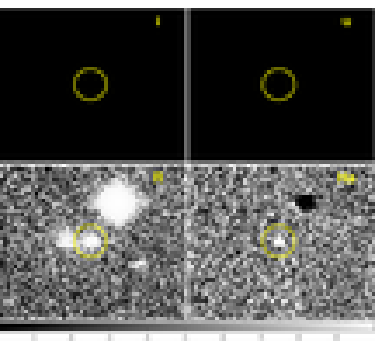}
\caption[]{Same as Fig..1 for galaxy ACO85J004245.68-092327.8}
\end{figure}

\begin{figure}
\centering
\includegraphics[width=7cm,angle=0]{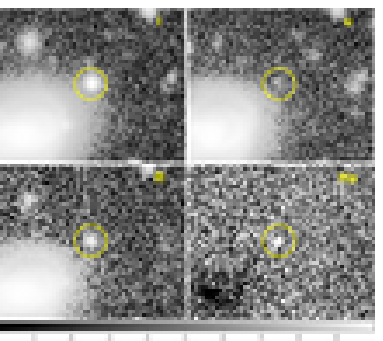}
\caption[]{Same as Fig..1 for galaxy ACO85J004247.85-092522.6}
\end{figure}

\clearpage

\begin{figure}
\centering
\includegraphics[width=6cm,angle=270]{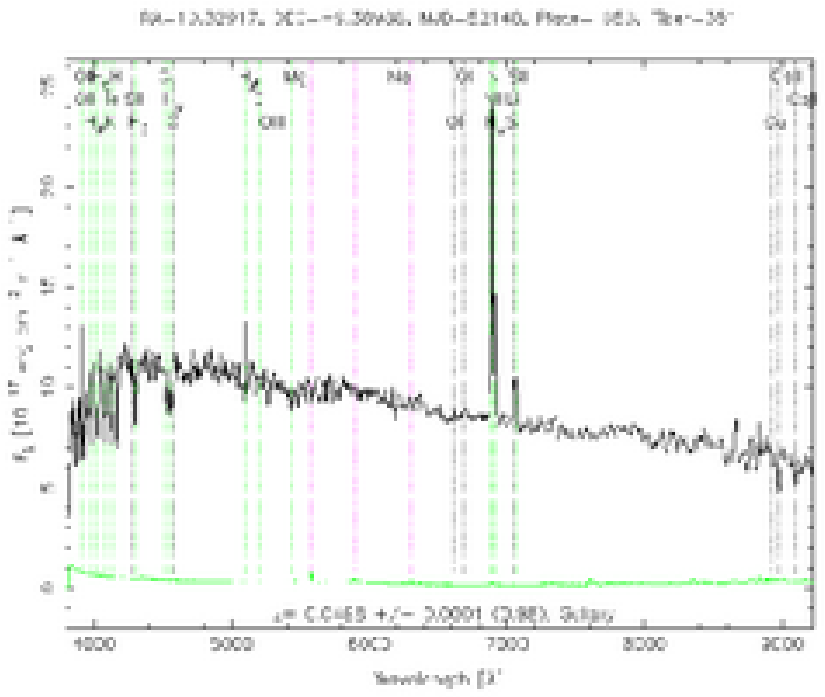}
\caption[]{SDSS spectrum for galaxy ACO85J004119.01-092323.50.}
\label{fig:sdss1}
\end{figure}

\begin{figure}
\centering
\includegraphics[width=6cm,angle=270]{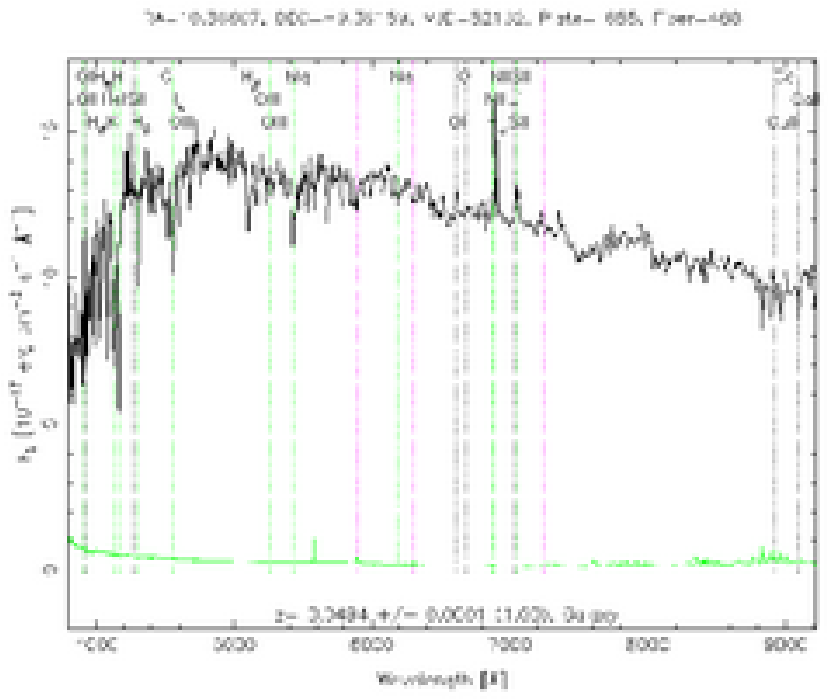}
\caption[]{SDSS spectrum for galaxy ACO85J004127.86-092329.54.}
\label{fig:sdss2}
\end{figure}

\begin{figure}
\centering
\includegraphics[width=6cm,angle=270]{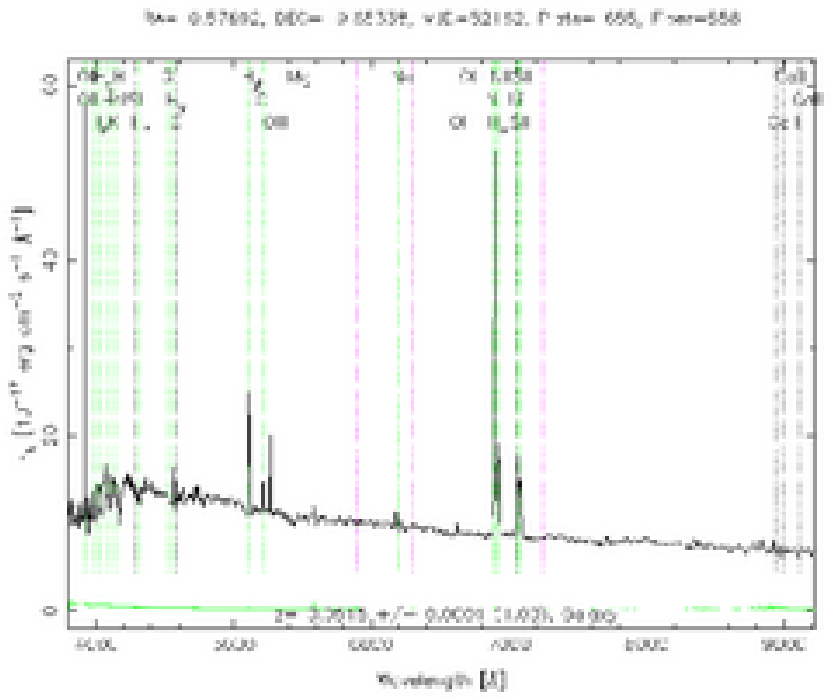}
\caption[]{SDSS spectrum for galaxy ACO85J004218.46-093912.10.}
\label{fig:sdss3}
\end{figure}

\begin{figure}
\centering
\includegraphics[width=6cm,angle=270]{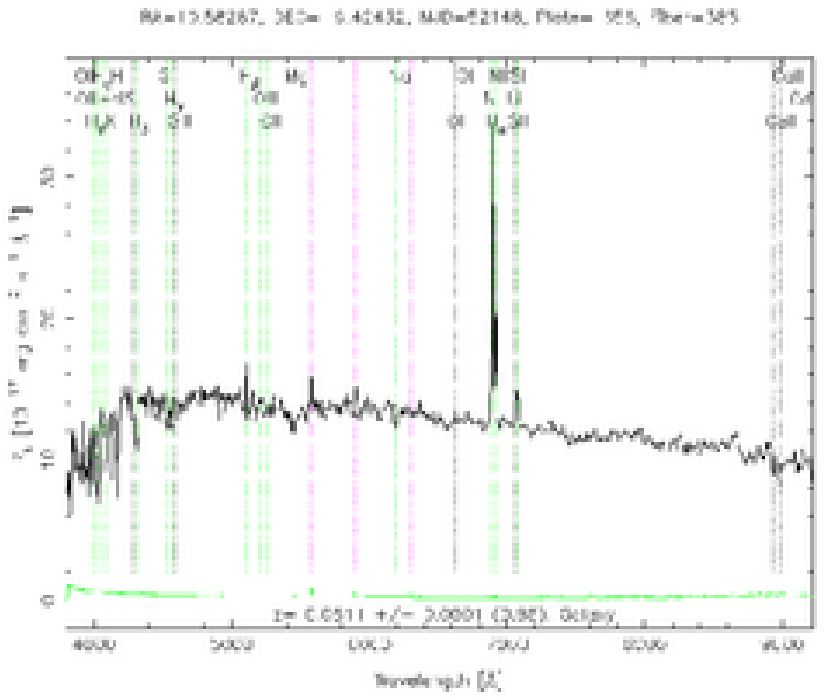}
\caption[]{SDSS spectrum for galaxy ACO85J004219.90-092527.55.}
\label{fig:sdss4}
\end{figure}

\begin{figure}
\centering
\includegraphics[width=6cm,angle=270]{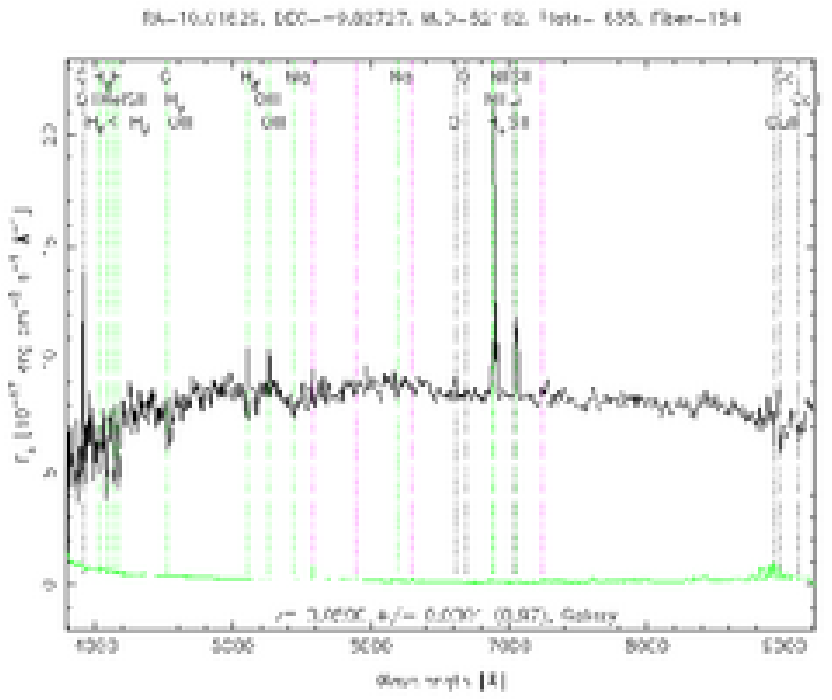}
\caption[]{SDSS spectrum for galaxy ACO85J004228.37-094938.28.}
\label{fig:sdss5}
\end{figure}

\begin{figure}
\centering
\includegraphics[width=6cm,angle=270]{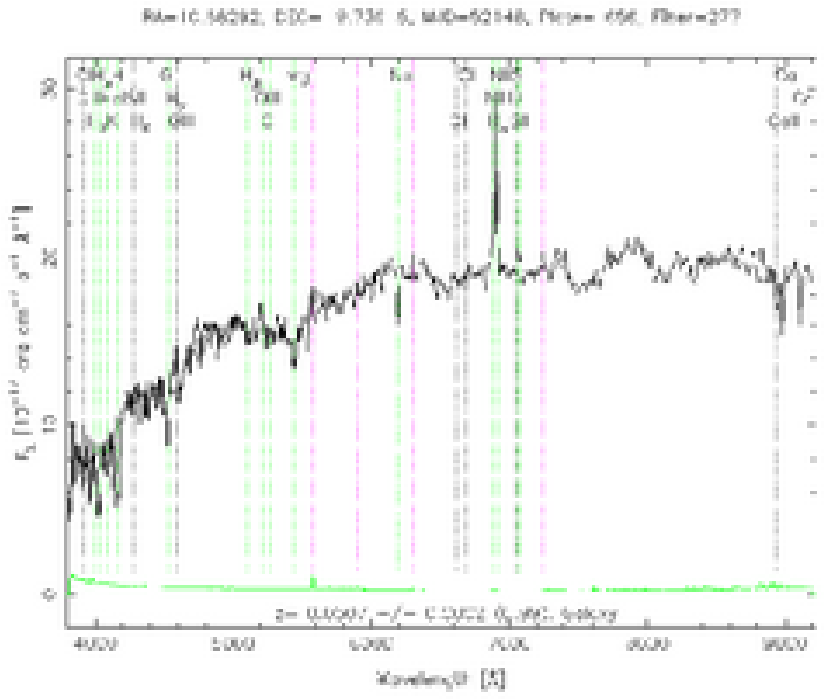}
\caption[]{SDSS spectrum for galaxy ACO85J004243.90-094420.83.}
\label{fig:sdss6}
\end{figure}


Table~3 gives the list of 6 galaxies not detected in our \ha\ image
due to the filter cut but with \ha\ emission in their SDSS
spectrum. It includes the following columns: (1)~ (1)~number, (2)~full
IAU name, (3)~SDSS spectroscopic redshift, (4)~SDSS r magnitude,
(5)~\ha equivalent width in \AA , (6)~error on the \ha equivalent
width in \AA.

\begin{table*}
\centering
\caption{Broad band magnitudes of the 6 galaxies with \ha\ emission in their SDSS spectra
but not detected in our narrow band filter due to its wavelength cut.}
\begin{tabular}{rrrrrr}
\hline
Number & Name  &  $z_{\rm SDSS}$ & $r_{\rm SDSS}$ & EW(\ha) & $\Delta$EW(\ha)\\
       &       &                 &                &  (\AA)  &  (\AA) \\
\hline
     1	& ACO85J004119.01-092323.50 &  0.0498 & 17.37 &   23.9 &    0.4 \\ 
     2	& ACO85J004127.86-092329.54 &  0.0494 & 17.43 &    4.5 &    0.3 \\ 
     3	& ACO85J004218.46-093912.10 &  0.0515 & 17.69 &   64.4 &    2.4 \\ 
     4	& ACO85J004219.90-092527.55 &  0.0511 & 17.42 &   23.4 &    0.3 \\ 
     5	& ACO85J004228.37-094938.28 &  0.0500 & 17.76 &   20.0 &    0.7 \\ 
     6	& ACO85J004243.90-094420.83 &  0.0507 & 16.28 &    7.7 &    0.2 \\ 
\hline
\end{tabular}
\end{table*}

\end{appendix}

\end{document}